\documentclass[showpacs,preprintnumbers,amsmath,amssymb,APSl,prd,nofootinbib,superscriptaddress, twocolumn]{revtex4-2}

\usepackage{dcolumn}
\usepackage{bm}
\usepackage{ifpdf}
\usepackage{hyperref}
\usepackage{bm}
\usepackage{xcolor,color,graphicx,graphics}
\usepackage[spanish,english]{babel}
\usepackage[latin1]{inputenc}
\usepackage[OT1]{fontenc}
\usepackage{latexsym,amssymb,amsmath,amsfonts}
\usepackage{makeidx}
\usepackage{epsfig,subfigure}
\usepackage{natbib}
\usepackage{epstopdf}
\usepackage{mathrsfs}
\hypersetup{colorlinks=true, linkcolor=blue, citecolor=green}
\usepackage{enumerate}
\usepackage{xcolor}
 \usepackage{multirow}
 
\definecolor{red}{rgb}{1,0,0}

\def\+{^\dagger}

\def\<{\leftarrow}
\def\>{\rightarrow}

\def\({\left(}
\def\){\right)}

\newcommand{\bi}{\begin{itemize}} 				\newcommand{\ei}{\end{itemize}}
\newcommand{\benu}{\begin{enumerate}} 		\newcommand{\enu}{\end{enumerate}}
\newcommand{\bd}{\begin{dinglist}{0}}     \newcommand{\ed}{\end{dinglist}}
\newcommand{\bfig}{\begin{figure}[htbp]}  \newcommand{\efig}{\end{figure}}
        			
\newcommand{\bc}{\begin{center}} 				  \newcommand{\ec}{\end{center}}
\newcommand{\be}{\begin{equation}} 				\newcommand{\ee}{\end{equation}}
\newcommand{\bsub}{\begin{subequations}}  \newcommand{\esub}{\end{subequations}}
\newcommand{\ba}[1]{\begin{array}{#1}} 		\newcommand{\ea}{\end{array}}
\newcommand{\bea}{\begin{eqnarray}}
\newcommand{\eea}{\end{eqnarray}}

\begin{document}
\title{Shadows from thin accretion disks of parametrized black hole solutions}

\author{Gonzalo J. Olmo}
\email{gonzalo.olmo@uv.es}
\affiliation{Departamento de F\'{i}sica Te\'{o}rica and IFIC, Centro Mixto Universidad de Valencia - CSIC. Universidad
de Valencia, Burjassot-46100, Valencia, Spain.}
\affiliation{Departamento de F\'isica, Universidade Federal do Cear\'a (UFC),\\ Campus do Pici, Fortaleza - CE, C.P. 6030, 60455-760 - Brazil.}

\author{Jo\~{a}o Lu\'{i}s Rosa}
\email{joaoluis92@gmail.com}
\affiliation{Departamento de F\'isica Te\'orica and IPARCOS,
	Universidad Complutense de Madrid, E-28040 Madrid, Spain}
\affiliation{Institute of Physics, University of Tartu, W. Ostwaldi 1, 50411 Tartu, Estonia}

\author{Diego Rubiera-Garcia} \email{drubiera@ucm.es}
\affiliation{Departamento de F\'isica Te\'orica and IPARCOS,
	Universidad Complutense de Madrid, E-28040 Madrid, Spain}

    \author{Alejandro Rueda}
\email{alerue02@ucm.es}
\affiliation{Departamento de F\'isica Te\'orica and IPARCOS,
	Universidad Complutense de Madrid, E-28040 Madrid, Spain}

\author{Diego S\'aez-Chill\'on G\'omez}
\email{diego.saez@uva.es} 
\affiliation{Department of Theoretical Physics, Atomic and Optics, and Laboratory for Disruptive Interdisciplinary Science (LaDIS), Campus Miguel Delibes, \\ University of Valladolid UVA, Paseo Bel\'en, 7,
47011 - Valladolid, Spain}
\affiliation{Departamento de F\'isica, Universidade Federal do Cear\'a (UFC),\\ Campus do Pici, Fortaleza - CE, C.P. 6030, 60455-760 - Brazil.}

\date{\today}
\begin{abstract}
We discuss the optical appearance from thin accretion disks in parametrized black holes, namely, solutions characterized by an arbitrarily large number of parameters without any regards to the theory of the gravitational and matter fields they come from. More precisely, we consider the leading-order terms of the spherically symmetric Johanssen-Psaltis (JP) and Konoplya-Rezzolla-Zhidenko (KRZ) parametrizations after imposing constraints from asymptotic flatness and solar system observations. Furthermore, we use the inferred correlation, by the Event Horizon Telescope Collaboration, between the size of the bright ring (which is directly observable) and the size of the central brightness depression (which is not) of M87 and Sgr A$^*$ central supermassive objects, to constrain the parameters of the leading-order JP and KRZ solutions. Using ten samples of the Standard Unbound distribution previously employed in the literature to reproduce certain scenarios of General Relativistic HydroDynamical simulations, we produce images of four samples of JP and KRZ geometries enhancing and diminishing the shadow's size, respectively. Via a qualitative and quantitative analysis of the features of the corresponding photon rings and, in particular, of their relative brightness, we argue that it should be possible to distinguish between such parametrized solutions and the Schwarzschild geometry via future upgrades of very long baseline interferometry. We furthermore consider images of some naked objects within these parametrizations, and also discuss the role of inclination in comparing images of different black holes.

\end{abstract}

\maketitle

\section{Introduction}

Generated images using ray-tracing and General-Relativistic MagnetoHydroDynamic (GRMHD) simulations of the light emitted by an accretion disk orbiting a black hole universally report the existence of a bright ring of radiation surrounding a central brightness depression \cite{Vincent:2011wz,Mizuno:2018lxz}. These two features of black hole images, commonly (but also somewhat misleadingly) referred to as the {\it photon ring} and the {\it shadow}, respectively,  offer a new channel to peer into the nature of black holes at the light of the imaging made, by the Event Horizon Telescope (EHT) Collaboration, of the supermassive objects at the center of the M87 \cite{EventHorizonTelescope:2019dse} and Milky Way (Sgr A$^*$) \cite{EventHorizonTelescope:2022wkp} galaxies. Indeed, these observations offer us a great opportunity to probe the strong-field regime of the gravitational interaction \cite{EventHorizonTelescope:2020qrl}, but at the same time represent a challenge involving a precise understanding of light deflection in black hole space-times \cite{Perlick:2021aok}, ray-tracing numerical codes \cite{Cardenas-Avendano:2022csp}, accretion disk physics and modeling \cite{Vincent:2020dij}, image reconstruction, technological upgrades, and so on \cite{Lupsasca:2024wkp}. Very Long Baseline Interferometry (VLBI) projects like the EHT above, its planned upgrade to the next-generation Event Horizon Telescope (ngEHT) \cite{Tiede:2022grp}, or the Black Hole Explorer \cite{Galison:2024bop,Lupsasca:2024xhq}, will provide a flow of data in the decades to come in the characterization of black hole images, supplying us with the suitable tools to tackle such a challenge.

On the theoretical side, a number of theorems guarantee the uniqueness of the Kerr solution of General Relativity (GR) in describing axi-symmetric, rotating black holes characterized solely by their mass and angular momentum \cite{Kerr:1963ud}, and forms the core of the Kerr hypothesis, namely, that every black hole in the Universe, no matter its size or origin, is of Kerr type. In order to test this hypothesis using black hole images from accretion disks (a field frequently dubbed by the community simply as ``shadows" \cite{Falcke:1999pj}) we not only need to check its compatibility with current observations but at the same time we can also pursue alternative geometries (of black hole type or not) in an effort to systematically characterize  every possible image before coming to meet the observations.

Referring to black holes, the uniqueness theorems pose a formidable barrier to any such alternatives, and which can be only overcome when matter fields (scalar, electromagnetic, non-abelian, etc) are allowed into the game. These are typically referred to as {\it hairy black holes}, with the additional matter fields playing the role of hair \cite{Herdeiro:2015waa}. The presence of matter fields allows also for the possibility of horizonless ultra-compact objects to exist, such as wormholes \cite{VisserBook} and boson  and Proca stars \cite{Liebling:2012fv}, among many others, and whose observational status was extensively discussed in \cite{Cardoso:2019rvt}. The imaging of both modified black holes and ultra-compact objects has been investigated far and wide, see e.g. \cite{Eichhorn:2021iwq,Konoplya:2019sns,Kumar:2020owy,Eichhorn:2022oma,Rosa:2022tfv,Pantig:2022gih,Hu:2022lek,Uniyal:2022vdu,Okyay:2021nnh,Yang:2022btw,Rosa:2023qcv,Rosa:2022toh,Rosa:2024eva,Rosa:2024bqv,Rosa:2023hfm,Wang:2023vcv,Macedo:2024qky,KumarWalia:2024yxn,Murk:2024nod,Liu:2024lve,Ovgun:2024zmt}. 

An alternative to the above approach is to accept the possibility that there could be new gravitational physics beyond the one described by GR. The main arguments for this need to enlarge our gravitational theories are both theoretical (unavoidable existence of space-time singularities, lack of ultraviolet completion of GR, or the information loss problem) and observational (the need for dark  sources never observed in the lab in order for the cosmological concordance model to meet observations). This is the field popularly known as {\it modified gravity}, and which has blossomed in the last decade (for reviews see e.g. \cite{DeFelice:2010aj,Olmo:2011uz,Clifton:2011jh,Nojiri:2017ncd,Shankaranarayanan:2022wbx}). However, a practical difficulty arises here in that typically such theories yield much more involved equations of motion preventing the finding of black holes/compact objects of interest by direct attack of them, while known algebraic short-cut and numerical methods within GR are usually not directly applicable. As a consequence, exact solutions of any such theories are scarce, particularly when dealing with axially symmetric solutions. 

An alternative procedure to direct-solving the field equations of modified gravity is the employ of {\it parametrized} solutions, namely, solutions consisting on an (arbitrarily large) number of parameters. Such a procedure is aimed to study large arrays of metrics while remaining agnostic to the specific theory of the gravitational and matter fields that could engender them\footnote{This can be achieved by applying a {\it reverse-engineering} procedure, by which the line element is first set and then the field equations are driven back to find the gravity plus matter fields threading the geometry. This procedure has been employed to great extend in the literature of black hole solutions \cite{Bambi:2023try}.}. One can then constrain the corresponding space of parameters via any set of observations, and later try to tie such constraints into the extra parameters of any theory capable to support such a the metric. In the last few years several such parametrized schemes have been proposed, such as the Johanssen-Psaltis (JP) \cite{Johannsen:2011dh} and {the Konoplya-Rezzolla-Zidenko (KRZ) one \cite{Konoplya:2016pmh} and its several restrictions \cite{Rezzolla:2014mua}. Noting that the imaging of any rotating non-Kerr black hole becomes a tough nut to crack \cite{Gralla:2019drh,Vincent:2020dij}, restricting the problem to spherical symmetry represents a first step towards the detailed characterization of such images. 

The main aim of this work is to simulate images of parametrized, spherically symmetric JP and KRZ black holes by a thin accretion disk. Our goal is to investigate the differences in their photon ring and shadow features as compared to the standard Schwarzschild solution of GR. To do so, we shall select the two lowest-order such solutions in their parametrized series which are compatible with asymptotic flatness and solar system constraints, leaving a single extra parameter. Furthermore, we shall work within the bounds for such an extra parameter derived from the observations of the shadow's size for the supermassive black holes of M87 and Sgr A*, as inferred by the EHT Collaboration. This technique was employed by Vagnozzi et. al. in \cite{Vagnozzi:2022moj} to a great success, placing constrains on dozens of spherically symmetric metrics proposed in the literature. Here we shall push the parameters of both the JP and KRZ solutions to their maximum negative (positive) allowed values, as providing a reduced (enlarged) shadow's size within the EHT bounds. Within each class of solutions we shall discuss the modifications to the features of the photon rings as well as those of the overall images.

This work is organized as follows: In Sec. \ref{C:II} we summarize the theoretical framework for the generation of images, including the null equations of motion in spherically symmetric geometries and the conceptual basis of the photon ring and central brightness depression features of black hole images and their associated quantities for observational purposes. In Sec. \ref{C:III} we build the lowest-order parametrized JP and KRZ geometries  according to the bounds they are constrained to satisfy, and select our four geometries whose generation of images is described in Sec. \ref{C:IV} using an optically and geometrically thin disk whose emission is modeled by ten samples of monochromatic profiles (in the reference frame of the disk) of the Standard Unbound distribution previously employed in the literature. The corresponding images are discussed in Sec. \ref{C:V}, where we elaborate on the chances to distinguish between them and those of the Schwarzschild black hole of GR using the features of the corresponding photon rings. A case of a horizonless compact object within the KRZ family is discussed in Sec. \ref{S:VI} for comparison with black hole images. Furthermore, inclined images for the JP and KRZ black holes are discussed in Sec. \ref{S:VII}. We conclude in Sec. \ref{C:VIII} with some final remarks.

\section{Theoretical framework for the generation of images} \label{C:II}

\subsection{Geodesic equations and ray-tracing procedure}

Black hole images revolve around the ray-tracing of null geodesic trajectories in the background geometry, which for the sake of our work corresponds to that of a spherically symmetric space-time described by the metric 
\begin{equation} \label{eq:linele}
ds^2=-A(r)dt^2+B(r)dr^2+r^2 d\Omega^2,
\end{equation}
where $A(r)$ and $B(r)$ are the two (independent) metric functions characterizing such a space-time, with $d\Omega^2=d\theta^2 + \sin^2 \theta d\phi^2$ the line element on the two-spheres. Consider a particle following a trajectory $x^{\mu}(\lambda)$ (with $\lambda$ the affine parameter labeling such a trajectory) with four-velocity $\dot{x}^{\nu}$ (here a dot denotes a derivative with respect to $\lambda$) in this geometry. The Lagrangian density for a null such particle reads as $2\mathcal{L}=g_{\mu\nu}\dot{x}^{\mu}\dot{x}^{\nu}=0$. Given the time-reversal invariance and spherical symmetry of the system, the Euler-Lagrange equations provide two conserved quantities, namely, $E=-A\dot{t}$ and (taking the motion in the plane $\theta=\pi/2$ without any loss of generality) $L=r^2 \dot{\phi}$, interpreted as the energy and angular momentum, respectively. Furthermore, building the Hamiltonian of the system allows to write the corresponding geodesic equations as (suitably re-absorbing a factor $L^2$ in the definition of the affine parameter)
\begin{equation} \label{eq:geo}
AB \dot{r}^2=\frac{1}{b^2}-V(r),
\end{equation}
where $b \equiv L/E$ is dubbed the impact parameter and
\begin{equation}
V(r)=\frac{A(r)}{r^2}
\end{equation}
is the effective potential of the system. Of particular interest are {\it circular orbits}, namely, those for which $\dot{r}=\ddot{r}=0$. The second condition is equivalent, via Eq.(\ref{eq:geo}), to the critical points of the potential, which explicitly read as (here $A_{ps}\equiv A(r_{ps})$)
\begin{equation}
r_{ps}A_{ps}'-2A_{ps}=0 ,
\end{equation}
whose fulfillment provides the radius of the {\it photon sphere}, $r=r_{ps}$. Furthermore, the first condition above provides the corresponding impact parameter via
\begin{equation} \label{eq:crips}
b_{c}=\frac{r_{ps}}{\sqrt{A_{ps}}}.
\end{equation}
Radial locations such that $\dot{r}=0$ but $\ddot{r} \neq 0$ are {\it turning points}, corresponding to the radial distance at which a light ray is deflected by the black hole. 

For the purposes of the ray-tracing procedure, it is convenient to rewrite the geodesic equation in Eq.(\ref{eq:geo}) in terms of the variation of the azimuthal angle, using the conserved quantities, as
\begin{equation} \label{eq:geoang}
\frac{d\phi}{dr}=\mp \frac{b}{r^2} \sqrt{\frac{AB}{1-\frac{b^2A}{r^2}}}.
\end{equation}
Assuming an observer located at asymptotic infinity, the ray-tracing procedure takes advantage of the time-reversal symmetry of the geodesic equation to backtrack light rays issued from the observer's screen  with an impact parameter $b$ employing the sign $(-)$ in the above equation, providing the deflection angle experienced by it. In turn, this allows us to classify the different trajectories according to the value of their impact parameter as compared to the critical one. In this sense, those light rays with $b>b_{c}$ find a turning point at some location $r_{tp} >r_{ps}$ and the integration of Eq. (\ref{eq:geoang}) is carried out from there back to the observer using the sign $(+)$. On the other hand, those with $b<b_{c}$ never find a turning point in their path, cross the photon sphere, and eventually hit the event horizon. The dividing line between these two cases, $b=b_{c}$, provides a divergent deflection angle in Eq. (\ref{eq:geoang}), in agreement with the circular nature of such orbits at $r=r_{ph}$.

\subsection{Photon rings}

Photons with $b \gtrsim b_{c}$ are of particular relevance from an observational point of view, given the fact that they suffer large deflections in their transit from the vicinity of the black hole to the observer. Indeed such photons can turn several times around the black hole from their emission point in the disk before reaching the observer. Labeling with an integer number $n$ the number of intersections with the equatorial plane of the disk, $n=0$ corresponds to the disk' direct emission, i.e., light rays issued from the disk directly to the observer, while $n=1,2, \ldots$ are highly-bent trajectories  intersecting, successively, the back and front of the disk (because of this, $n$ is usually dubbed as the number of half-turns). Therefore, the optical appearance of a black hole is dominated by a bright ring of radiation corresponding to the direct emission and a number of replications of it labeled by $n$ and nested on it. This sequence comprises the so-called {\it photon ring}, composed of local enhances of the image as given by the additional contributions to the luminosity provided by those photons that travel several half-turns around the disk. 

From the above discussion, it is clear that the photon ring is associated to nearly-bound orbits, that is, trajectories that are slightly displaced from the photon sphere. Consider then a photon that starts its trip at a slightly larger radial distance than the photon sphere radius, $r_o=r_{ps} + \delta r$. The trajectory of such a photon} can be analyzed by perturbing the geodesic equation in Eq. (\ref{eq:geo}) with the result that after a number of half-turns $n$ it will be located at a radial distance given by
\begin{equation}
r=r_o e^{\gamma_{ps} n},
\end{equation}
where $n=[\phi/\pi]$ is identified with the number of half-turns, while the explicit expression for $\gamma_{ps}$ reads as (see e.g. \cite{Kocherlakota:2023qgo}, cast here in a different language)
\begin{equation} \label{eq:lyaexpp}
\gamma_{ps}=\pi \frac{1}{A'_{ps}b_{c} ^{1/2}} \left(A_{ps}^{'2}-2A_{ps} A_{ps}'' \right)^{1/2} ,
\end{equation}
which is dubbed as the lensing {\it Lyapunov exponent}, a quantity setting the instability scale of the system 
\cite{Cardoso:2008bp} and which depends solely on the background geometry. For instance, this exponent equals $\pi$ for the Schwarzschild black hole.

Photon rings are of great relevance in black hole imaging given the tight correlation between their formation, which is due to the background geometry, and their interferometric shape, furthermore blended with the details of the accretion disk physics \cite{Lara:2021zth}, in actual observations  \cite{Johnson:2019ljv,Cardenas-Avendano:2023dzo}. A key aspect regarding photon rings and the Lyapunov exponent comes from the fact that successive photon ring images on the observer's screen satisfy, when viewed face-on, the approximate relations \cite{Kocherlakota:2023qgo}
\begin{equation} \label{eq:scaleprs}
\frac{b_{n+1} - b_{c}}{b_n -  b_{c}} \approx \frac{\omega_{n+1}}{\omega_n} \approx \frac{I_{\nu;n+1}}{I_{\nu;n}} \approx e^{-\gamma_{ps}},
\end{equation}
where $\{\eta_n,\omega_n,I_{\nu;n}\}$ are the $n^{th}$ image radii, width, and flux density, respectively. Therefore, future interferometric prospects, which are expected to measure the $n=1$ photon ring (and hopefully the $n=2$ too) could allow us to run tests on the background geometry via the self-similar scaling of higher-order images above and measurement of the Lyapunov exponent $\gamma_{ps}$. It should be stressed, however, that because a photon with a given impact parameter will not traverse the same region of the disk on each turn, inhomogeneities in the emission profile will induce deviations in the exponential decay of the luminosity with respect to the predictions based on the Lyapunov exponent above. Therefore, while such an index is a useful quantity on a theoretical basis, it must be confronted with actual scenarios of the disk.

\subsection{Shadow}

The second main feature of black hole images corresponds to the central brightness depression generated by the much shorter length-path of those photons hitting the event horizon. Therefore, and following our discussion above, such a brightness depression approaches, in the limit $n \to \infty$, a critical curve corresponding to the projection, on the observer's screen, of the photon sphere, and associated to the usual Falcke's black hole shadow \cite{Falcke:1999pj}. However, for accretion disk models which have gaps in their emission regions (which includes thin and thick disks, but excludes purely spherical models \cite{Vincent:2022fwj}), the fact that there will be emission from inside the photon sphere will produce rings inside the critical curve and will make the central brightness depression to be reduced in size, up to a  minimum theoretical value corresponding to the gravitationally red-shifted image of the event horizon (the so-called inner shadow  \cite{Chael:2021rjo}).

The central brightness depression/black hole shadow carries observational opportunities of its own, and indeed it has led a fruitful discussion on whether it allows one to discriminate between two compact objects via observations \cite{Lima:2021las}. However, the shadow size is not a direct observable of the EHT Collaboration observations given the fact that the latter cannot measure luminosity contrasts below $\sim 10\%$ of the peak luminosity. Instead, the EHT methodology relies on a correlation between the observed angular size of the bright ring of radiation, and the theoretically-computed size of the shadow \cite{EventHorizonTelescope:2022xqj}. This is possible thanks to the fact that the mass-to-distance ratio of Sgr A$^*$ has been studied during decades via stellar cluster dynamics, which in turn allows us to infer the angular radius of the ring of radiation. Additionally, a calibration factor must be introduced accounting for various sources of uncertainties in the bright ring size acting as a proxy for the shadow's size, and which includes those associated to measurement, fitting, model, and emissivity of the plasma, representing one of the criticisms on the validity of this procedure. 

The net result of the above discussion is that the fractional deviation in the EHT's inferred shadow's radius $r_{sh}$ of angular size $\theta$, and the one of the Schwarzschild black hole, $r_{sh,Sch}=3\sqrt{3}M$ of angular size $\theta= 3 \sqrt{3} \theta_g$, as defined by
\begin{equation}
\delta \equiv \frac{r_{sh}}{r_{sh,Sch}}-1 = \frac{b_{c}}{3\sqrt{3}M}-1,
\end{equation}
can be found by combining observational data on the ratio $M/D$ from the Keck and VLTI observations treated as uncorrelated measurements, resulting in \cite{EventHorizonTelescope:2022xqj}
\begin{equation}
\delta \approx -0.060 \pm 0.065.
\end{equation}
This is translated in constraints on the estimated size of the shadow. For our purposes, we take the $2\sigma$-constraints given by
\begin{equation} \label{eq:bounsha}
4.21 \leq \frac{r_{sh}}{M} \leq 5.56,
\end{equation}
which must be compared with the prediction for the shadow's radius of the Schwarzschild geometry, as given by $r_{sh}/M =3\sqrt{3} \approx 5.196$.

\section{Parametrized black hole solutions} \label{C:III}

In the literature there are many samples of spherically symmetric solutions found within the context of modified gravity (either via direct attack of the field equations, through short-cut techniques, or from the application of numerical methods), though axially symmetric solutions are far more scarce. A different approach to this subject is to consider generic solutions that are characterized by an arbitrarily large number of parameters, which can be solutions of different theories of gravity without any need to find them explicitly, being instead {\it postulated}. Therefore, any constraints on such parametrized black hole metrics to represent current observations (in particular, those of the shadow's size by the EHT above), potentially becomes a constraint on any theory of gravity capable of engendering them.

The lowest-order coefficients of a parametrized solution can be fixed according to weak-field limit tests such as those carried out in the solar system. Expanding the metric of a spherically symmetric object of a metric theory of gravity in series of $M/r$ (where $M$ is the ADM mass of the space-time) one gets \cite{Will:2014kxa}
\begin{eqnarray}
ds^2 &=&-\left(1-\frac{2M}{r} + 2(\beta - \gamma)\frac{M^2}{r^2} + \ldots \right)dt^2 \nonumber \\
&+& \left(1+2\gamma \frac{M}{r} + \ldots \right)dr^2 +r^2 d\Omega^2, \label{Eq:PPN}
\end{eqnarray}
where $\beta$ and $\gamma$ are dimensionless parameters. Lunar Laser Ranging (LLR) experiment \cite{Williams:2004qba} together with the Cassini experiments \cite{Bertotti:2003rm} set the constraint 
\begin{equation}
\vert \beta - \gamma \vert \leq 2.3 \times 10^{-4},
\end{equation}
implying tight constraints on the parameters of any such solution, as we shall see at once. For the sake of this work we consider two parametrizations which have raised great interest in the literature (for an alternative parametrization, dubbed as Rezzolla-Zhidenko \cite{Rezzolla:2014mua} and an orthogonal analysis to our own,  see \cite{Kocherlakota:2023qgo}).

\subsection{JP parametrization}

The Johanssen-Psaltis (JP) parametrization introduced in \cite{Johannsen:2011dh} is based on a spherically symmetric line element of the form
\begin{equation}
ds^2=-f_S (1+h_{JP}) dt^2 + \frac{(1+h_{JP})}{f_S}dr^2 +r^2 d\Omega^2,
\end{equation}
where $f_S=1-2M/r$ is Schwarzschild metric function (which fixes the functions $A$ and $B$ in the line element in Eq. (\ref{eq:linele})), and $h_{JP}(r)$ can be expressed as an infinite series of the form
\begin{equation}
h_{JP}(r)=\sum_{k=0}^{\infty} \epsilon_k \left(\frac{M}{r}\right)^k,
\end{equation}
where $\epsilon_k$ are a set of coefficients parameterizing the metric. Here $\epsilon_0=0$ must be imposed in order to guarantee asymptotic flatness of the space-time, the condition $\epsilon_1=0$ is required for $M$ to represent the correct asymptotic (ADM) mass of the space-time, while $\vert \epsilon_2 \vert \leq 4.6 \times 10^{-4}$ by LLR experiment constraints means that this parameter can be approximated to zero at all effects. This way, the first non-trivial term in this parametrization becomes
\begin{equation}
h_{JP}(r)=\frac{M^3}{r^3},
\end{equation}
and we denote $\epsilon_3 \equiv \epsilon$ for simplicity. Performing in such a case a series expansion of the metric function for large distances, $r \to \infty$,  we get 
\begin{equation}
-g_{tt} \approx 1-\frac{2M}{r} + \epsilon \frac{M^3}{r^3} + \mathcal{O}(r^{-4}),
\end{equation}
and this way we regard every term beyond the $1/r^3$ as negligible\footnote{We point out that this argument is correct if we deal with compact objects like black holes because, for objects like the Sun, $M/R_\odot \sim 10^{-6}$, implying that $M^3/r^3$ is, at least, $10^{-6}$ times smaller than $M^2/r^2$.}. An interesting feature of these lowest-order JP black holes is that the event horizon is located at the same Schwarzschild radius as the GR case, $r_h=2M$. This is of obvious interest for the purpose of the cast images given the fact that the surface delineating the central brightness depression would be at the same location in both cases, facilitating the comparison of images on as an equal footing as possible. Note that for $\epsilon <-8$ these lowest-order JP black holes have a second surface outside the event horizon that holds a space-like singularity (as given by the blow up of curvature scalars) so we shall not consider that range of parameters.

One can study the evolution of the shadow's radius $r_{sh}$ for the JP parametrization, getting to the conclusion that the compatible values of JP black holes with the EHT bounds in Eq.  (\ref{eq:bounsha}) lie (approximately) within the range
\begin{equation} \label{eq:JPrange}
\epsilon \in [-5,11.4].
\end{equation}
the lowest (largest) bound corresponding to the maximal (minimal) allowed shadow's radius by the EHT Collaboration. 

\subsection{KRZ parameterization}

The Konoplya-Rezzolla-Zhidenko (KRZ) parametrization was introduced in Ref. \cite{Konoplya:2016pmh} and for the sake of this work we take a restriction of it for the non-rotating case, suitably cast via the line element
\begin{equation}
ds^2=-f_{KRZ}dt^2 + \frac{dr^2}{f_{KRZ}} +r^2 d\Omega^2,
\end{equation}
where
\begin{equation}
f_{KRZ}=1-\frac{2M(1+h_{KRZ})}{r},
\end{equation}
and
\begin{equation}
h_{KRZ}=\frac{1}{2} \sum_{k=0}^{\infty}  \eta_k \left(\frac{M}{r} \right)^k.
\end{equation}
As opposed to the JP parametrization, the KRZ one is always asymptotically flat (i.e. for any values of its coefficients), yet we take the condition $\eta_0=0$ to ensure consistence of the parameter $M$ with the asymptotic mass, and $\eta_1=0$ to satisfy the LLR constraint. This way, the first non-negligible term in this parametrization is
\begin{equation}
h(r)=\frac{1}{2} \eta \frac{M^2}{r^2},
\end{equation}
and we denoted $\eta \equiv \eta_2$. Performing again series expansions in the limit $r \to \infty$ we get 
\begin{equation}
-g_{tt} \approx 1-\frac{2M}{r} - \eta \frac{M^3}{r^3} + \mathcal{O}(r^{-4}),
\end{equation}
which is the same expansion as in the JP parametrization but with the sign of the new term reversed. 

\begin{figure*}[t!]
\includegraphics[width=4.3cm,height=4.3cm]{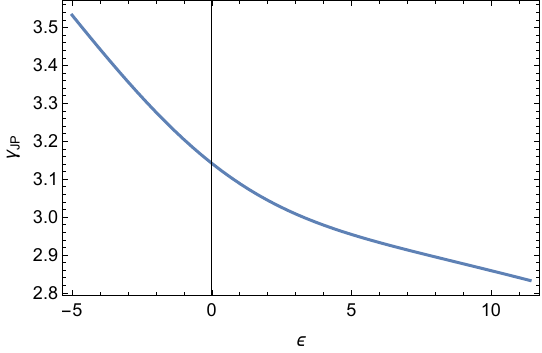}
\includegraphics[width=4.3cm,height=4.3cm]{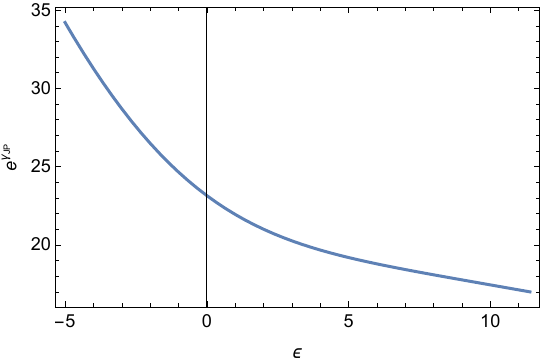}
\includegraphics[width=4.3cm,height=4.3cm]{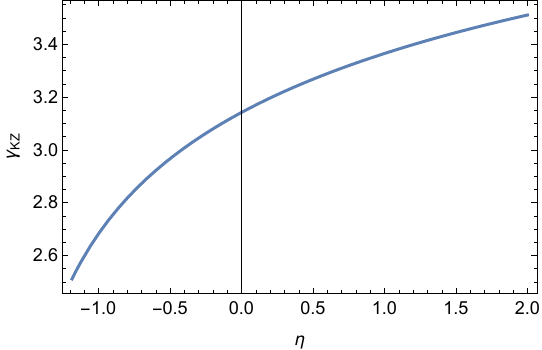}
\includegraphics[width=4.3cm,height=4.3cm]{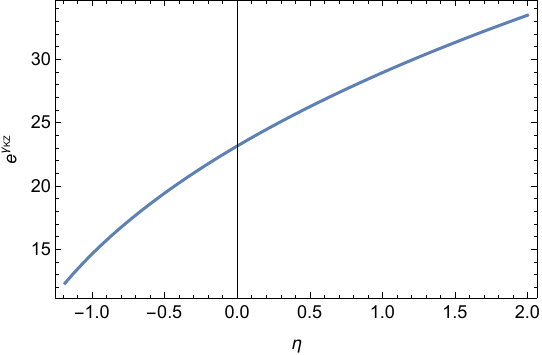}
\caption{The Lyapunov exponent and its exponential version of nearly-bound geodesics for the JP (two left figures) and KRZ (two right figures) solutions as a function of their respective parameters. The range of parameters is taken to fulfil the intervals in Eqs. (\ref{eq:JPrange}) and (\ref{eq:KRZrange}). In their respective ends, the exponential-Lyapunov exponent (associated to the ratio of intensities between photon rings, see Eq.(\ref{eq:scaleprs})) varies more than a factor two. }
\label{fig:Lya}
\end{figure*}

The lowest-order, non-trivial KRZ black holes above have their event horizons at different locations than their Schwarzschild counterpart. Furthermore, such a horizon disappears for $\eta < -32/27$, before the highest bound in Eq. (\ref{eq:bounsha}) can be reached in the shadow's size, leaving a kind of naked object. Indeed, assuming black holes consistent with such bounds' limit the space of parameters of KRZ black holes is restricted to
\begin{equation} \label{eq:KRZrange}
\eta \in [-32/27, 2],
\end{equation}
the lowest (largest) bound corresponding to the minimal (maximal) allowed shadow's radius by the EHT Collaboration, i.e., the opposite as in the JP parametrization.

\section{Imaging parametrized black holes} \label{C:IV}

In this section we describe our setting to generate images of parametrized JP and KRZ black holes, including the picks of geometrical configurations, the assumptions for the disk's model, and the numerical code employed. 

\subsection{Choice of configurations}

On the background geometry side, we shall use the following four parametrized solutions:
\begin{itemize}
\item JPn: JP black holes with $\epsilon=-5$.
\item JPp: JP black holes with $\epsilon=11.4$.
\item KRZn: KRZ black holes with $\eta=-32/27$.
\item KRZp: KRZ black holes with $\eta=2$.
\end{itemize}

The reason for our choices is that these are the parametrized black holes which push the shadow's constraints to their limits (upwards and downwards), this way enhancing our chances to see differences in their cast images as compared among themselves, and also to the Schwarzschild black hole. To see this, in Fig. \ref{fig:Lya} we depict the Lyapunov exponent in Eq. (\ref{eq:lyaexpp}) as well as its exponential version, given the impact of the latter in the features of successive photon rings as stressed in Eq. (\ref{eq:scaleprs}),  for the JP and KRZ solutions as functions of their respective parameters. Their ends (for both signs of their parameters) in this figure signal their maximum and minimum values within the shadow's bounds, corresponding to the JPn, JPp, KRZn, and KRZp solutions. As seen in this plot, the exponential version of the Lyapunov exponent varies more than a factor two between both ends, which will have a reflection in the features of the corresponding photon rings in their cast images.

\subsection{Disk's model}

The radiative transfer (Boltzmann) equation weaves together intensity, emissivity, and absorptivity of the fluid making up the accretion disk \cite{Gold:2020iql}. Typically, General Relativistic MagnetoHydroDynamic (GRMHD) simulations need to fix such features (besides many other features of the fluid) in order to simulate images of the black hole under specific scenarios for the accretion flow.  Here, and aligned with many other studies in the literature, we shall significantly short-cut the complexity of this process, since our interest is to compare the features of the images of parametrized black holes between themselves and also against those of the Schwarzschild one in as a controlled setting as possible. We therefore assume a disk which is\footnote{It should be pointed out that in our simplified modeling we shall also neglect Doppler beaming and relativistic shift effects, both tightly tied to accretion disk's rotation and responsible for the observed asymmetric brightness of M87 and Sgr A$^*$ by the EHT Collaboration. While these effects can in fact be neglected for face-on images which are the main consideration of this work (which would appear as circular and uniformly bright) this effect should play a role in the inclined images considered later in Sec. \ref{S:VI}, and could actually be employed as a way to compute the disk's orientation \cite{Medeiros:2021apx}. For simplicity of our analysis and computational times, we shall not incorporate these effects in our setting.}
\begin{itemize}
\item Optically thin (i.e. absolute zero opacity).
\item Infinitesimally-thin geometrical shape. 
\end{itemize} 

The first assumption allows light rays to revolve several half-times around the disk, which is required to have photon rings in the observer's screen; otherwise photons would be re-absorbed when crossing again the accretion disk, and only the disk's direct emission would be available. While, in practical terms, optical thinness is not absolute, and bearing in mind that every photon that lingers for too long within it will be eventually re-absorbed after a finite number of half-turns, the observational capabilities of future projects of VLBI  technology are not expected to be able to resolve more than the $n=1$ (and possibly the $n=2$) photon ring. Therefore, for the sake of our analysis we shall conform up to the $n=2$ ring\footnote{In the influential paper \cite{Gralla:2019xty} the authors refer to the contributions $n=0$, $n=1$, and $n=2$ emissions as direct, lensing, and photon ring emissions, respectively. We shall keep the word direct emission for $n=0$, but shall refer to the contributions $n=1$ and $n=2$ as first and second photon rings instead.}.

As for the second assumption, it might seem to be far too restrictive given the fact that real disks have a non-vanishing thickness. However, it has been discussed in the literature that the infinitely-thin approximation quite resembles (qualitatively and quantitatively) the images of realistic simulations with thick disks \cite{Vincent:2022fwj}, while the fully spherical symmetric accretion flow is regarded as a setting too constrained to match the features of M87 and Sgr A$^*$ observations. 

\begin{figure*}[t!]
\includegraphics[width=3.5cm,height=3.5cm]{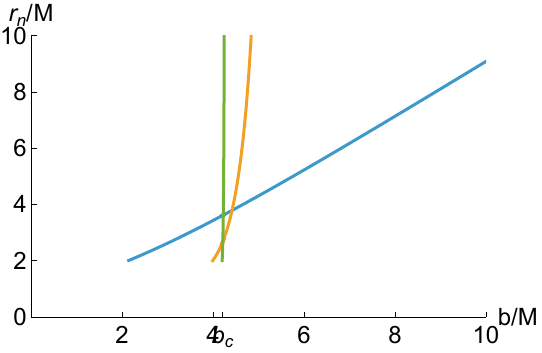}
\includegraphics[width=3.5cm,height=3.5cm]{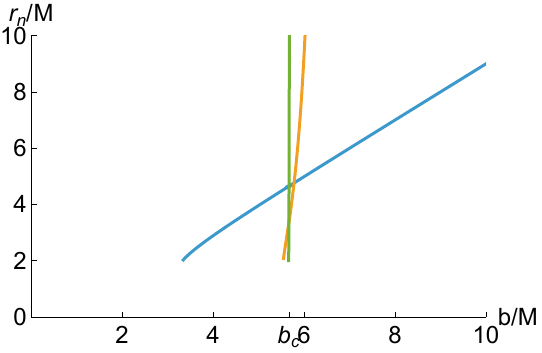}
\includegraphics[width=3.5cm,height=3.5cm]{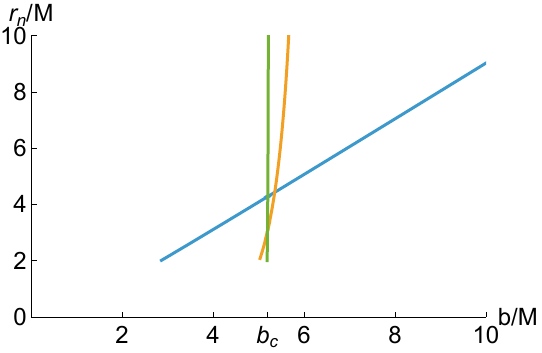}
\includegraphics[width=3.5cm,height=3.5cm]{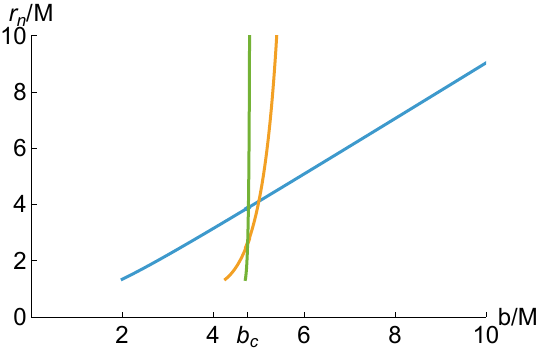}
\includegraphics[width=3.5cm,height=3.5cm]{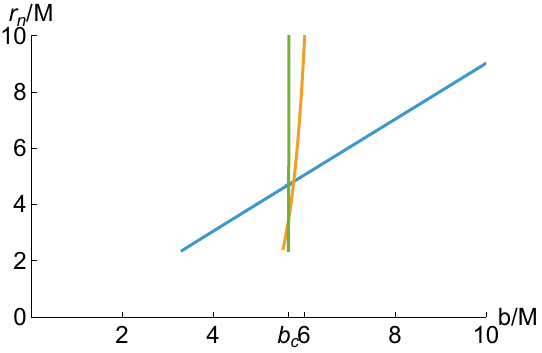}
\caption{The transfer function $r_n$ as a function of $b$ (in units of $M$) for (from left to right) JPn, JPp, Schwarzschild, KRZn, and KRZp geometries. On each plot we depict three curves corresponding to the $n=0$ (blue), $n=1$ (orange) and $n=2$ (green) contributions to the image. The two geometries producing the transfer functions in the far left and and far right enhance the shadow's size, while the two left and right next to the central (Schwarzschild) curves diminish it.}
\label{fig:trans}
\end{figure*}

Under these conditions, Liouville's theorem tells us about the conservation of the invariant intensity ratio, that is
\begin{equation} \label{eq:Lio}
\frac{I_{\nu_e}}{\nu_e^3} = \frac{I_{\nu_o}}{\nu_o^3},
\end{equation}
where $\nu_e$ and $\nu_o$ are the photon's frequency in the emission and observer's frames, respectively, while $I_{\nu_e}$ and $I_{\nu_o}$ are the corresponding intensities.  To generate our images, we assume a monochromatic emission in the disk's frame, that is, $I_{\nu_e} \equiv \delta(\nu-\nu_e) I(r)$. This is different from what actual EHT observations do, in the sense that they work at a fixed (230Gz) frequency in the observer's frame. However, our purpose in this work is not to match real EHT images, but to compare the differences between the expectations based on a Schwarzschild black hole and those of the parametrized JP and RZ black holes when pushed to their limits in their compatibility with the shadow's size within as a simplified setting as possible to precisely extract its photon ring features.

Bearing this discussion in mind, Eq.(\ref{eq:Lio}) can be written as
\begin{equation}
I_{\nu_0}= g^3 I_{\nu_e},
\end{equation}
where 
\begin{equation}
g=\frac{\nu_o}{\nu_e}= \frac{A^{1/2}(r)}{A^{1/2}(r \to \infty)},
\end{equation}
where we have used Eq.(\ref{eq:linele}) to relate each frequency, while the asymptotic character of the space-time allows to fix $A(r \to \infty) \to 1$, so $g=A^{1/2}(r)$. Integrating over the observed frequency we find
\begin{equation}
I_{ob}(r)= \int d \nu_o I_{\nu_o} = \int (g d \nu_e) (g^3 I_{\nu_e}) = g^4 I(r),
\end{equation}
where in the last equality we have used the monochromatic nature of the intensity in the observer's frame.

Next we need to add up all the photons that, in the optically-thin assumption, contribute to the image. Given the expected exponential suppression of luminosity between successive photon rings, Eq.(\ref{eq:scaleprs}), for our purposes  we cut our simulations at the $n=2$  and thus write the final observed intensity as
\begin{equation}
I_{ob}= \sum_{n=0}^2 \xi_n A^2(r) I(r)\Big \vert_{r=r_n(b)}
\end{equation}
In this equation $\xi_n$ is dubbed the {\it fudge factor}, and its role is to correct our computation of the intensity  by taking into account the fact that our model neglects the non-vanishing thickness of the disk, which actually appears in GRMHD simulations of time-averaged images. In agreement with previous analyses in the literature, we shall assume (though this choice is not universal) this factor to be $\xi_0=1$ for the direct emission, $\xi_1=\xi_2=1.5$ for the first and second photon rings, and $\xi_n=0$ for higher-order ($n > 2$) photon rings. It should be stressed that this choice does not change the relative luminosity between the $n=1$ and $n=2$ photon rings, but instead the boost of luminosity of both photon rings as immersed in the direct emission. On the other hand, the intensity is evaluated at the radii $r=r_n(b)$, which is dubbed as the transfer function. It corresponds to the locations on the thin disk at which a photon with a given impact parameter $b$ hits it, found via the backwards ray-tracing procedure of individual and bunches of light rays. These locations appear as two-dimensional plots for three curves, corresponding to light rays that crossed it just once (the direct emission) or more (the photon ring emissions), all of them represented in the same plot.  This function therefore {\it transfers} the information on impact factors of issued light rays on the observer's screen with radial distances at where the disk is emitting, a critical step towards imbuing luminosities in the latter for the generation of images in the former.

In Fig. \ref{fig:trans} we depict the transfer function $r_n(b)$ for the direct ($n=0$, blue), $n=1$ (orange), and $n=2$ (green) photon ring emissions for (from left to right) the JPn, JPp, Schwarschild, KRZp, and KRZn geometries. Each of these curves corresponds to the variation in the radial location of each emission within its impact parameter window, which is narrowed as higher-order photon rings are considered. Furthermore, each successive curve has a greater slope, and given the fact that the latter is associated to the degree of demagnification for each curve on the corresponding images (see the account on the analytic analysis on this and other subjects of Ref. \cite{Perlick:2021aok}), this correlates with what we already know of the exponentially decreasing contribution of successive light trajectories to the image. On the other hand, comparing the shapes of these curves among different background geometries, one sees that there are non-negligible differences in the locations, spatial separation to each other, and slopes. This will have an impact on the features of the corresponding images, as we shall see below.

\subsection{Choice of emission profiles}

\begin{figure}[t!]
\includegraphics[width=9.1cm,height=5.5cm]{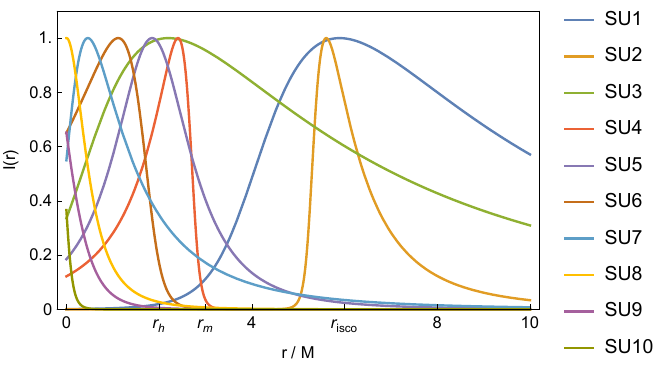}
\caption{The intensity $I(r)$ for the SU profiles employed in this work, set for the Schwarzschild space-time (in units of $M$), normalized to their maximum intensity. In this plot $r_{ISCO}, r_{ps}, r_h$ denote the location of the innermost stable circular orbit, the photon sphere unstable orbit, and the event horizon, respectively (values taken for the Schwarzschild black hole in units of $M=1$, i.e. $r_{ISCO}=6,r_{ps}=3,r_h=2$).}
\label{fig:profiles}
\end{figure}

The next ingredient of our analysis is the choice of emission profiles. Again we anchor our choices in the reproduction of GRMHD simulations in the literature, which currently gravitate around the family of suitable adaptation of Standard Unbound (SU) models 
\begin{equation}
I_{SU}(r,\mu,\theta,\gamma)=\frac{e^{-\frac{1}{2} [\gamma + \text{arcosinh}(\frac{r-\mu}{\theta})]^2}}{\sqrt{(r-\mu)^2+\sigma^2}} \ .
\end{equation}
As we can see, these models are characterized by three parameters that control the location of the emission peak, $\mu$,  its width, $\theta$, and its asymmetry, $\gamma$. Within this family of emission profiles, the equatorial approximation is assumed, that is, all of the emission originates from the equatorial plane, while the effects on the luminosity of the photon rings caused by the fact that light rays can spend additional time in the off-equatorial, high-emission region of the thick disk can be accounted for using the fudge factors mentioned above. 
Furthermore, it is typically assumed that the emitting source follows circular Keplerian orbits until the innermost stable circular orbit (ISCO), after which they plunge into the black hole following Cunninghan's prescription \cite{Cunningham:1975zz}. These approximations and assumptions have been proven useful in reproducing the actual structure of photon rings in GRMHD simulations via surveys of different SU profiles, see e.g. \cite{Gralla:2020srx,Paugnat:2022qzy,Cardenas-Avendano:2023dzo}. Such surveys demonstrate that both the effective region of emission, i.e. the location of the peak (from the event horizon to the ISCO) and how spread the emission is, are critical aspects on the finer details of  the simulations. In this sense, those profiles whose emission increases monotonically as the horizon is approached are regarded as more suitable to reproducing M87 observation \cite{Gralla:2020srx}, while those peaking near the ISCO with little emission inside can be regarded as challenging models against the currently accepted models.

For the sake of our work we shall take a pool of SU models interpolating between the two cases above, all of which reproduce the two main features in black hole images, namely, the photon rings and the shadow. In Fig. \ref{fig:profiles} we depict the  intensity (normalized to their maximum values) for ten different choices of these parameters\footnote{The explicit values for the parameters used here correspond to those displayed in the Table appearing in Ref. \cite{DeMartino:2023ovj}.}, and organize the resulting sequence of emission models according to (roughly) the location of the peak of emission from the farthest (SU1 model) to the nearest (SU10 model). Such models were extracted from a pool of hundreds of simulations in the literature, including several popular ones like those employed by Gralla, Lupsasca and Marrone in \cite{Gralla:2020srx} (in the non-rotating limit considered here), and picked in order to find qualitatively and quantitatively different images. We classify these models into three main categories. Models SU1 and SU2  peak near the innermost stable circular orbit (ISCO) radius of time-like particles with very different spread (far larger for SU1). Models SU3, SU4, and SU5 peak between the event horizon and the photon sphere radius. Furthermore we can classify these models according to how spread its emission is on its external part (i.e. outside the event horizon); in this sense there are ``sharp" profiles (SU2, SU4) and ``wide" profiles (SU1, SU3, SU5). Finally, models SU6 to SU10 all show a monotonically decreasing behavior outside the event horizon radius. Furthermore, their corresponding intensities at the event horizon are lower as one goes through this sequence of models. We also point out that the models SU2, SU7 and SU8 were already employed in \cite{daSilva:2023jxa} (under the names GLM3, GLM1, and GLM2, respectively) to characterize photon rings in a bunch of spherically symmetric geometries. 

It should be stressed that the fact that for JP and RZ parameterized solutions the quantities characterizing the images are different as compared to the Schwarzschild black hole (photon sphere radius and critical impact parameter, though the event horizon is the same for the JP one) troubles their comparison on equal-footing. This is a common and seemingly unavoidable difficulty within this approach to this problem.

\subsection{Generation of images}

There are several geometrical quantities characterizing black hole images from a theoretical point of view, namely, the event horizon radius, the photon sphere radius and the critical impact parameter, as well as the Lyapunov exponent of photon rings. In Table \ref{Tab:geom} we display the values of such quantities for our four picks of solutions introduced at the beginning of this Section. Recall that two of them enhance the shadow's radius (JPn and KRZp), and the other two reduce it (JPp and KRZn). For completeness we include the corresponding values of these geometrical features for the Schwarzschild geometry, as the benchmark all these solutions are compared to.

\begin{table}[]
\begin{tabular}{|c|c|c|c|c|c|}
\hline
 & JPn & JPp & Sch & KRZn  & KRZp \\ \hline
Parameter & -5.0 & 11.4 & 0 & -32/27 & 2  \\ \hline
$r_{h}$ & 2 & 2 &  $2$ & 1.333 & 2.359 \\ \hline
$r_{ps}$ & 3.313 & 2.634 &  $3$ & 2.541 & 3.425 \\ \hline
$b_{c}$ &  5.666 & 4.213 &  $3\sqrt{3} \approx 5.196$ & 4.758 & 5.659 \\ \hline
$\gamma_{ps}$ & 3.532  &2.832  & $\pi \approx 3.141$ & 2.511 &  3.510 \\ \hline
\end{tabular}
\caption{Theoretical quantities charactering the features of the JPn, JPp, Sch, KRZn, and KRZp solutions, namely, horizon radius $r_h$, photon sphere radius $r_{ps}$ and critical impact parameter $b_{c}$ in Eq.(\ref{eq:crips}), and Lyapunov exponent $\gamma_{ps}$ in Eq.(\ref{eq:lyaexpp}). All quantities in units of $M=1$.}
\label{Tab:geom}
\end{table}

From the computational side, we run our simulations with a version of our GRAVITYp code based on Mathematic$@$, using ten thousand light rays covering a range of impact parameters, $b \in [0,10]$, and a further set of ten thousand light rays closely tracking the photon ring, $b \in [(1-10^{-2})b_{c}, (1+10^{-2}) b_{c}]$. The observer's screen is located at a distance of $r_o=1000M$, a distance at which all geometries are flat to any practical effect. We consider two sets of simulations, a first set where we consider face-on (i.e. zero inclination between the axis of the disk and the observer's line of sight) images, and in Sec. \ref{S:VI} a second set where we consider inclined observations at $80^\circ$ angles.

\section{Interpretation and discussion of face-on black hole images} \label{C:V}

\subsection{Role of the emission models on the images}

\begin{figure*}[t!]
\includegraphics[width=4cm,height=3.2cm]{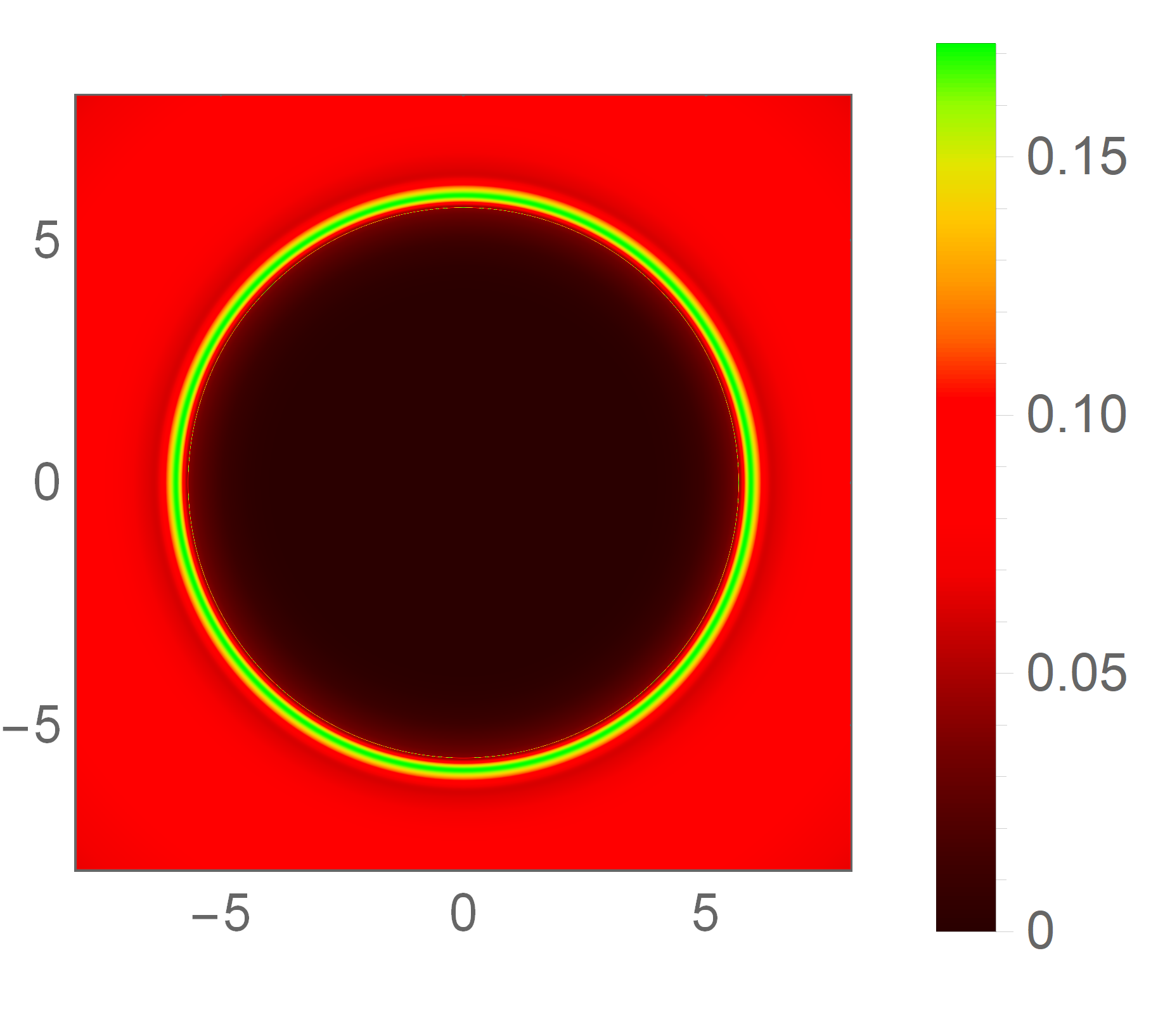}
\includegraphics[width=4cm,height=3.2cm]{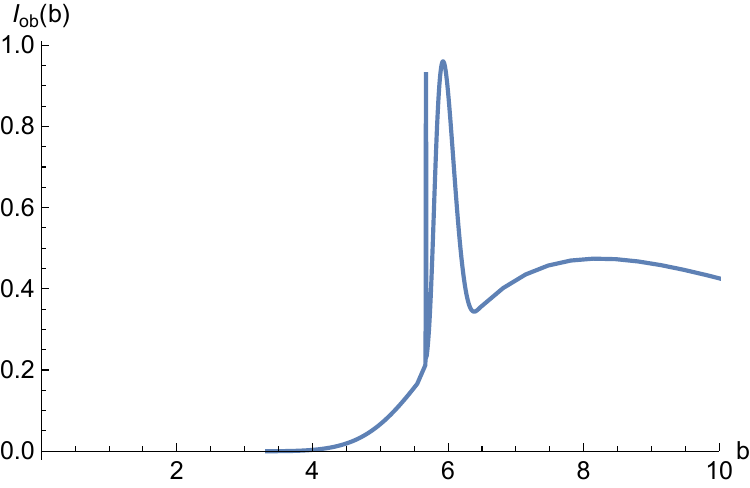} 
\includegraphics[width=4cm,height=3.2cm]{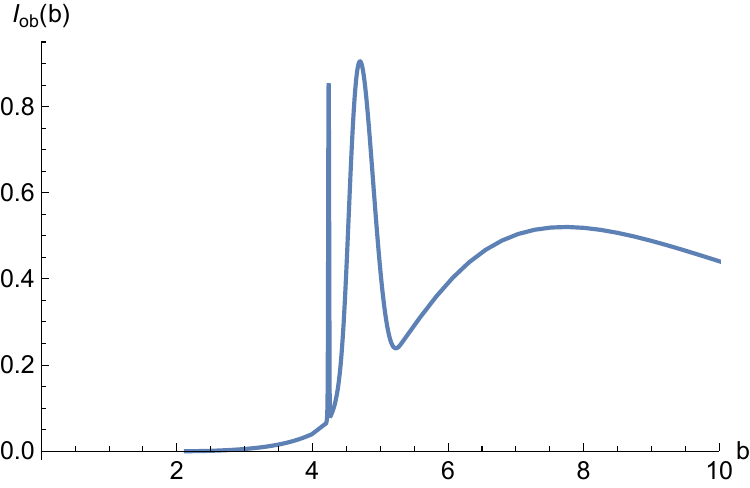}
\includegraphics[width=4cm,height=3.2cm]{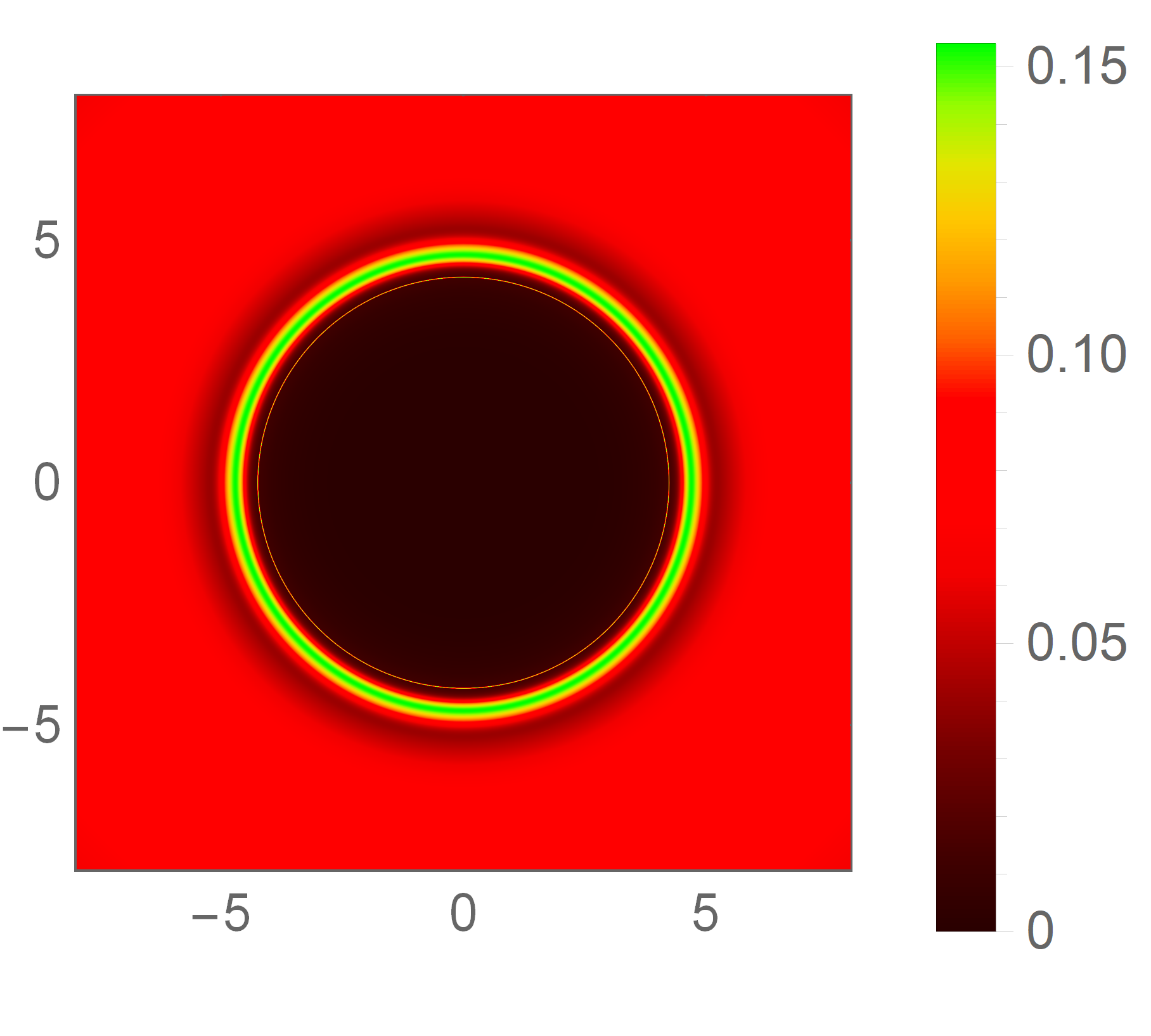} \\
\includegraphics[width=4cm,height=3.2cm]{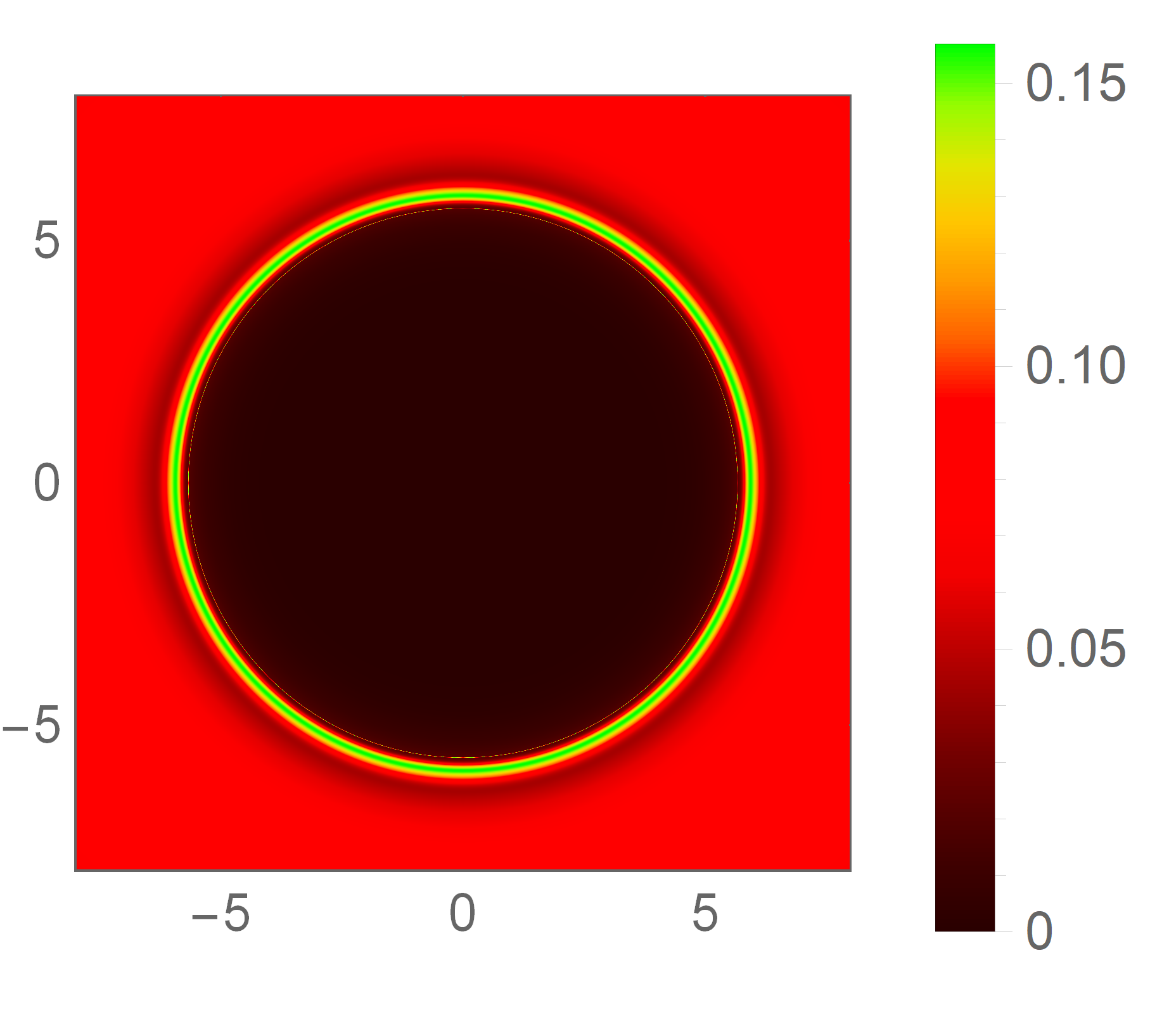}
\includegraphics[width=4cm,height=3.2cm]{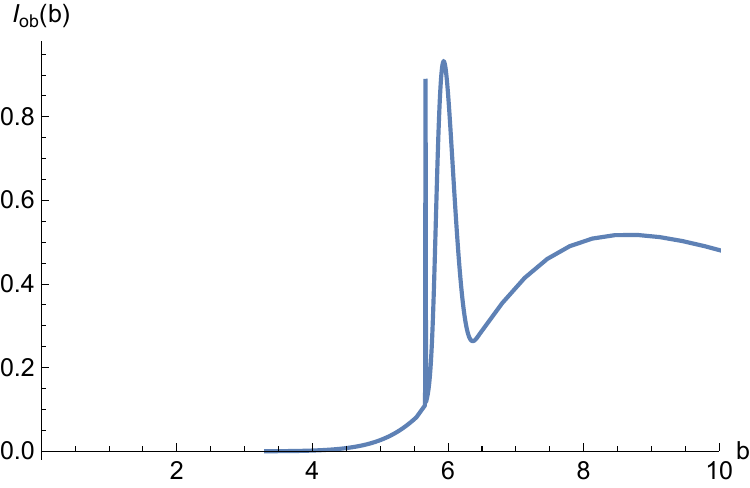}  
\includegraphics[width=4cm,height=3.2cm]{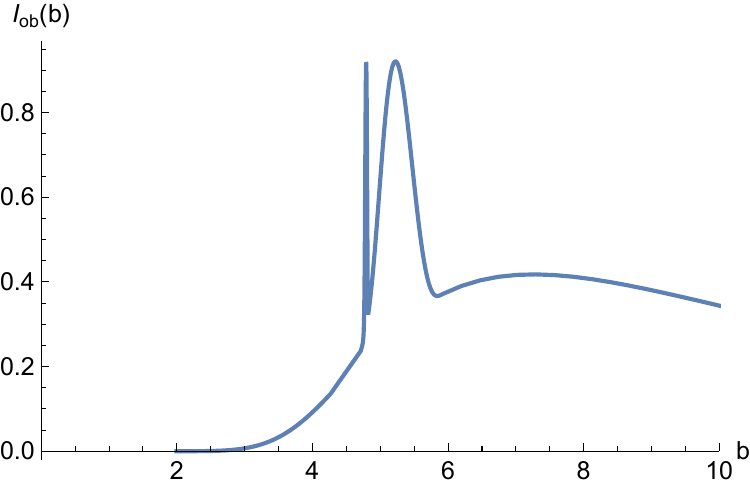}
\includegraphics[width=4cm,height=3.2cm]{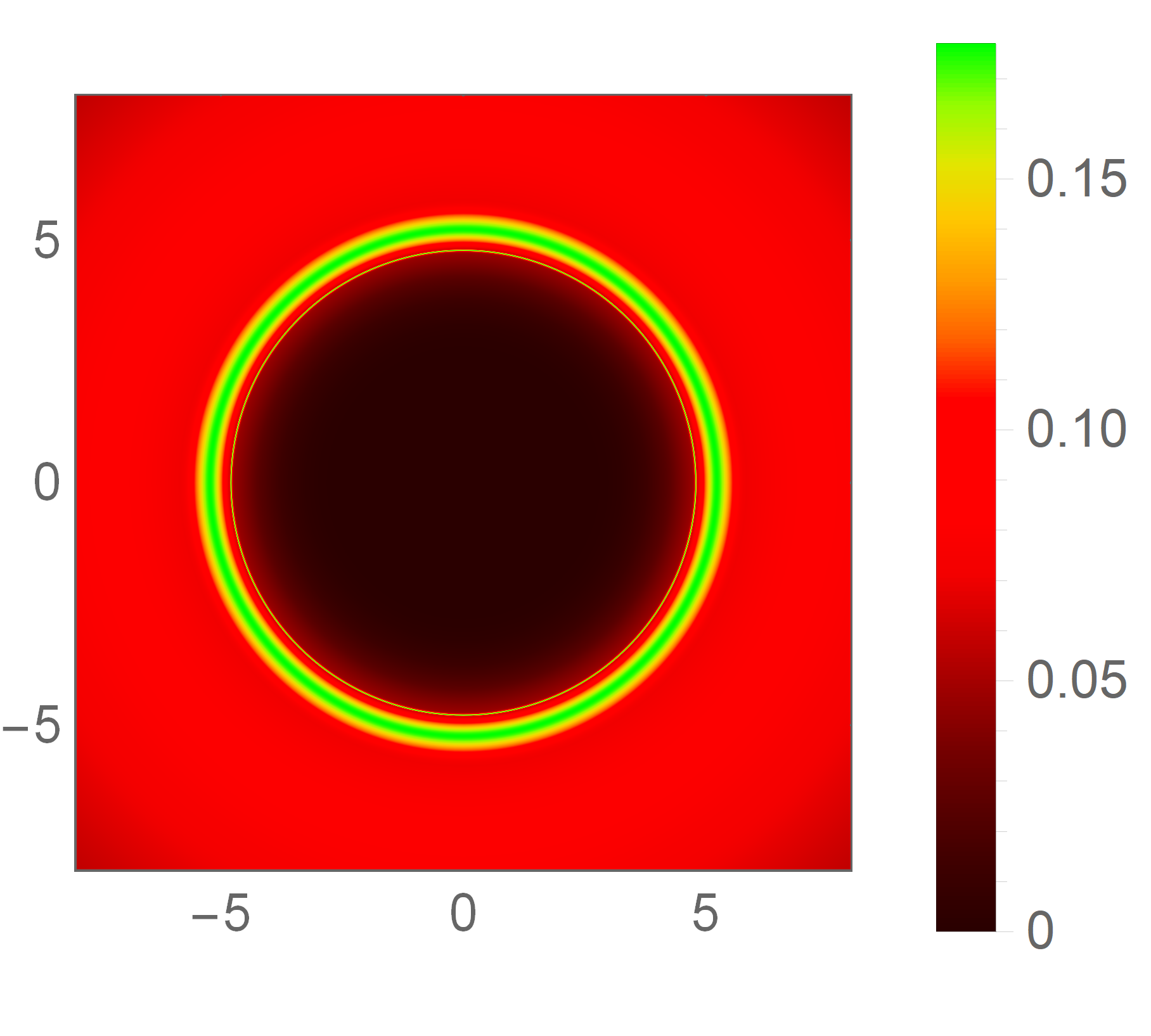}
\caption{The optical images (left and right figures) and the observed intensity $I_{ob}(b)$ (middle figures) for the JPn (left top figures), JPp (right top figures), KRZp (bottom left figures) and KRZn (bottom right figures) for the SU1 emission model.}
\label{fig:SU1}
\end{figure*}

\begin{figure*}[t!]
\includegraphics[width=4cm,height=3.2cm]{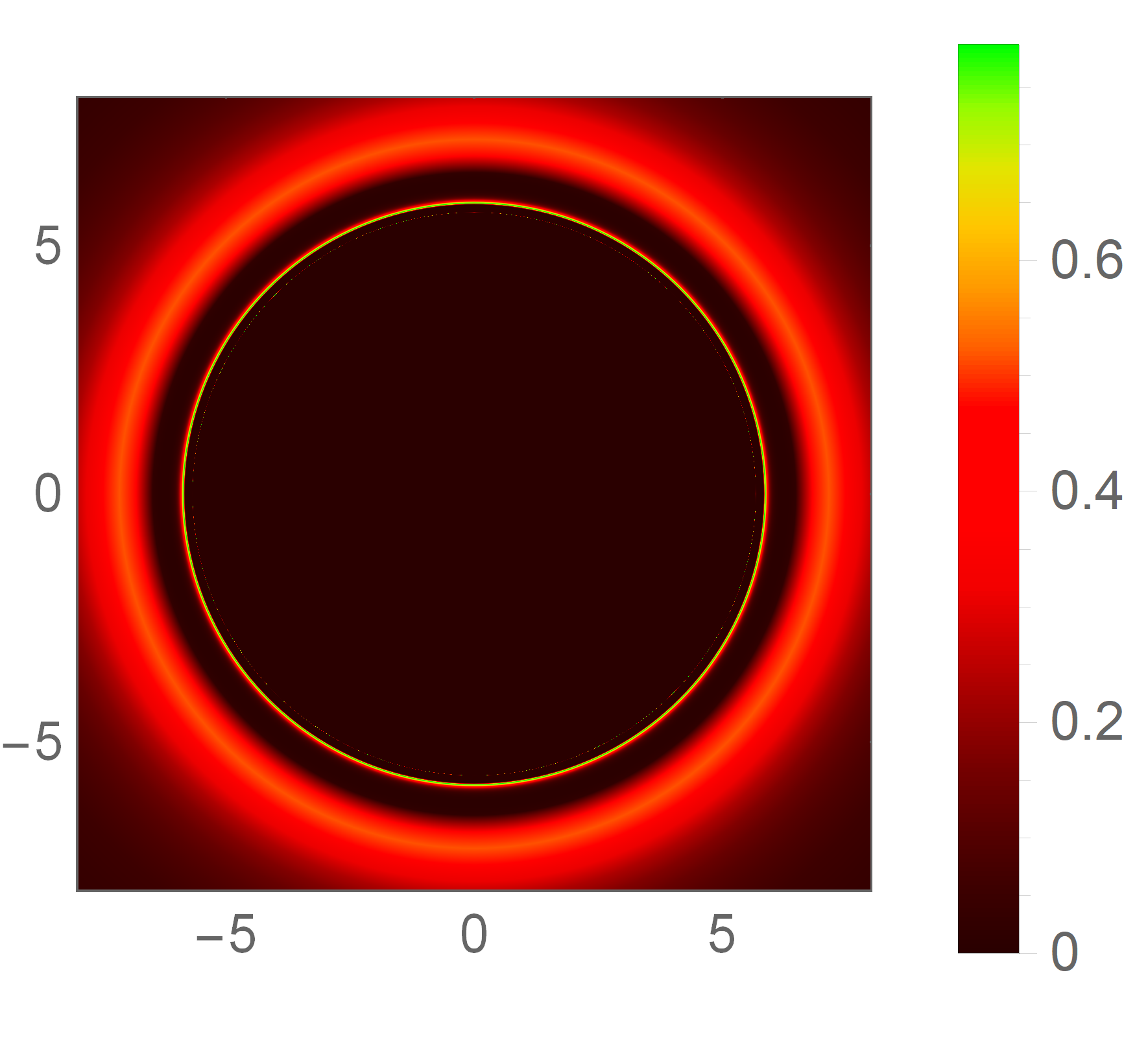}
\includegraphics[width=4cm,height=3.2cm]{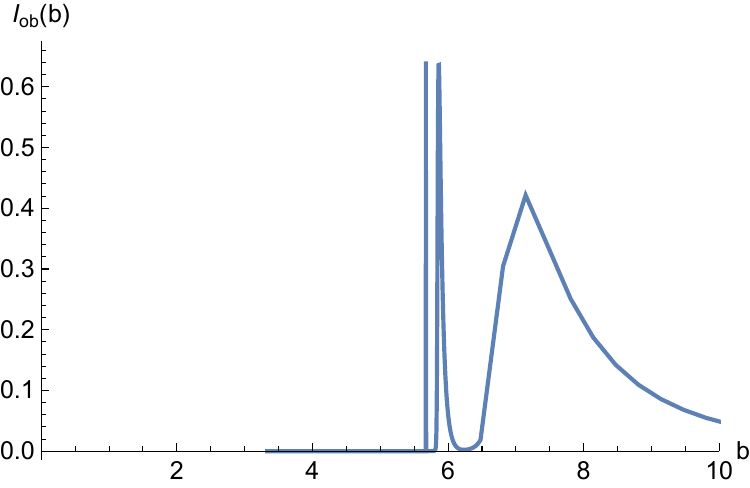} 
\includegraphics[width=4cm,height=3.2cm]{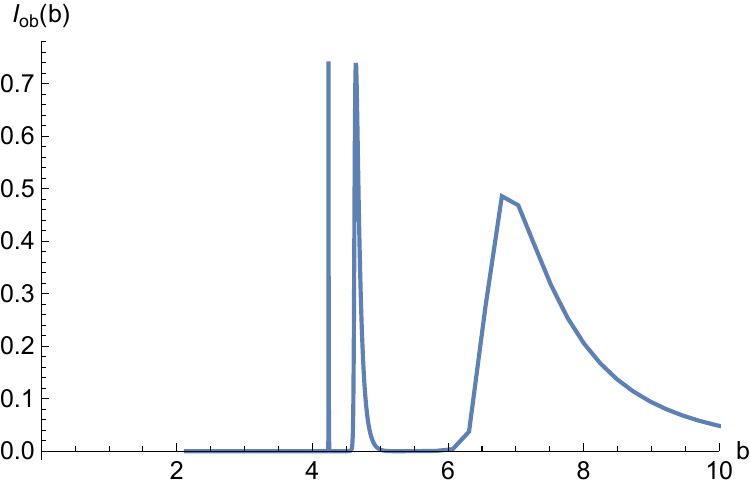}
\includegraphics[width=4cm,height=3.2cm]{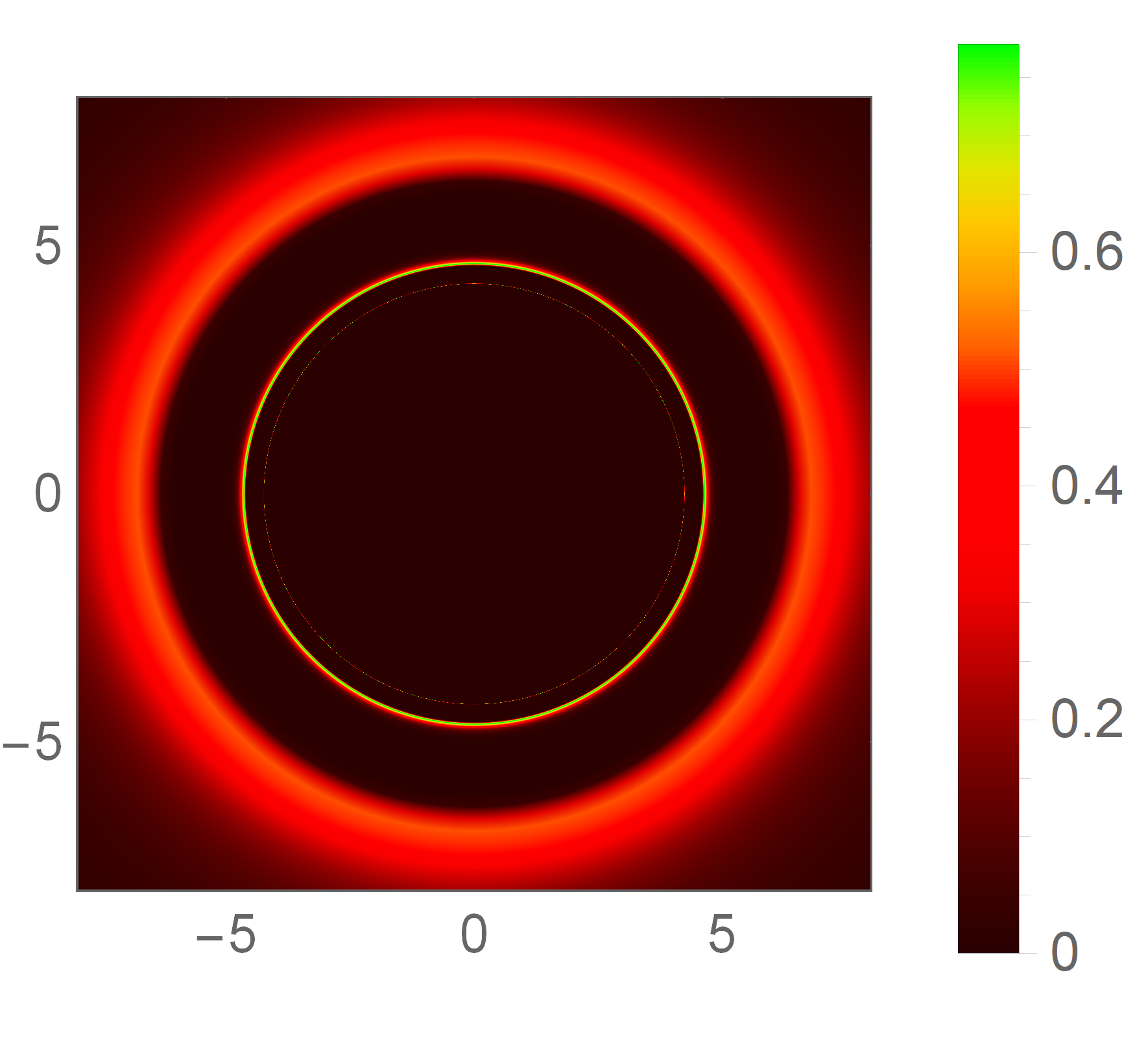} \\
\includegraphics[width=4cm,height=3.2cm]{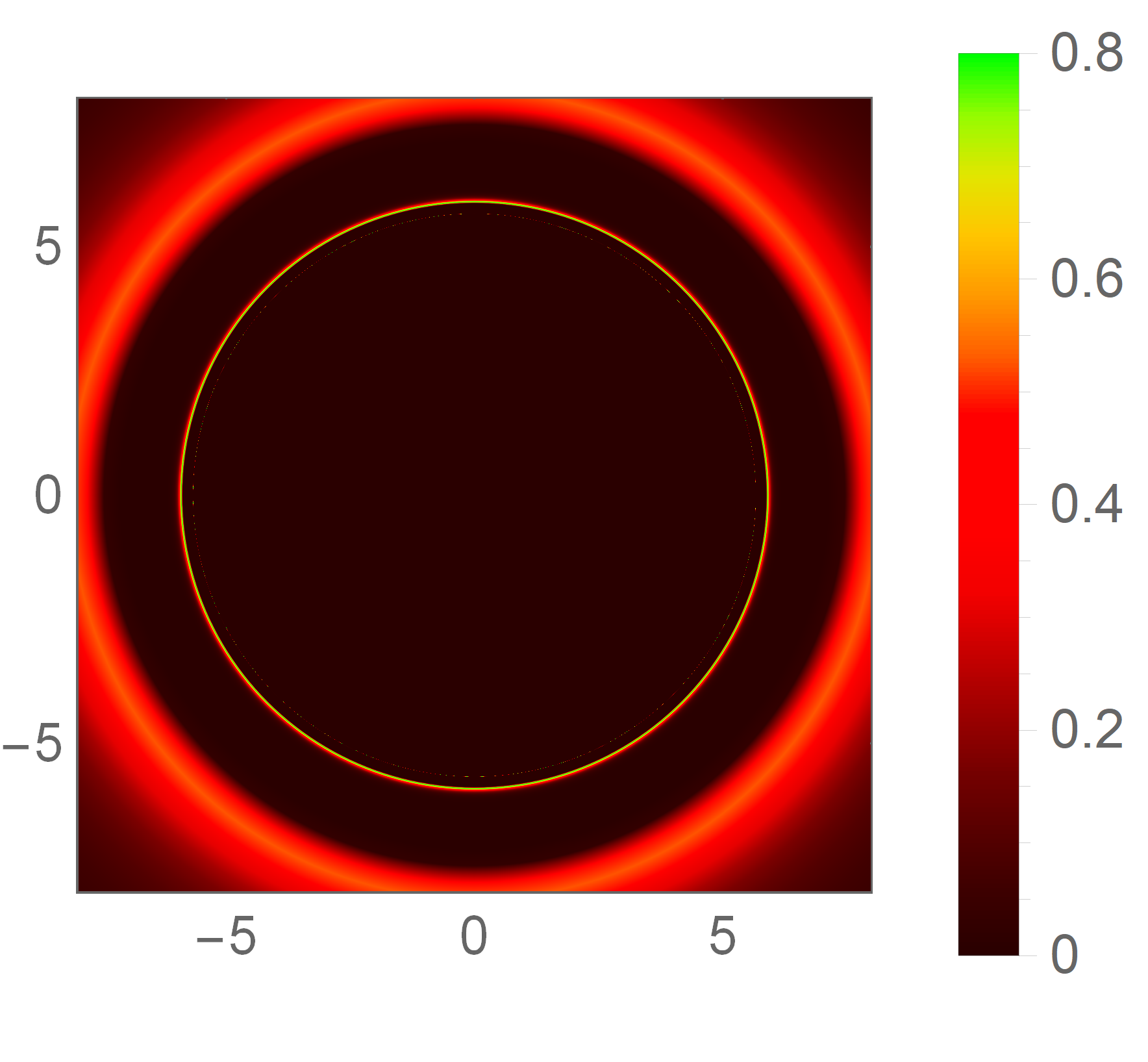}
\includegraphics[width=4cm,height=3.2cm]{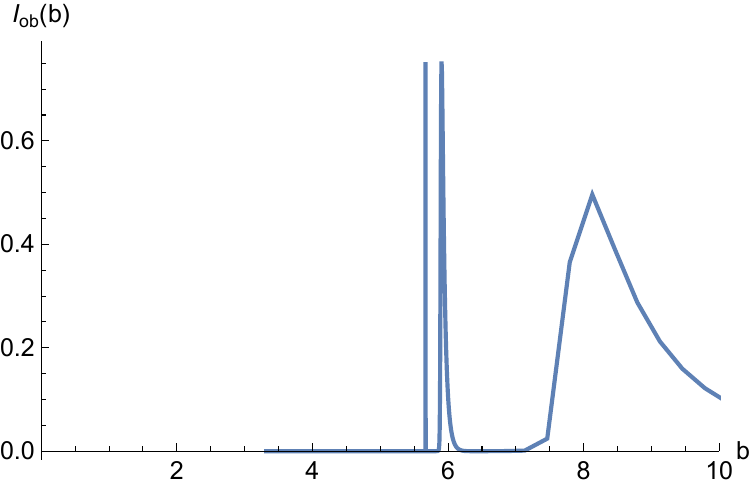}  
\includegraphics[width=4cm,height=3.2cm]{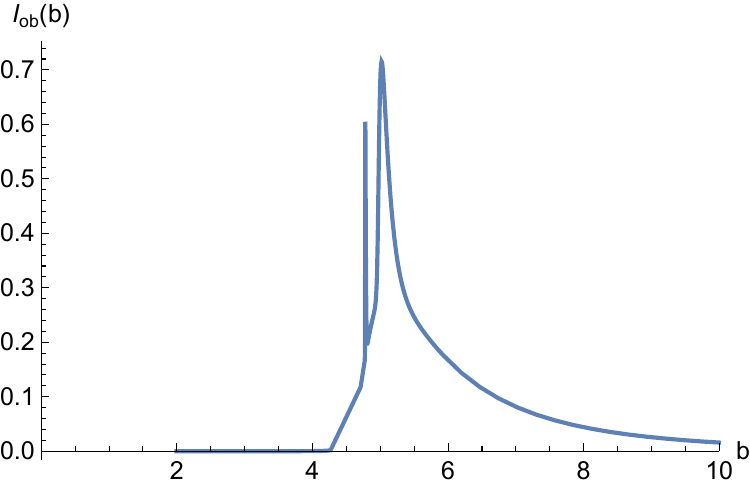}
\includegraphics[width=4cm,height=3.2cm]{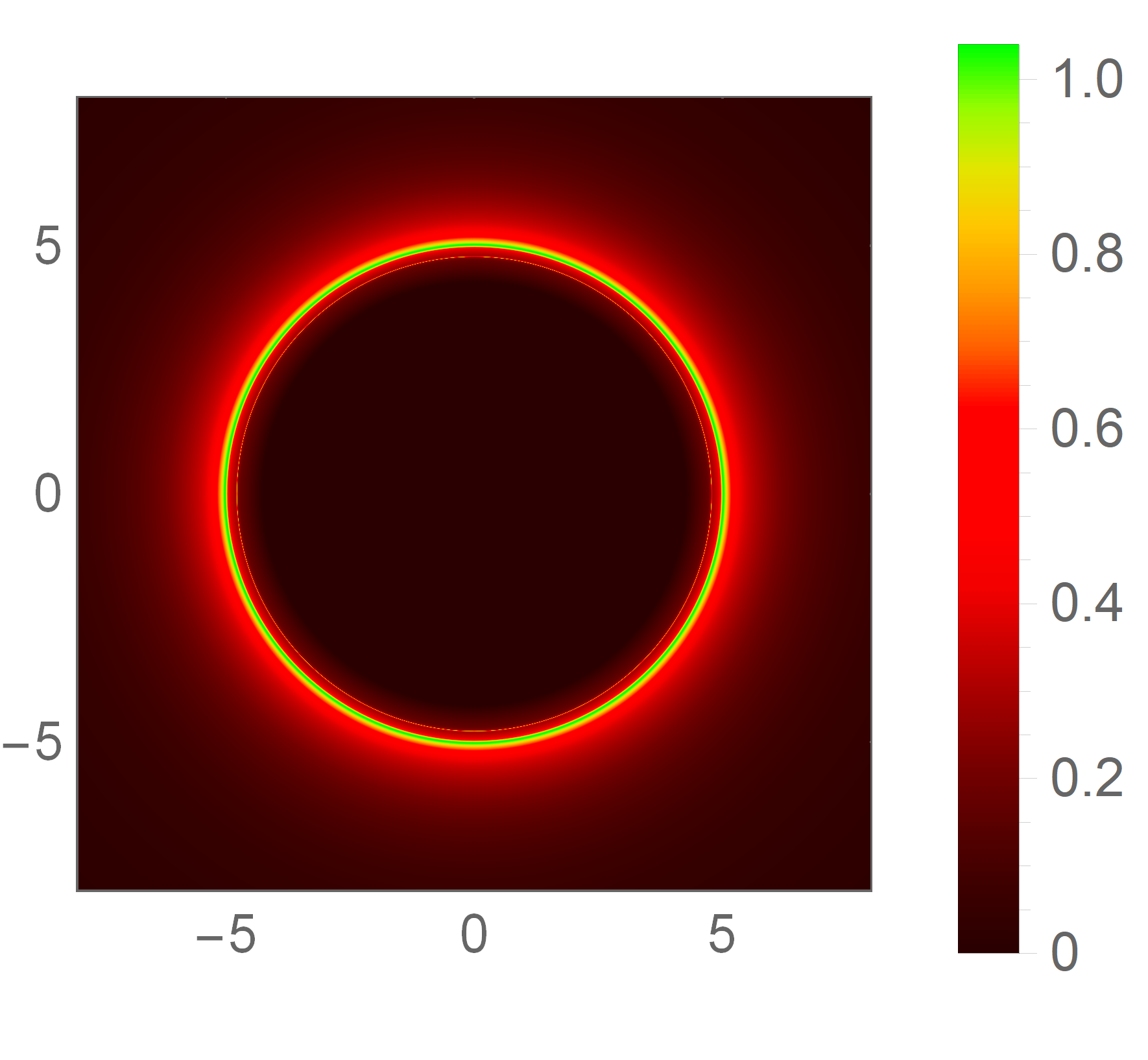}
\caption{The optical images (left and right figures) and the observed intensity $I_{ob}(b)$ (middle figures) for the JPn (left top figures), JPp (right top figures), KRZp (bottom left figures) and KRZn (bottom right figures) for the SU2 emission model.}
\label{fig:SU2}
\end{figure*}

\begin{figure*}[t!]
\includegraphics[width=4cm,height=3.2cm]{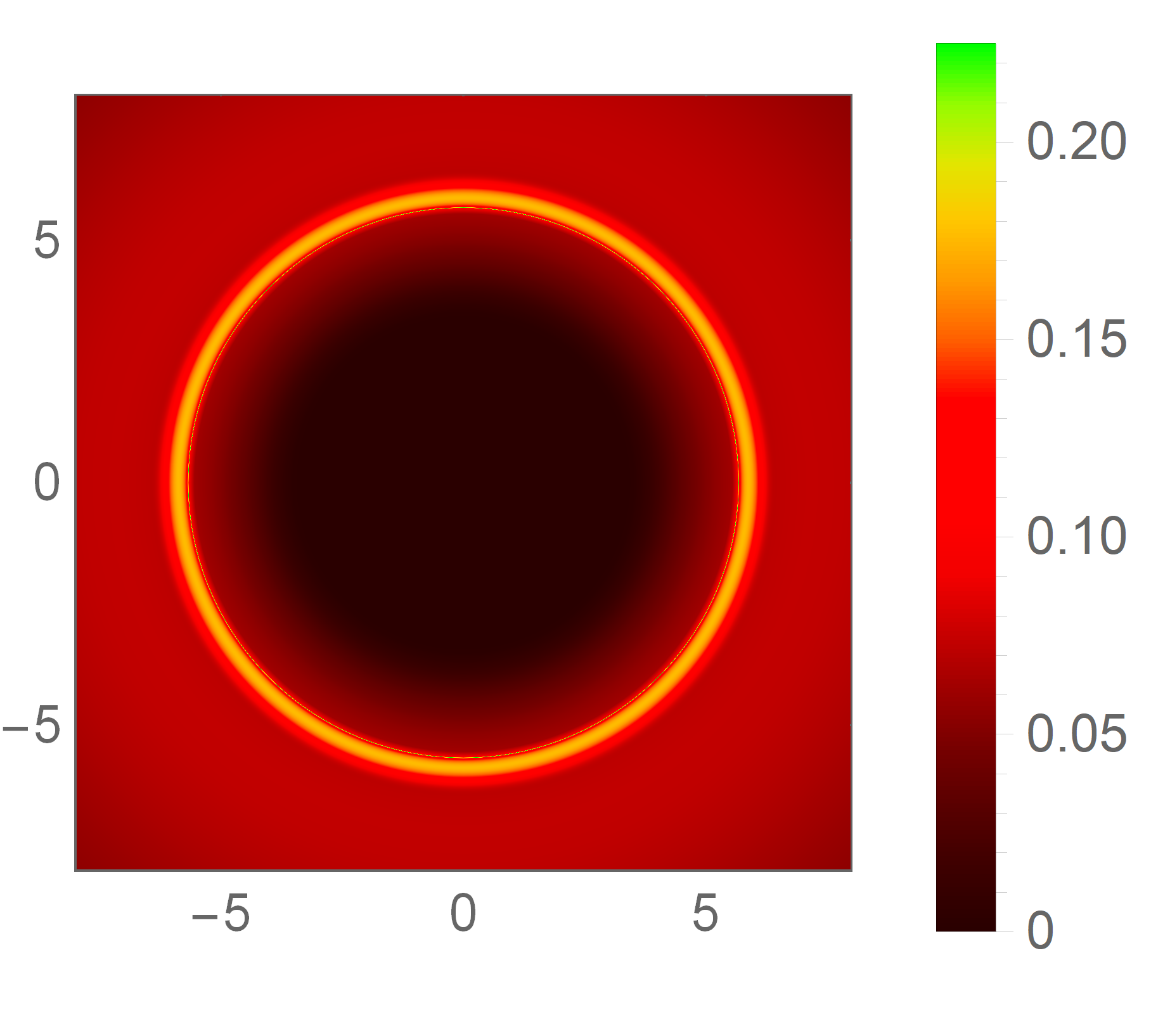}
\includegraphics[width=4cm,height=3.2cm]{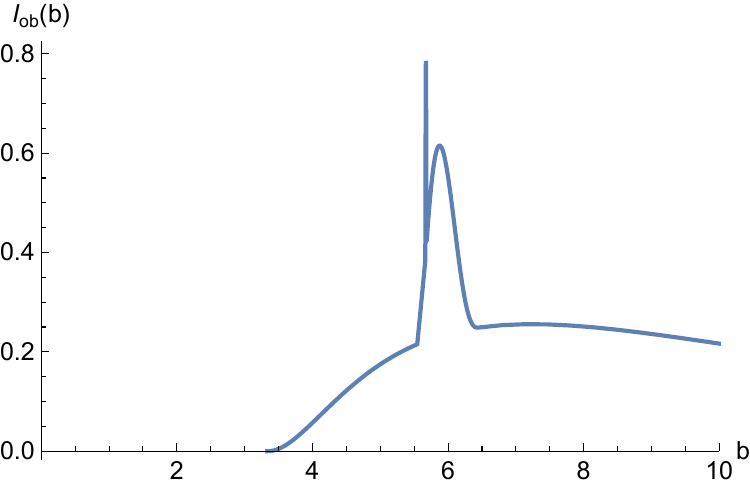} 
\includegraphics[width=4cm,height=3.2cm]{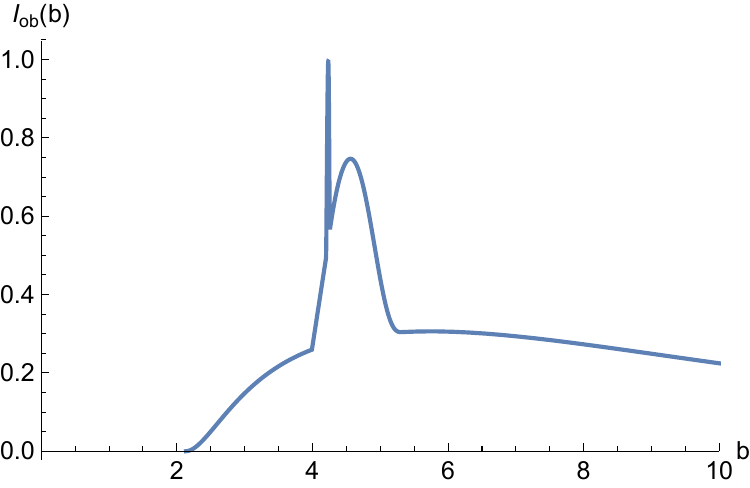}
\includegraphics[width=4cm,height=3.2cm]{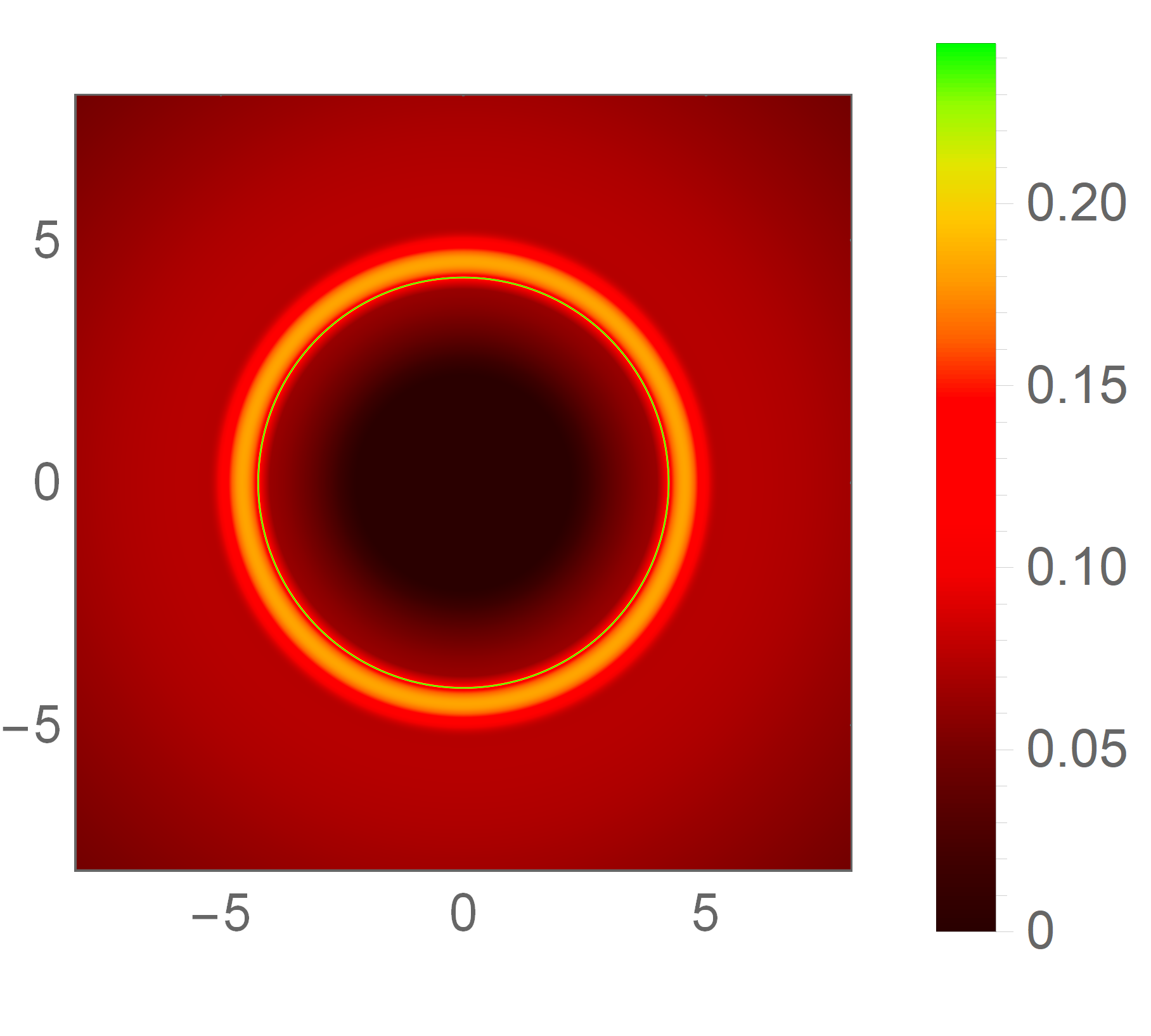} \\
\includegraphics[width=4cm,height=3.2cm]{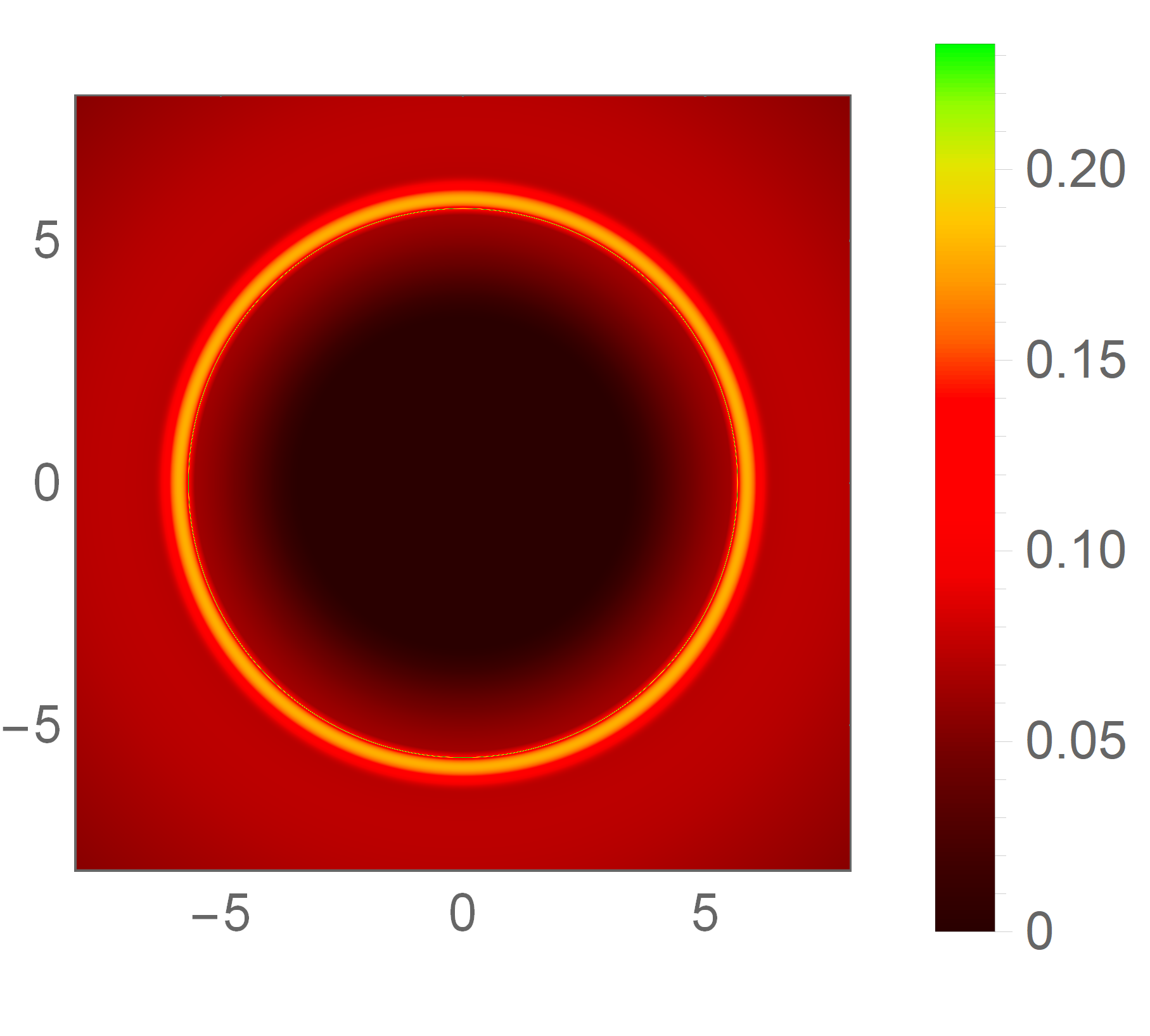}
\includegraphics[width=4cm,height=3.2cm]{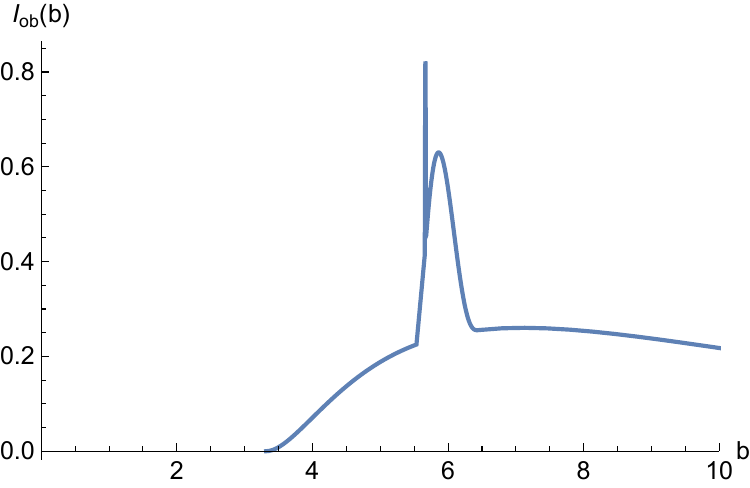}  
\includegraphics[width=4cm,height=3.2cm]{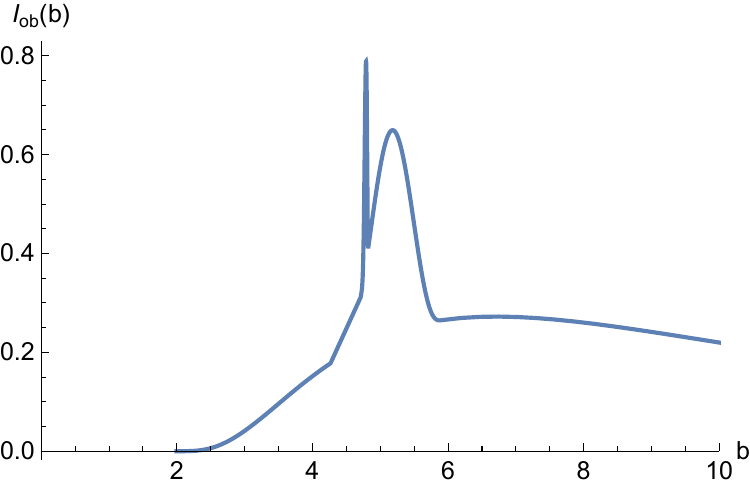}
\includegraphics[width=4cm,height=3.2cm]{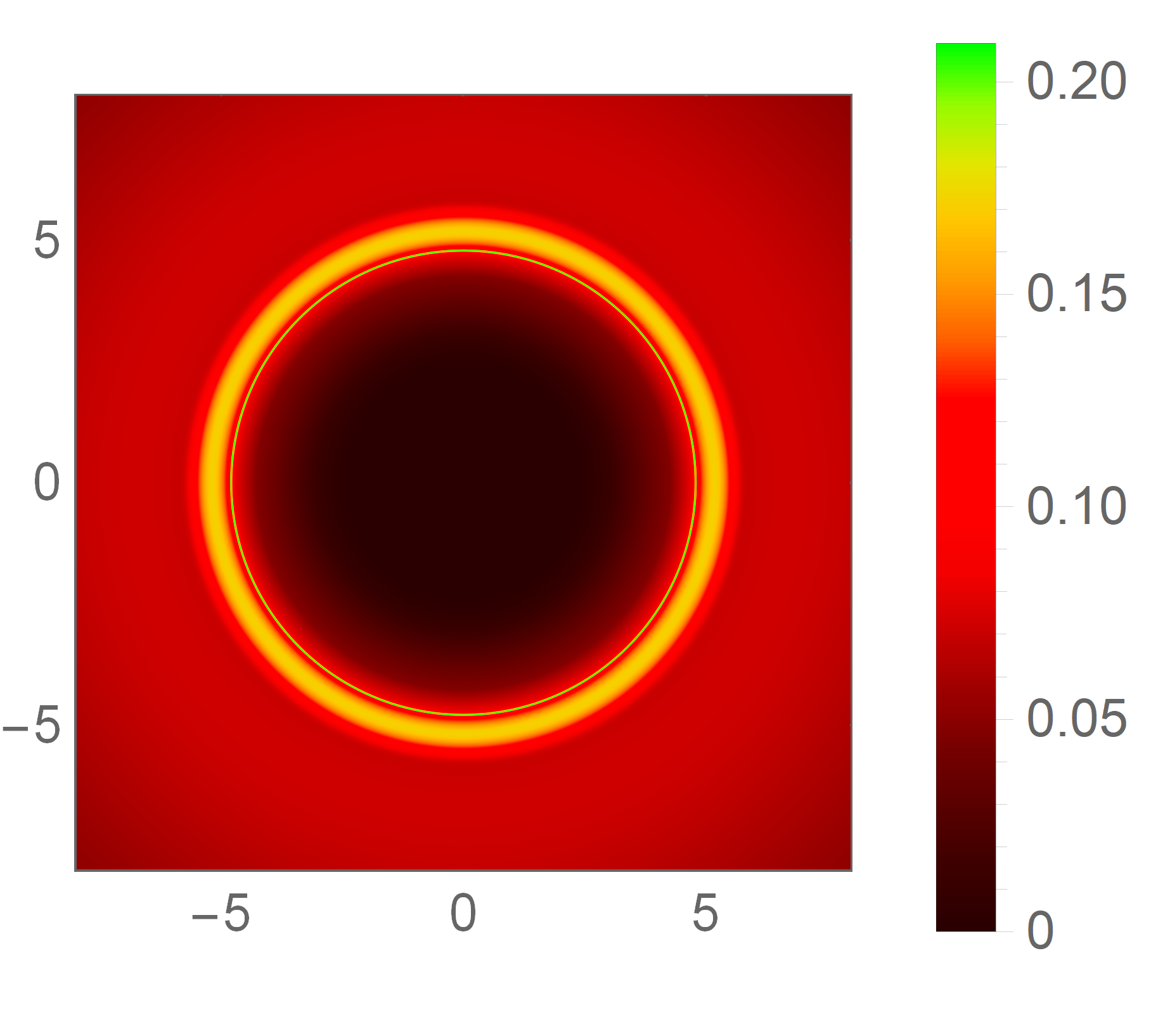}
\caption{The optical images (left and right figures) and the observed intensity $I_{ob}(b)$ (middle figures) for the JPn (left top figures), JPp (right top figures), KRZp (bottom left figures) and KRZn (bottom right figures) for the SU3 emission model.}
\label{fig:SU3}
\end{figure*}

\begin{figure*}[t!]
\includegraphics[width=4cm,height=3.2cm]{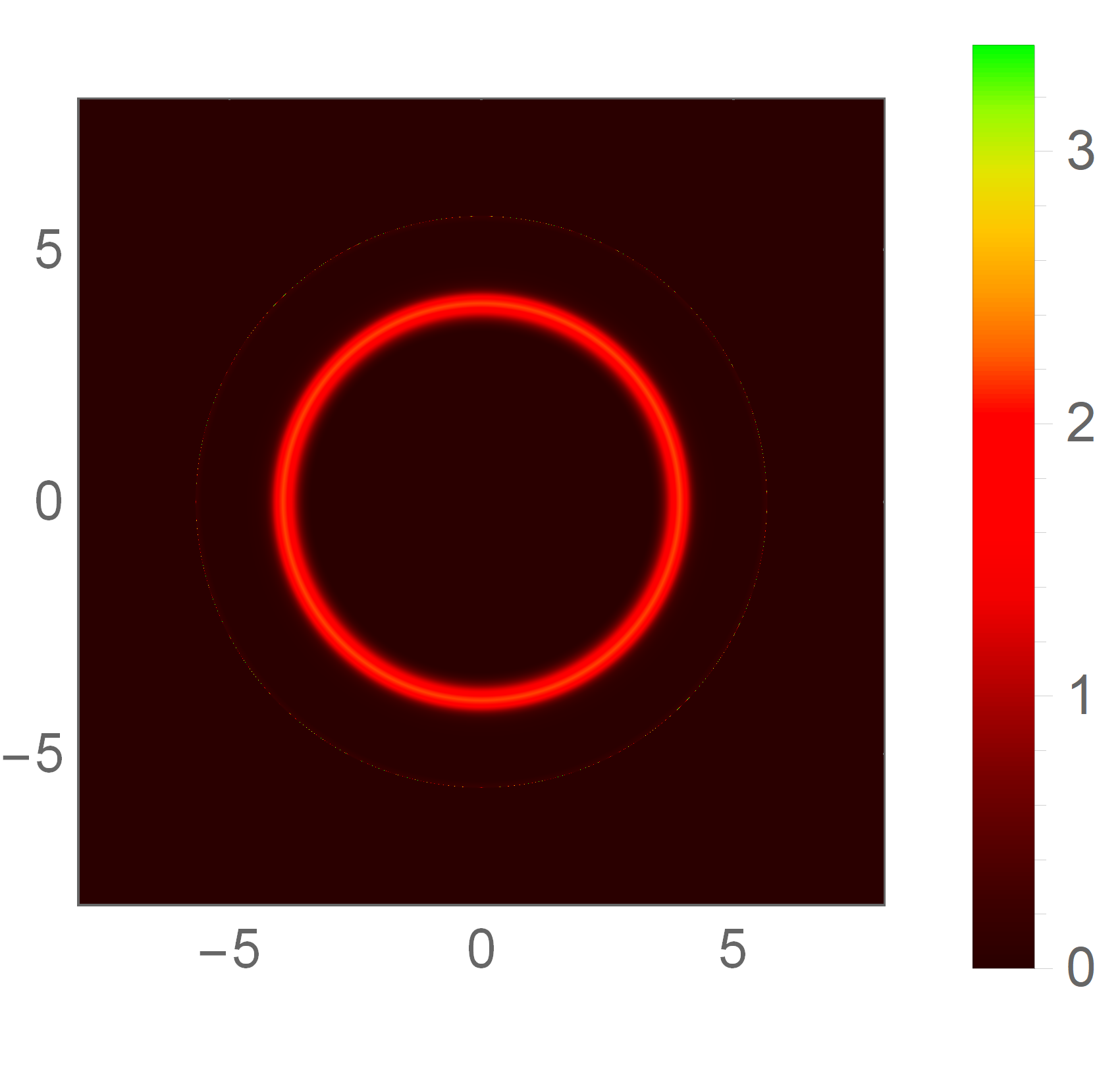}
\includegraphics[width=4cm,height=3.2cm]{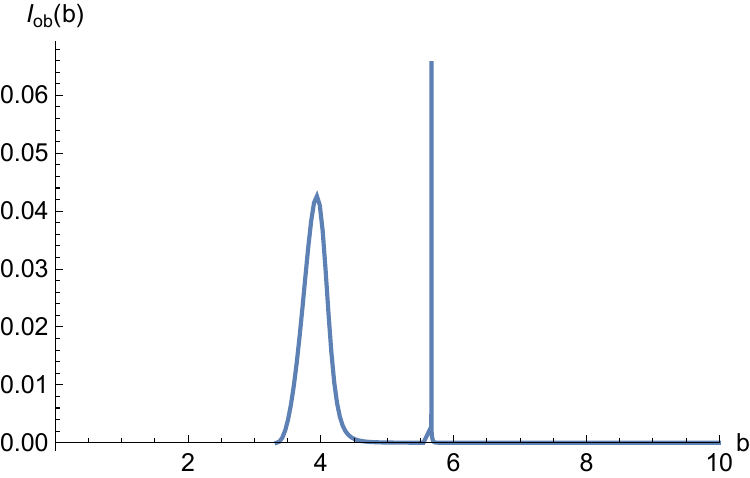} 
\includegraphics[width=4cm,height=3.2cm]{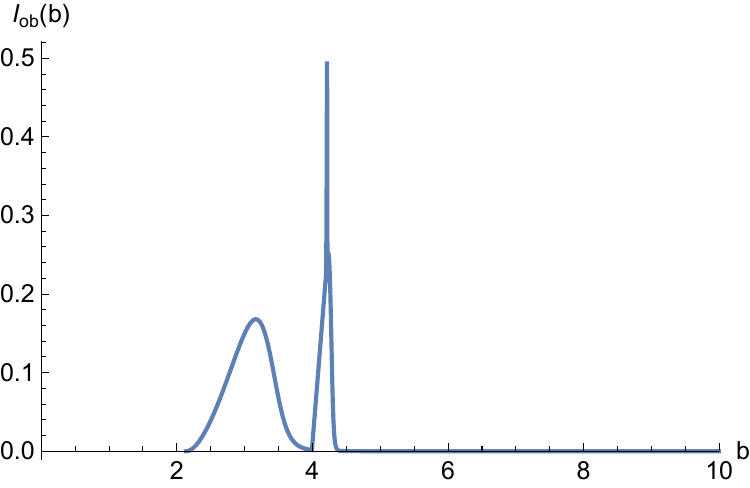}
\includegraphics[width=4cm,height=3.2cm]{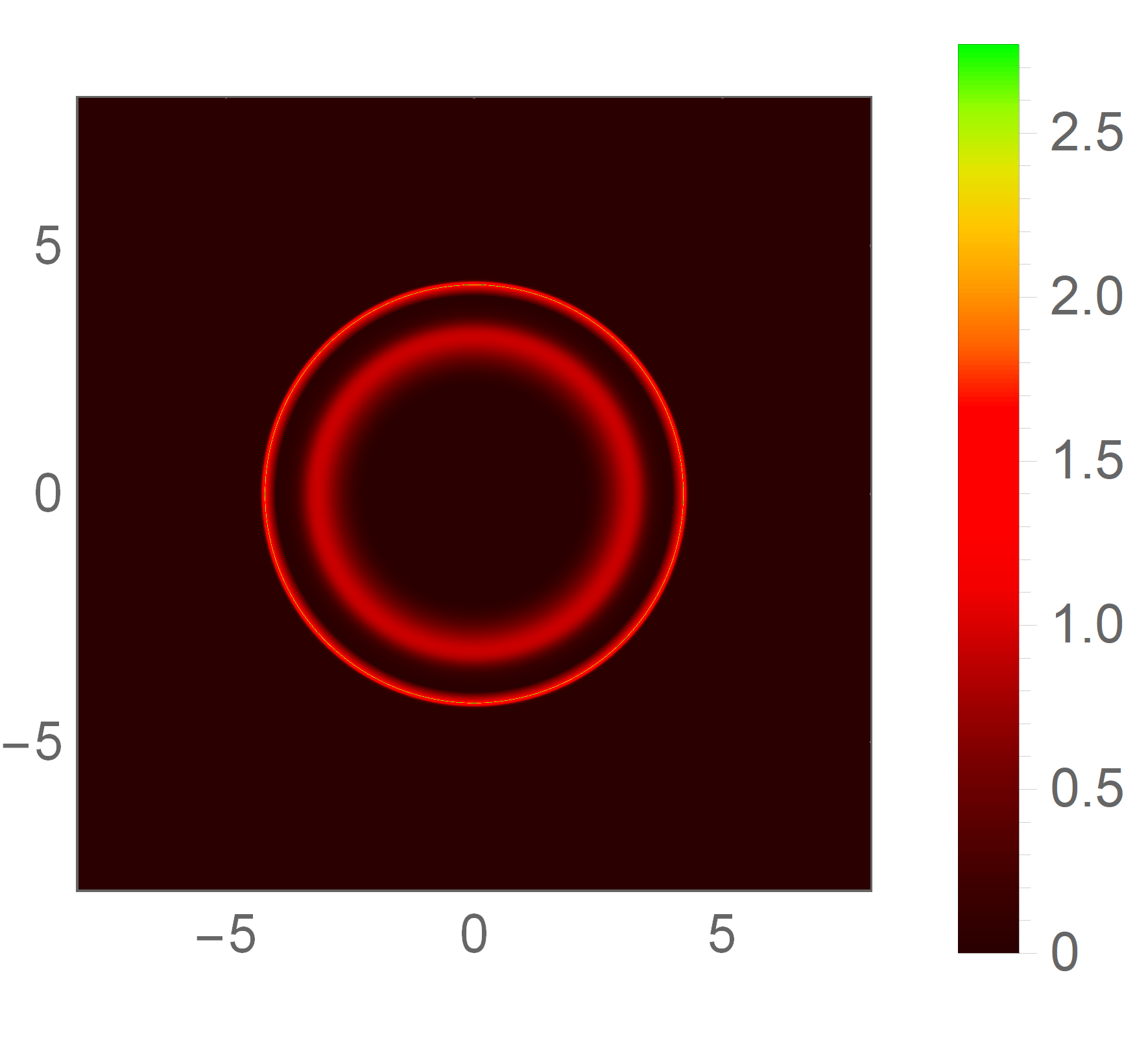} \\
\includegraphics[width=4cm,height=3.2cm]{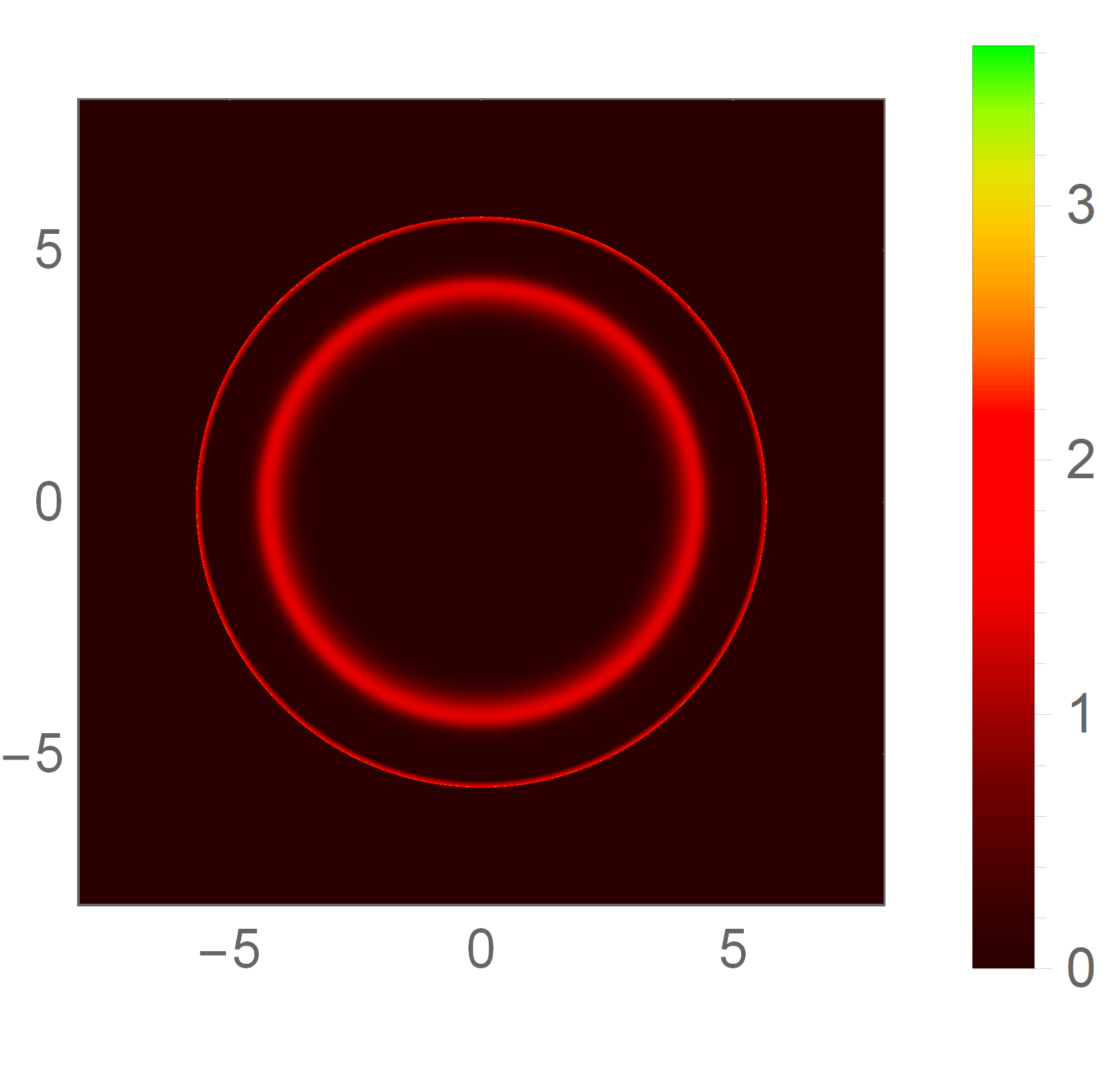}
\includegraphics[width=4cm,height=3.2cm]{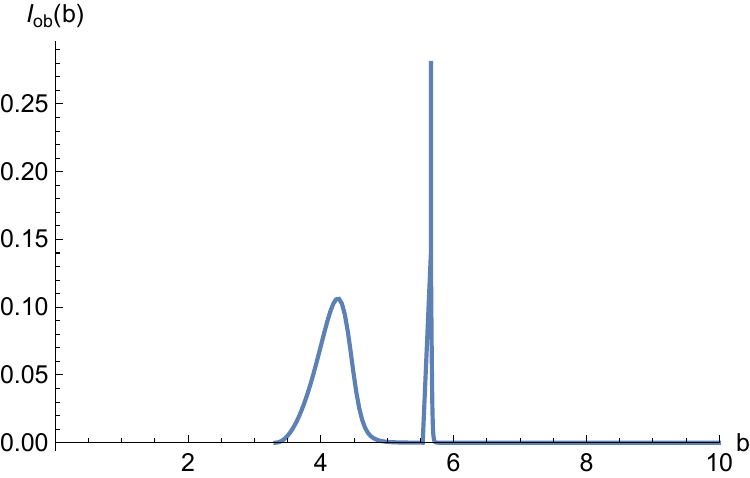}  
\includegraphics[width=4cm,height=3.2cm]{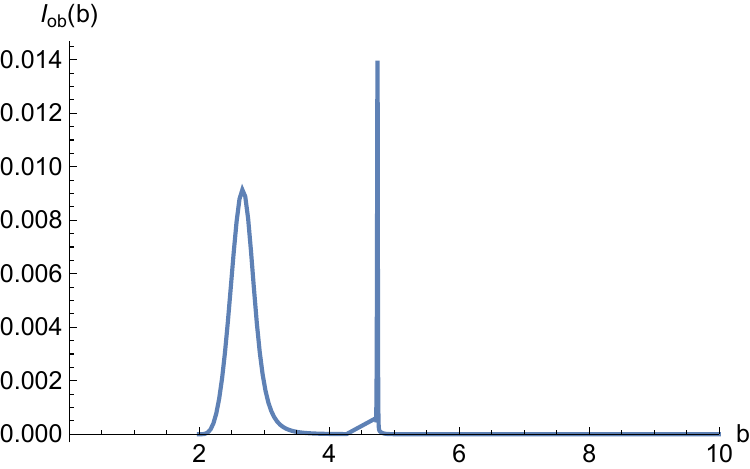}
\includegraphics[width=4cm,height=3.2cm]{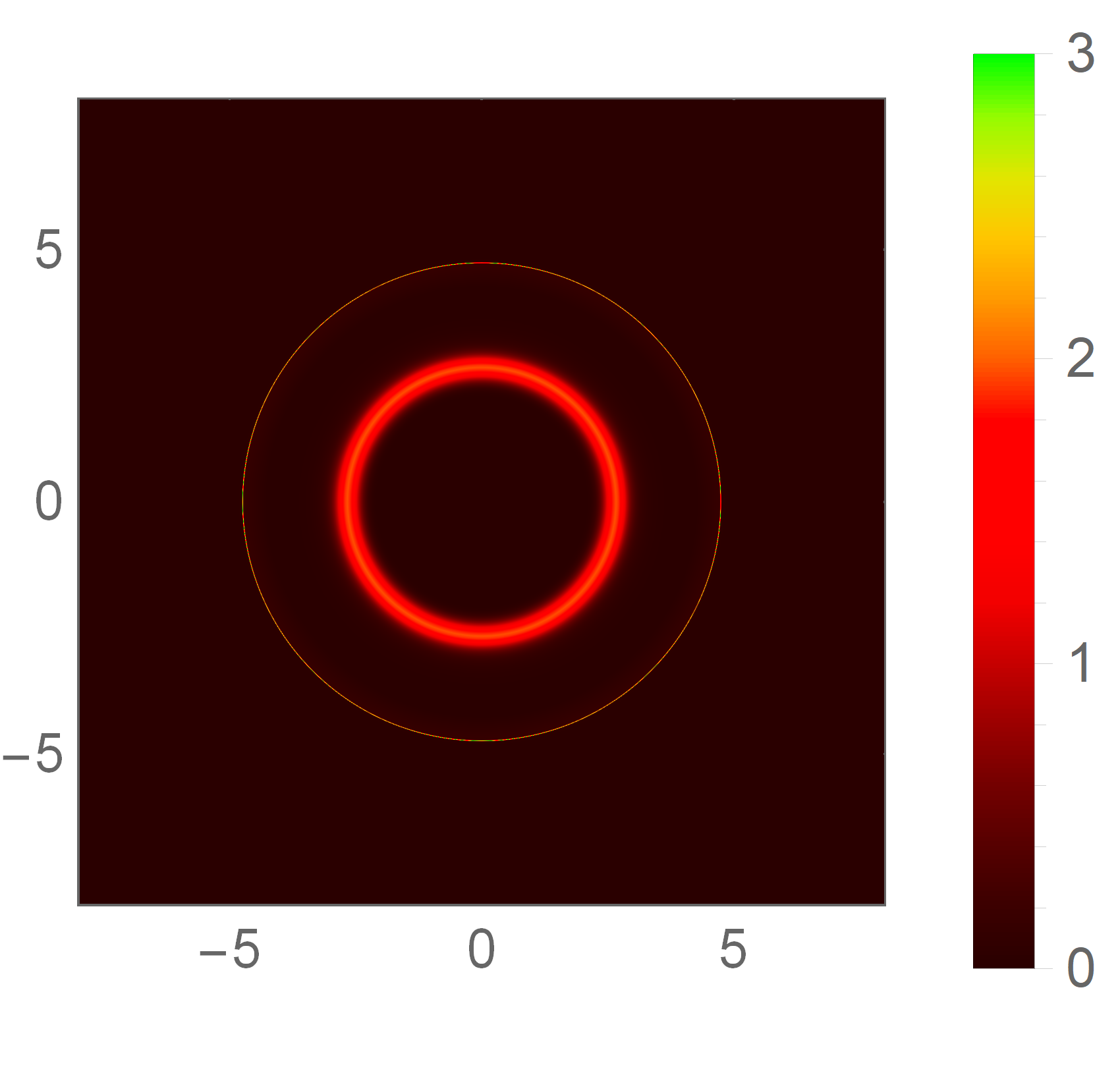}
\caption{The optical images (left and right figures) and the observed intensity $I_{ob}(b)$ (middle figures) for the JPn (left top figures), JPp (right top figures), KRZp (bottom left figures) and KRZn (bottom right figures) for the SU4 emission model.}
\label{fig:SU4}
\end{figure*}

\begin{figure*}[t!]
\includegraphics[width=4cm,height=3.2cm]{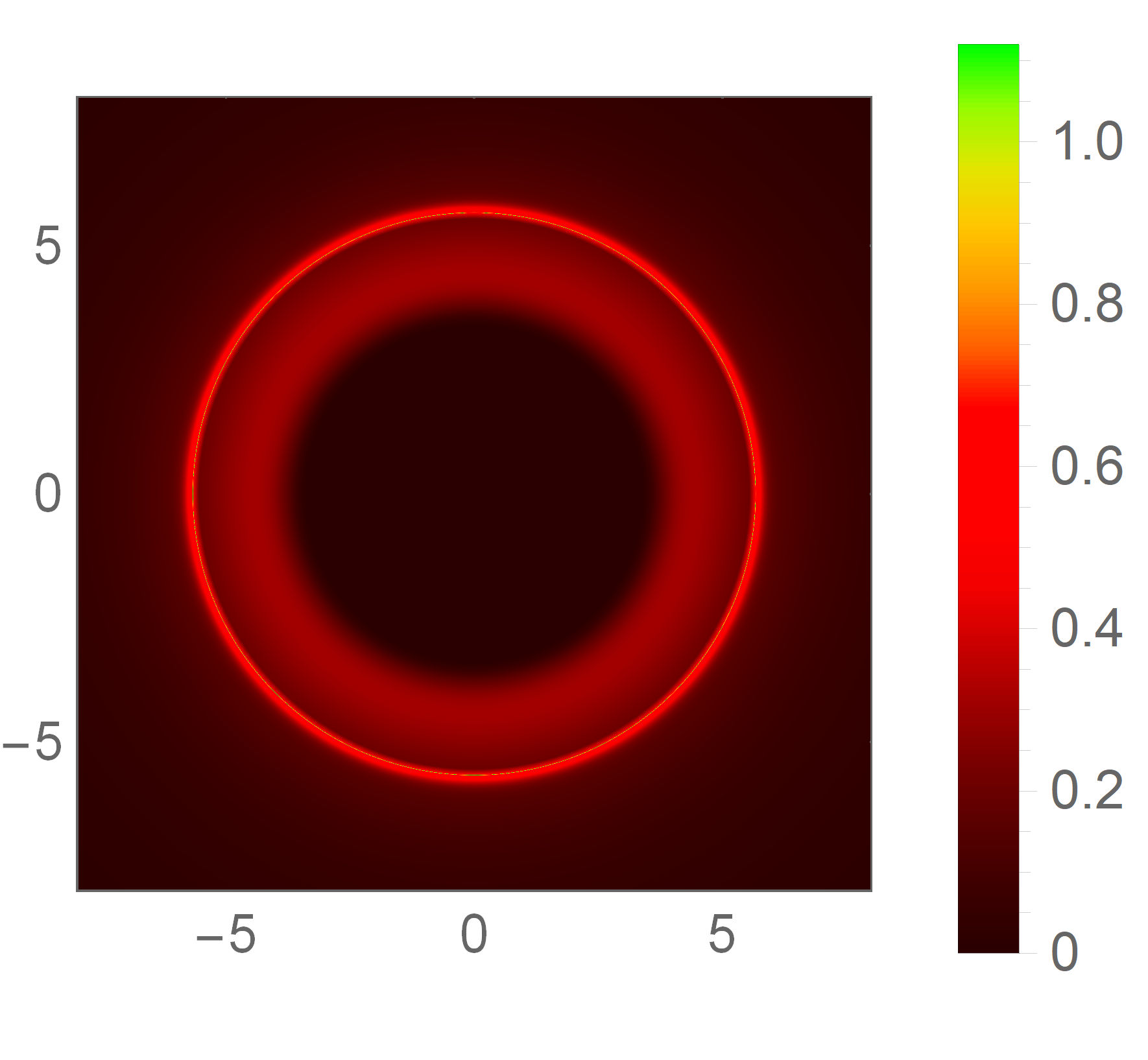}
\includegraphics[width=4cm,height=3.2cm]{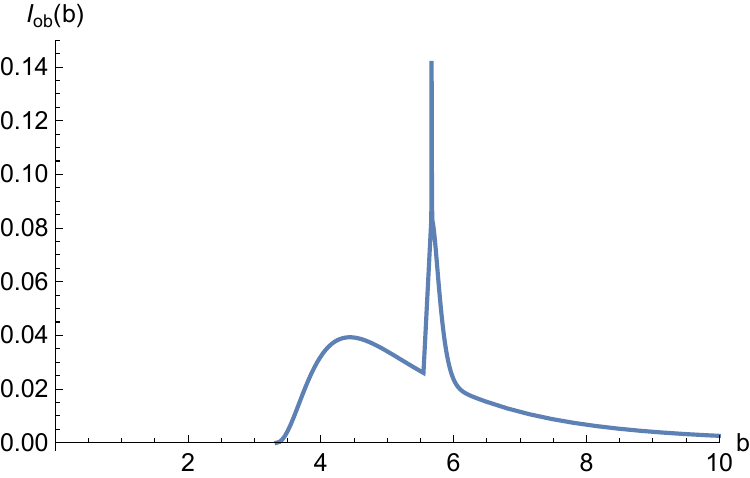} 
\includegraphics[width=4cm,height=3.2cm]{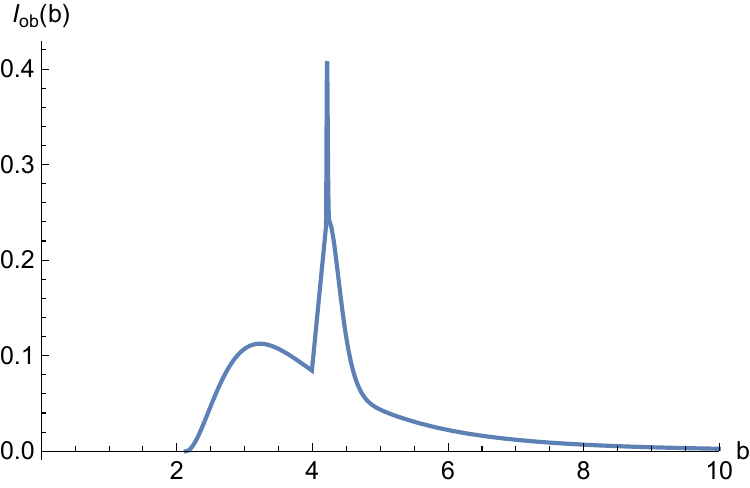}
\includegraphics[width=4cm,height=3.2cm]{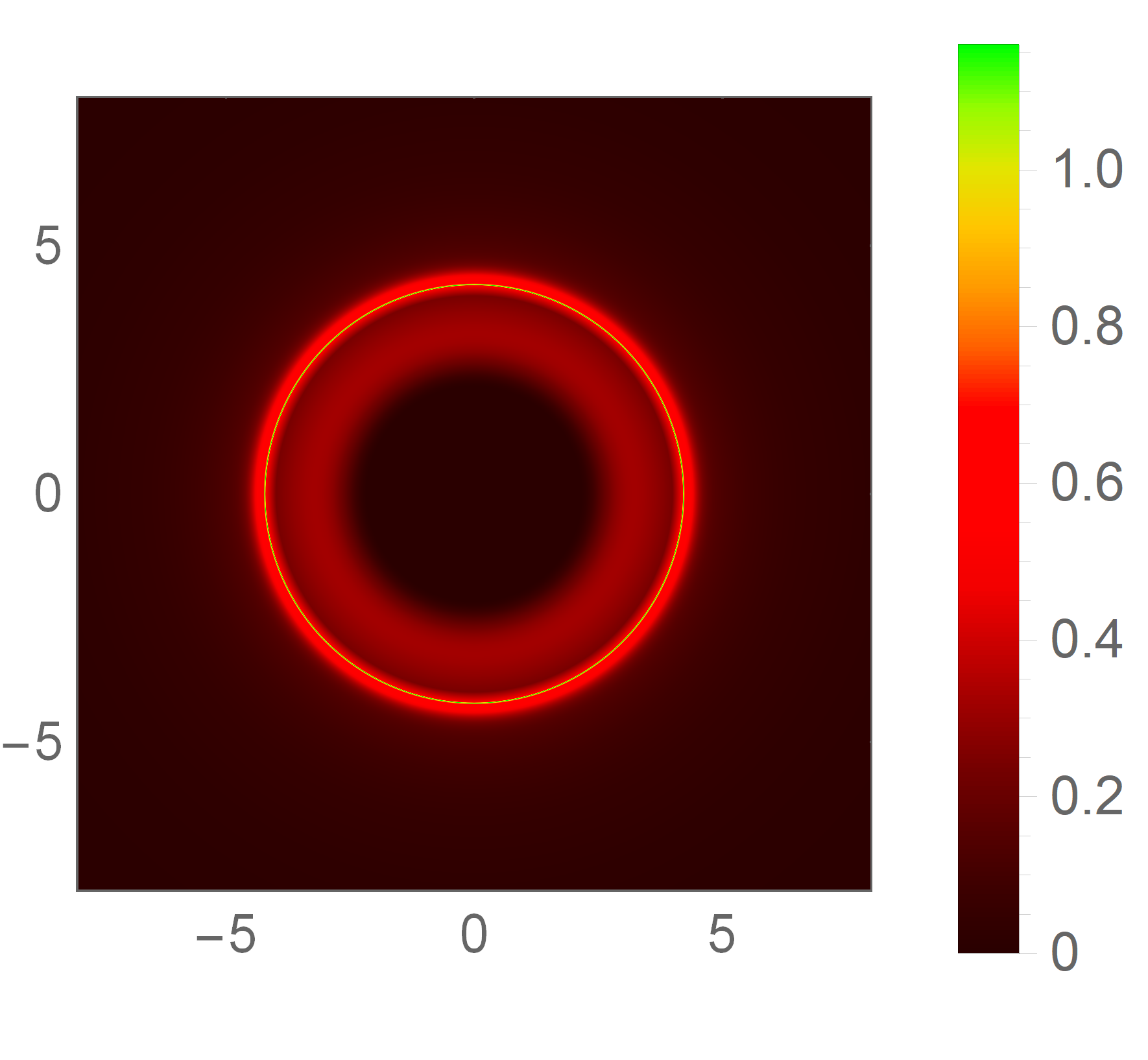} \\
\includegraphics[width=4cm,height=3.2cm]{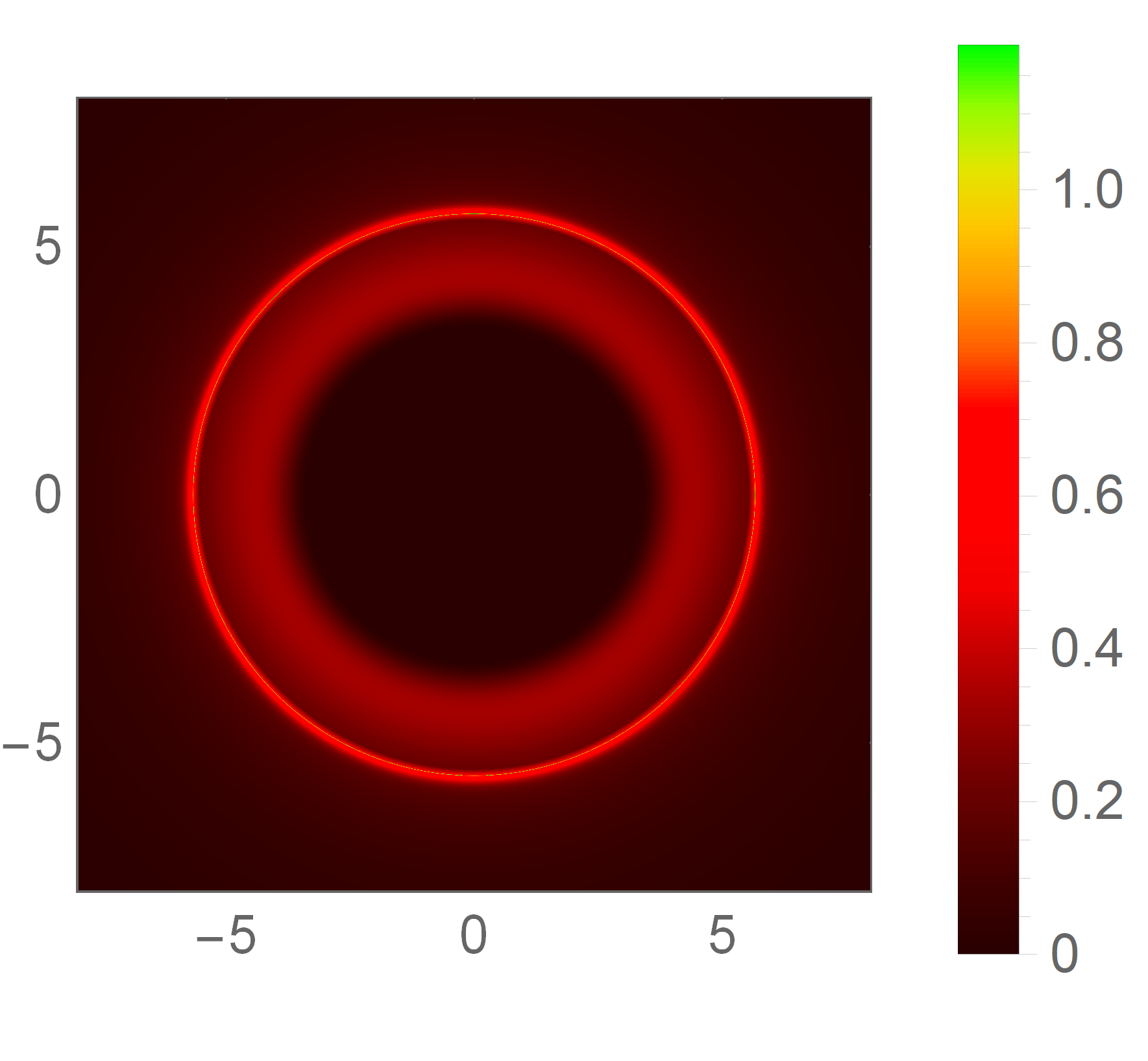}
\includegraphics[width=4cm,height=3.2cm]{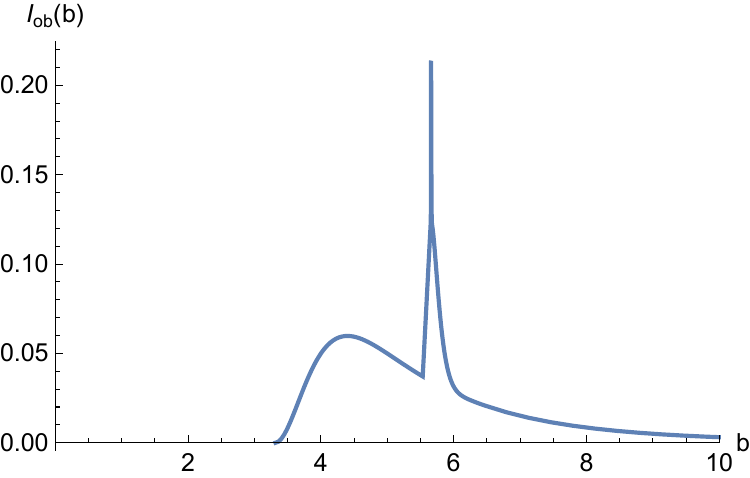}  
\includegraphics[width=4cm,height=3.2cm]{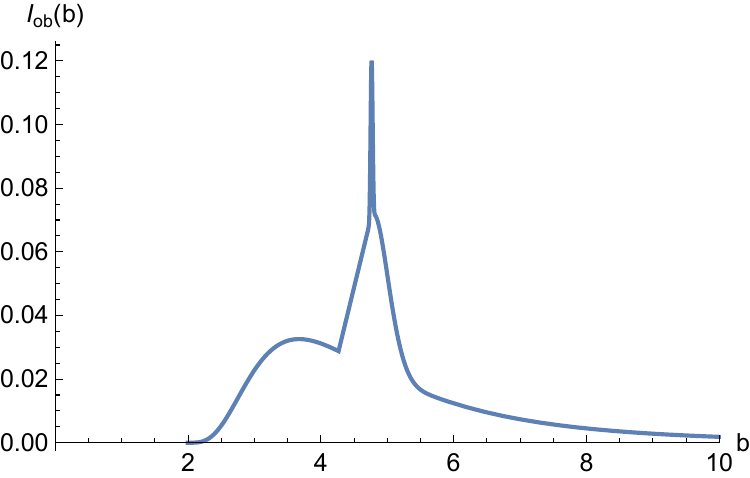}
\includegraphics[width=4cm,height=3.2cm]{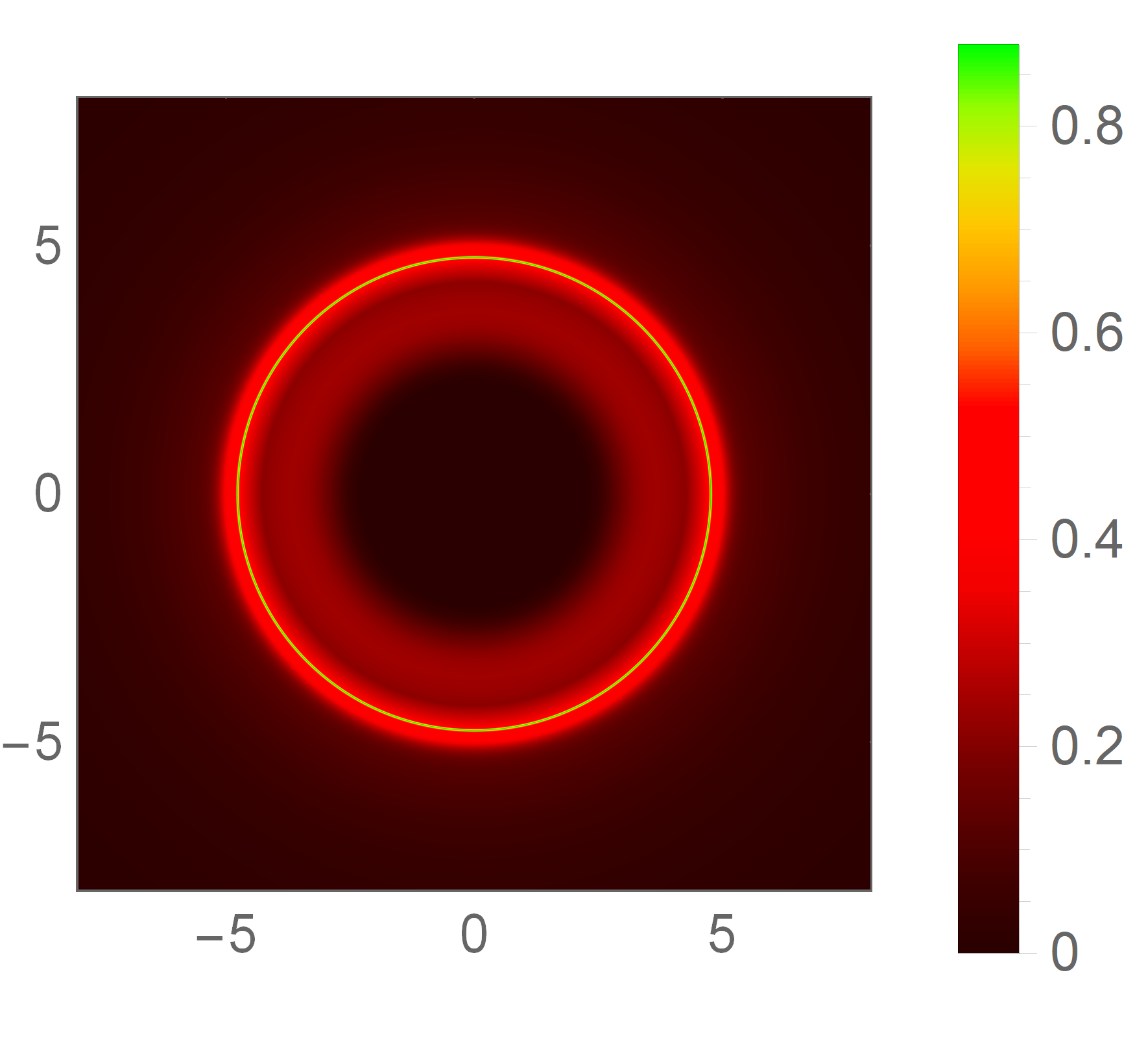}
\caption{The optical images (left and right figures) and the observed intensity $I_{ob}(b)$ (middle figures) for the JPn (left top figures), JPp (right top figures), KRZp (bottom left figures) and KRZn (bottom right figures) for the SU5 emission model.}
\label{fig:SU5}
\end{figure*}

\begin{figure*}[t!]
\includegraphics[width=4cm,height=3.2cm]{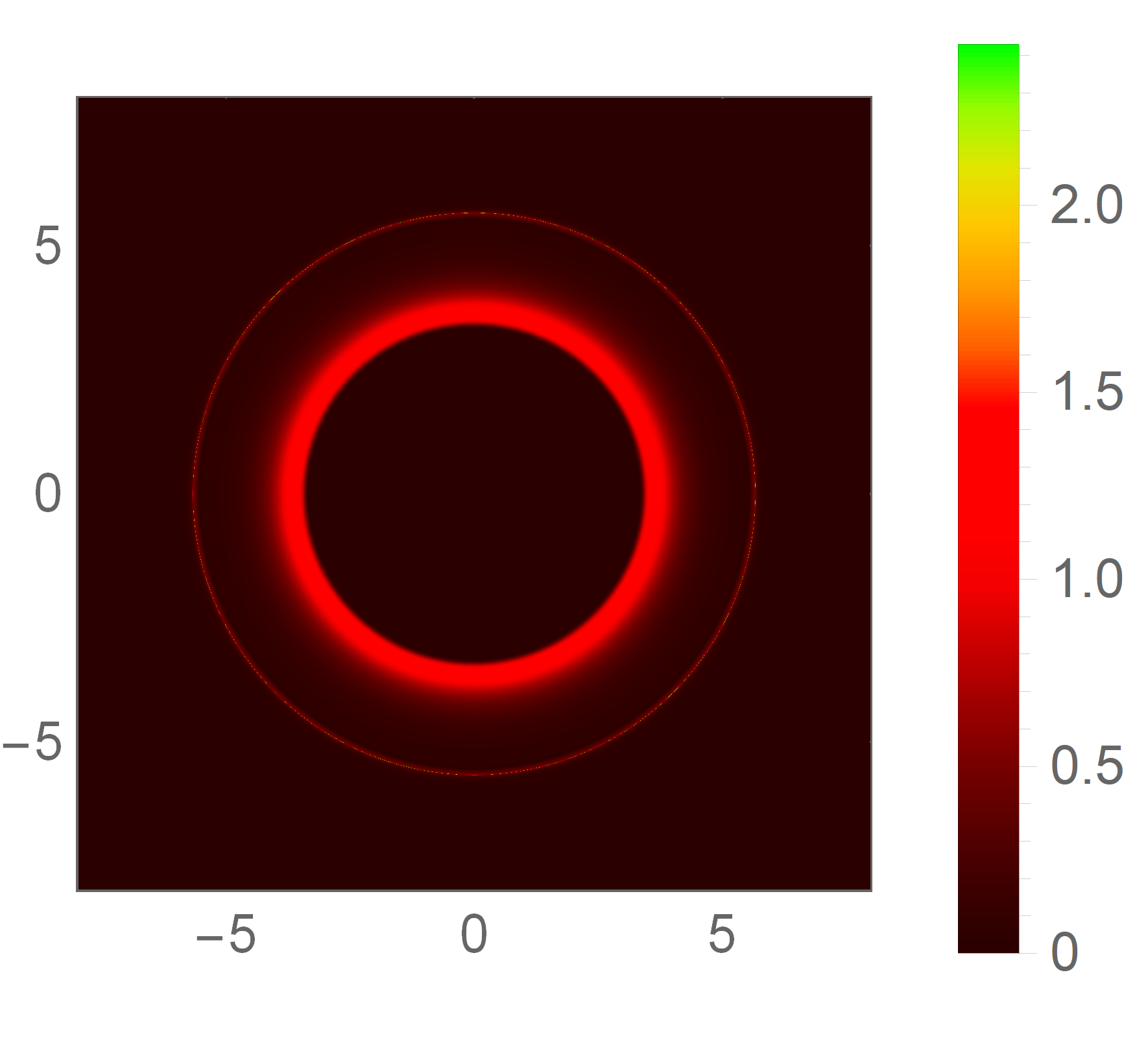}
\includegraphics[width=4cm,height=3.2cm]{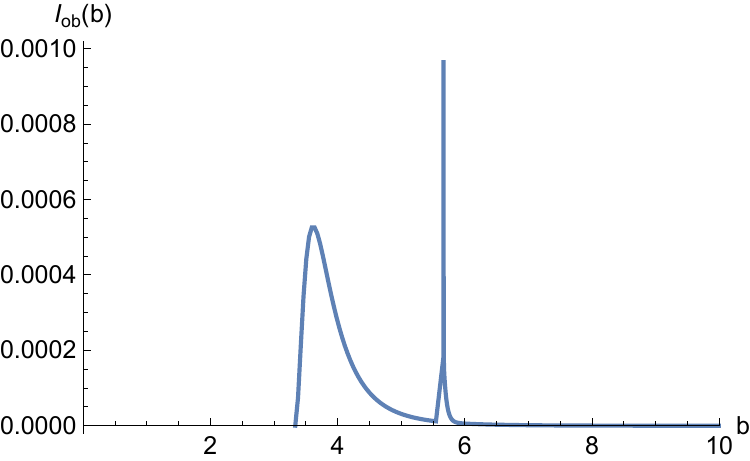} 
\includegraphics[width=4cm,height=3.2cm]{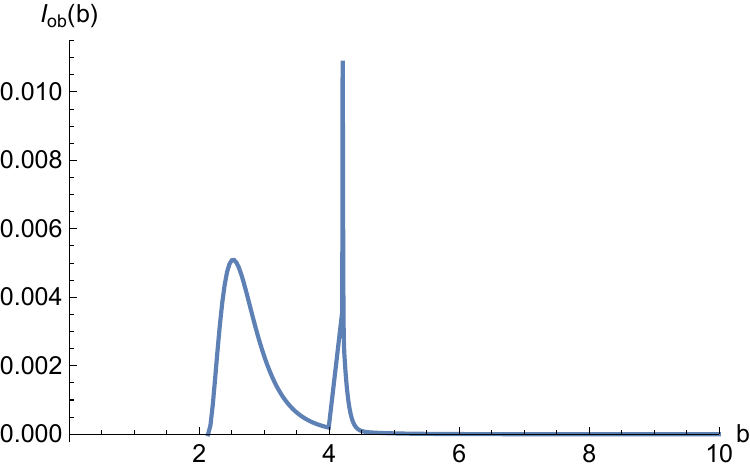}
\includegraphics[width=4cm,height=3.2cm]{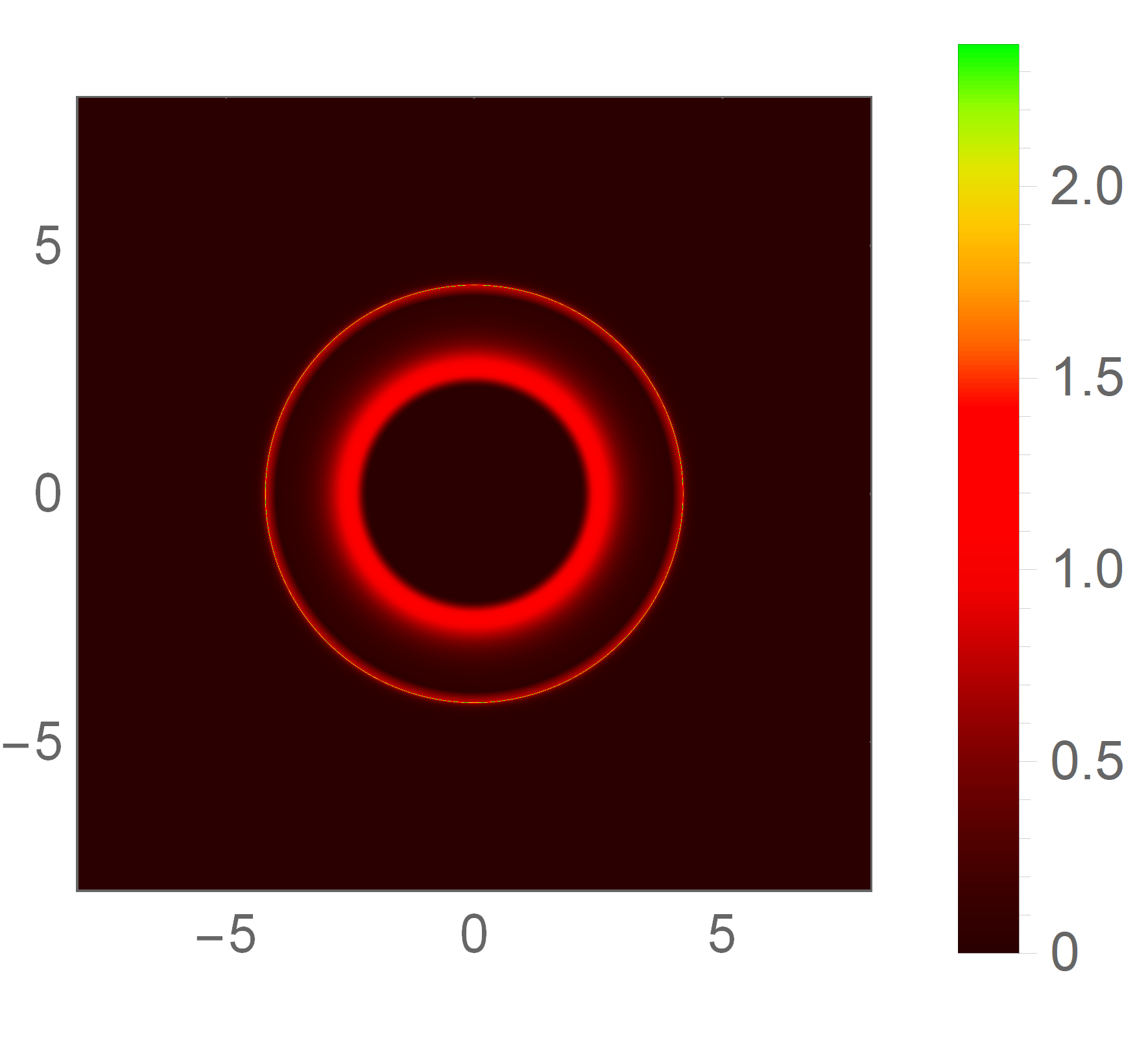} \\
\includegraphics[width=4cm,height=3.2cm]{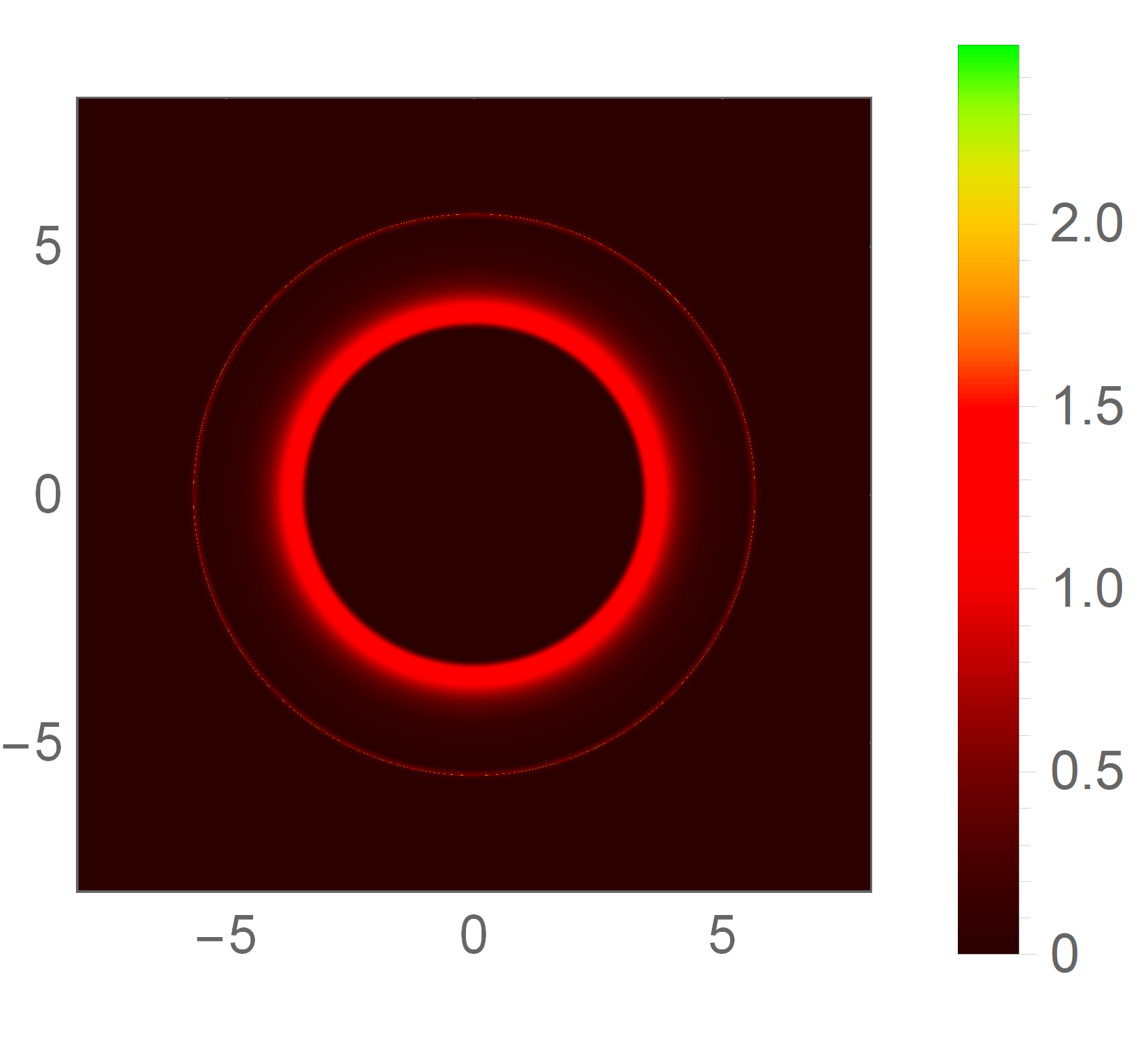}
\includegraphics[width=4cm,height=3.2cm]{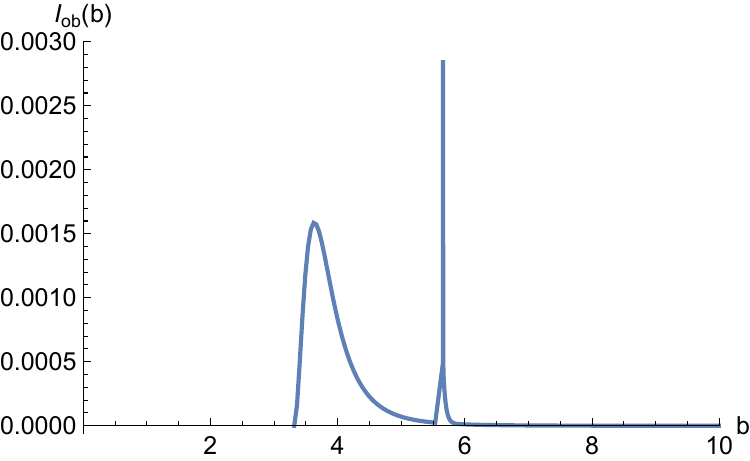}  
\includegraphics[width=4cm,height=3.2cm]{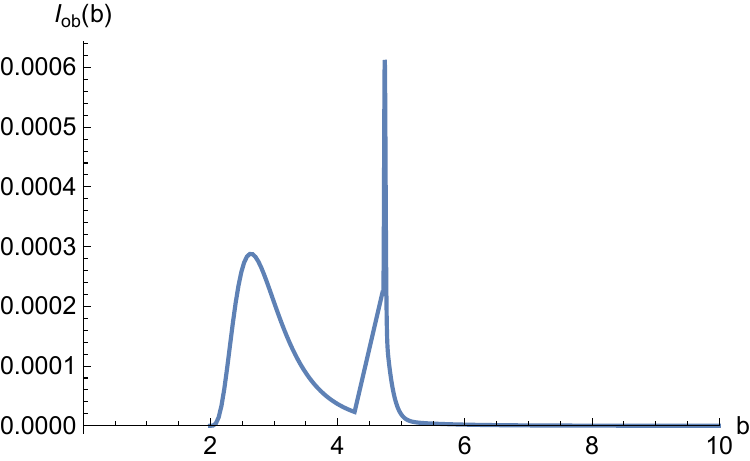}
\includegraphics[width=4cm,height=3.2cm]{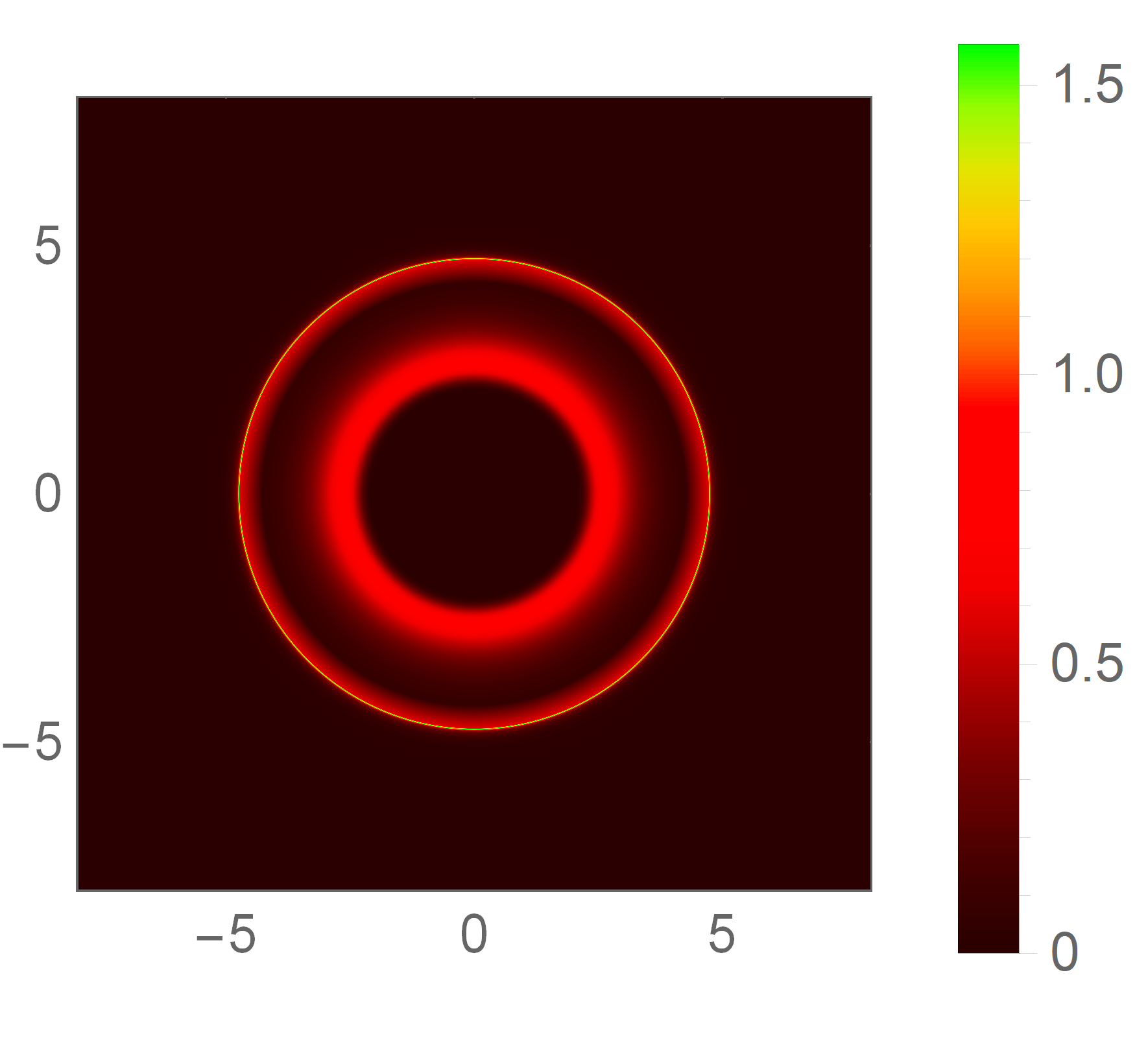}
\caption{The optical images (left and right figures) and the observed intensity $I_{ob}(b)$ (middle figures) for the JPn (left top figures), JPp (right top figures), KRZp (bottom left figures) and KRZn (bottom right figures) for the SU6 emission model.}
\label{fig:SU6}
\end{figure*}

In Figs. \ref{fig:SU1}, \ref{fig:SU2}, \ref{fig:SU3}, \ref{fig:SU4}, \ref{fig:SU5}, \ref{fig:SU6} we depict the result of our imaging of the JPn (two top left figures), JPp (two top right figures), KRZp (two bottom left figures), and KRZn (two bottom right figures) for the SU1, SU2, SU3, SU4, SU5, and SU6 models, respectively, for face-on simulations. Each pair of figures displays the full image and the observed intensity, respectively. 

Every full image displays the expected ring of radiation surrounding the central brightness depression of black hole images, in agreement with the corresponding figure of the observed intensity. However, there are significant differences, both qualitative and quantitative, between models. Such differences are mostly driven by the pick of the emission model and, to a minor degree (but not negligibly), by the choice of background geometry.

The qualitatively similar effect of the emission model on the black hole images can be roughly arranged into the three same categories above, according to the region where the peak of emission is located:

\begin{itemize}

\item Models whose emission peaks far outside the photon sphere  but near the ISCO (SU1 and SU2) have a $n=2$ photon ring nearing the outer edge of the shadow (SU1) or directly inside it (SU2). How spread is the emission onward (``spread" for SU1 but ``sharp" for SU2) determines how far the direct emission goes in the full image, yielding a much more extended ring of radiation in SU1 than in SU2.

\item Models peaking in the region between the event horizon and the photon sphere (SU3, SU4, SU5) have still their $n=2$ photon ring located near the critical curve with a significant direct emission inside it in all cases. The ``spread" model SU3 has the direct emission stretched to a large distance (similarly as SU1), the ``sharp" model SU4 has an additional brightness deficit between the internal direct emission and the photon ring, and SU5 has its photon rings inserted in the direct emission.

\item Models with a monotonically decreasing emission profile outside the event horizon (SU6 to SU10) have in all cases a $n=2$ photon ring overlapped with the direct emission, but with somewhat different features. SU6 has a spiked contribution of the direct emission in the interior region and a photon ring nearing the end of such emission. On the other hand, SU7, SU8, SU9 and SU10 models display a quite similar shape, with the boost of the photon ring occurring when there is still significant emission (as compared to its maximum). The only difference between them is qualitative in terms of the total intensity, a reflection of the emission profile's features.

\end{itemize}



\subsection{Features of the photon rings}

Once we have discussed and accepted the major role played by the accretion disk in generating the main features of the images, we turn now our attention to the differences in the structure and features of the photon rings and the direct emission for the different background geometries considered in this work\footnote{For a complementary work on photon ring signatures within spherically symmetric space-times see \cite{Wielgus:2021peu}.}. Looking at the set of figures \ref{fig:SU1}, \ref{fig:SU2}, \ref{fig:SU3}, \ref{fig:SU4}, \ref{fig:SU5}, \ref{fig:SU6}, we have arranged the solutions that enhance the shadow (JPn and KRZp, dubbed as ``enhancing") on the left (top and bottom figures, respectively) and those who diminish it (JPp and KRZn, dubbed as ``diminishing") on the right for a better visual comparison. We next proceed to make both a qualitative and quantitative description of the features of the photon rings for each set of figures.

From a qualitative point of view we observe neat differences between enhancing and diminishing solutions for every SU model. In the SU1 model (Fig. \ref{fig:SU1}), photon rings are clearly wider and more luminous in the diminishing geometries than in the enhancing ones, but in both cases the photon rings appear near the outer edge of the shadow. In the SU2 model (Fig. \ref{fig:SU2}) there is a much larger separation between the direct emission and the first and second photon rings, thus leaving a wider brightness deficit in the region between them.  The SU3 model (Fig. \ref{fig:SU3})  shows quite a similar imaging in the enhancing and diminishing solutions, if only the interior part of the direct emission spreads much farther inside. In the SU4 model (Fig. \ref{fig:SU4}), for the JP geometry the photon rings appear much closer to the direct emission in the diminishing solutions than in the enhancing ones, while the total intensity of both emissions are far more larger in the former than in the latter; for the KRZ geometry the visual appearance in the enhancing and diminishing solutions is more alike, except for a relatively more intense direct emission and farther distance between direct and photon ring emissions in the latter than in the former. When we reach the SU5 model (Fig. \ref{fig:SU5}), the optical appearance of enhancing/diminishing solutions become quite similar in the structure of their direct/photon ring emissions.  As we go further down the SU6/SU7/SU8/SU9/SU10 sequence this tendency becomes more acute, the reason why we have opted for only displaying the optical appearance for the SU6 model (Fig. \ref{fig:SU6}) as representative of this class of models and thus alleviate the number of images.

To characterize the photon ring luminosity from a quantitative point of view, we first realize that the expectations based on the Lyapunov exponent, as given by the scaling relation in Eq. (\ref{eq:scaleprs}) are clouded by two facts. The first one is that this exponent corresponds to the $n \to \infty$ limit while future VLBI projects cannot hope but to measure (under very optimistic expectations) the features of the $n=2$ photon ring. The second one is that, as mentioned, the Lyapunov exponent is a characterization of photon rings' luminosity under a fully homogeneous emission model, which does not correspond to the actual properties of the disk\footnote{Alternatively, one can rely on the width of successive photon rings as follows from the scaling relation in Eq.  (\ref{eq:scaleprs}). This has the drawback that, from an observational point of view, current measurements by the EHT Collaboration are limited to a $\sim 10 \%$ of its peak luminosity, implying difficulties on the determination of both endings of each ring \cite{Staelens:2023jgr}.}.

Bearing the above two fundamental difficulties in mind, here we consider the ratio of luminosities (i.e. the integrated luminosity below each curve) between the $n=1$ and $n=2$ photon rings. On the theoretical side, the corresponding Lyapunov exponent can be directly read off from the ray-tracing procedure between the $n=1$ and $n=2$ photon rings, providing a number $\gamma_2$ which deviates from the actual $n \to \infty$ Lyapunov exponent by less than $1\%$ for all  background geometries considered here. Such a number provides an expectation of a factor $\propto e^{\gamma_2}$ of luminosity suppression between such photon rings at homogeneous emission model, which must now be compared with the results of actual SU model simulations.

Following this procedure, in Table \ref{Table:Lya} we depict the relative luminosity between the $n=1$ and $n=2$ photon rings for the ten SU models in each background geometry, both enhancing (JPn, KRZp) and diminishing (JPp, KRZn) ones, and for completeness, we run the same exercise for the Schwarzschild geometry. On the one hand, we observe that the theoretical expectations based on the employ of the second-degree Lyapunov exponent ($\propto e^{\gamma_2}$), at fixed background geometry, are moderately modified upwards (for models SU1, SU2, SU3) but less significantly modified downwards or almost left unchanged (for every other SU model). On the other hand, at fixed SU model, we see that enhancing solutions (JPn and KRZp) clearly produce a moderately dimmer $n=2$ ring (up to a factor 3/2) than in the Schwarzschild geometry for all SU models (an effect slightly exaggerated in the emission models peaking near the event horizon) but similar for both JPn and KRZp geometries. Conversely, enhancing solutions (JPp and KRZn) display a larger boost of luminosity on their photon rings, which can raise up to a factor 2 for the KRZn configuration with emission models at the event horizon.

\begin{table*}[t!]
\begin{tabular}{|c|c|c|c|c|c|c|c|c|c|c|c|c|}
\hline
Geometry/Disk & $\gamma_2$  & $e^{\gamma_2}$  & SU1  & SU2  & SU3 & SU4 & SU5 & SU6 & SU7 & SU8 & SU9  & SU10  \\ \hline
Schwarzschild & 3.15 & 23.35 & 28.94 & 27.83 & 27.21 & 21.57 & 23.34 & 21.16 & 24.74 & 23.45 & 22.81 & 22.20 \\ \hline
JPn & 3.53 & 34.32  & 40.37 & 38.72 & 38.92  & 31.99  & 34.42 & 31.90 & 36.11 & 34.61 & 33.83 & 33.12 \\ \hline
KRZp & 3.51 & 33.63 & 40.31 & 39.31 & 38.11 & 31.87 & 33.59 & 31.12 & 35.34 & 33.89 & 33.15 & 32.48  \\ \hline
JPp & 2.85  & 17.29 &  24.57 & 23.90  & 21.39 & 16.10 & 17.43 & 15.27 & 18.68 & 17.42 & 16.83  & 16.26  \\ \hline
KRZn & 2.53 & 12.63 & 17.06& 15.52 & 16.13 & 10.26 & 12.87 & 10.60 & 13.93 & 12.70 & 12.09 & 11.40 \\ \hline
\end{tabular}
\caption{The ratio $\gamma_2 \equiv I_1/I_2$ of the luminosities between the $n=1$ and $n=2$ photon rings for the ten SU models and the four background geometries (JPn, KRZp, JPp, KRZn). For completeness we also include the values corresponding to the Schwarzschild geometry. In this table $\gamma_2$ denotes the theoretical expectations based on the second-degree Lyapunov exponent (and its exponential decrease of luminosity) for each such geometry, something directly read off from the ray-tracing procedure. For every background geometry, the actual luminosity ratio is moderately modified upwards for the SU1/SU2/SU3 models, while other models display less significant modifications (both downwards and upwards).}
\label{Table:Lya}
\end{table*}

\begin{figure*}[t!]	\includegraphics[width=8.8cm,height=5.5cm]{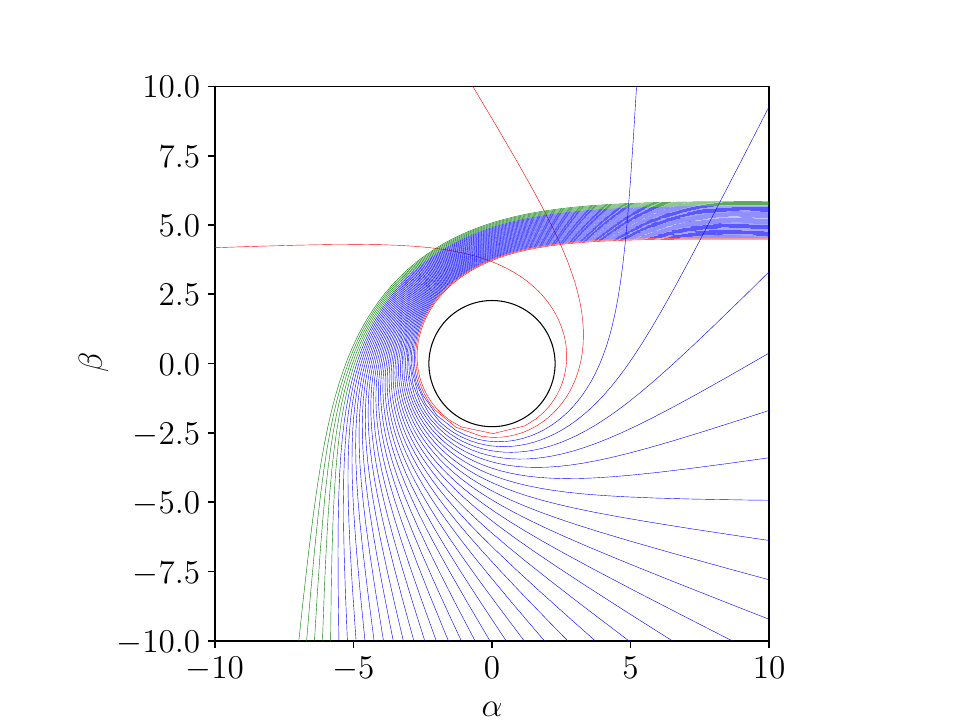} 	\includegraphics[width=8.8cm,height=5.5cm]{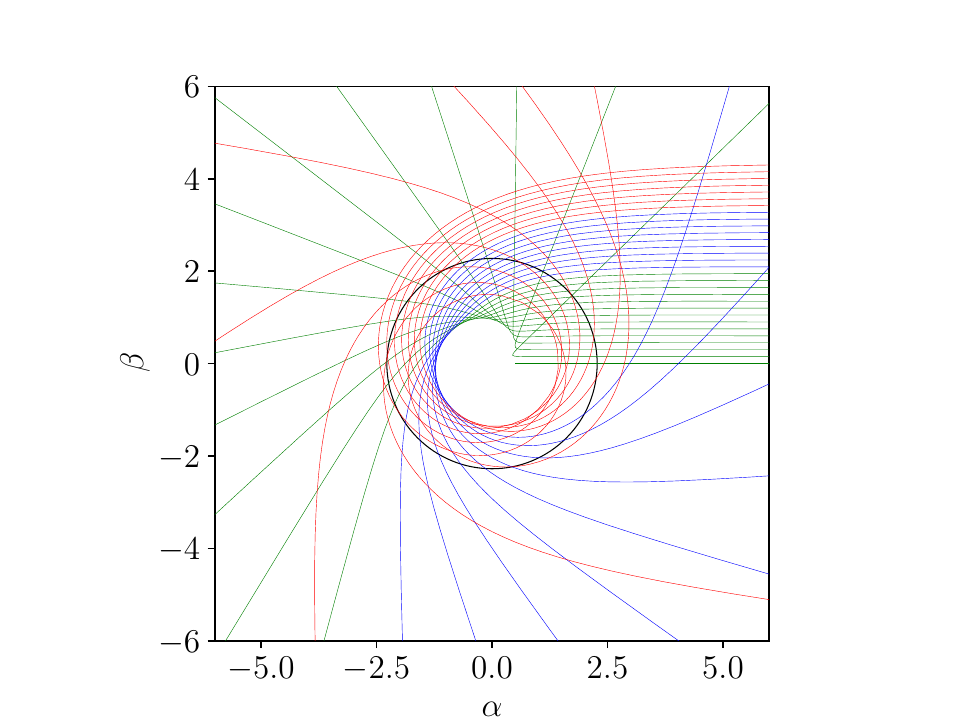}
	\caption{Ray-tracing for KRZ naked singularities with $\eta=-1.5$ for values of $b \in [b_{c},6]$ (left figure) and  
    $b \in [0,b_{c}]$ (right figure). Trajectories that cross the equator one or zero times are depicted in green, twice in blue and three times (or more) in red. The black circle indicates the circular orbit at $r_{ps}$. In this plot we have used coordinates  $\alpha = \frac{b}{M}\cos\phi$ and $\beta = \frac{b}{M}\sin\phi$. 
    }
	\label{fig:rt_naked}
\end{figure*}

The bottom line of the above qualitative and quantitative discussion is that parametrized solutions of the kind considered here and enhancing (diminishing) the shadow's radius produce significantly dimmer (brighter) rings than in the Schwarzschild geometry at every SU emission model, thus suggesting that if the promises of VLBI technologies come ever into reality \cite{Ayzenberg:2023hfw}, the observation of the $n=1$ and $n=2$ photon rings' features may allow us to distinguish between these parametrized black hole geometries (and the Schwarzschild one).

\section{Imaging a geometry with a naked singularity} \label{S:VII}

In section \ref{C:III} we obtained a boundary for $\eta=-32/27$ in the KRZ geometries for the solution to correspond to a black hole (recall Eq. \eqref{eq:KRZrange}). Such a value does not saturate the EHT constraints on the shadows of M87 and Sgr A$^*$, meaning that if we push further the parameter $\eta$ beyond that point we end up with a horizonless compact object, i.e. a naked singularity. For completeness of our analysis we shall now consider a naked KRZ geometry with $\eta = -1.5 $ to see what  the similarities and differences are when we remove the event horizon of these parametrized solutions.

%

\begin{figure}[t!]
\includegraphics[width=8cm,height=5.2cm]{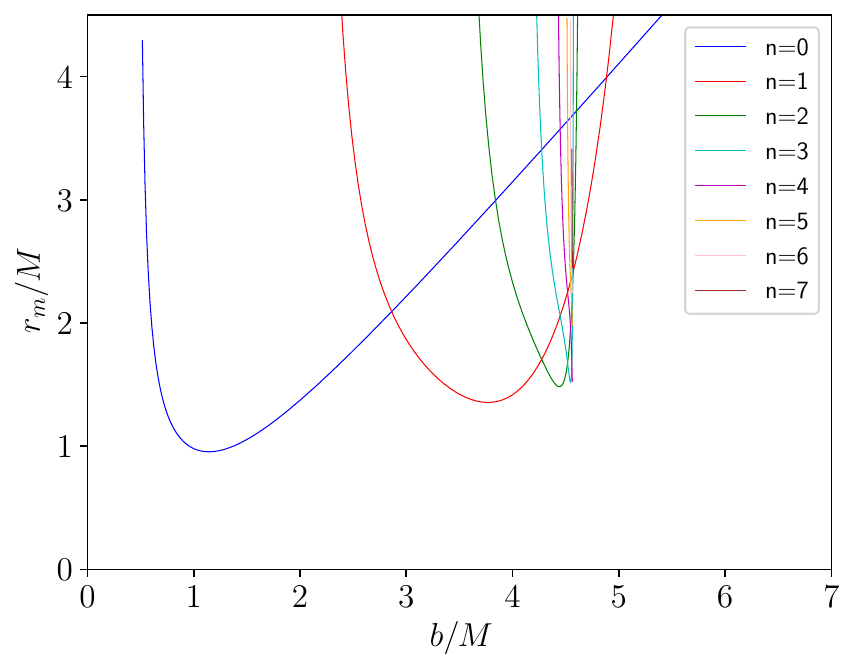} 
\caption{
The transfer function $r_n$ as a function of $b$ (in units of $M$) for our KRZ geometry with a naked singularity of Sec. \ref{S:VII}. We see the presence of eight curves corresponding to the $n=0,1,2,3,4,5,6,7$ contributions to the image. The inner part of each curve is not present in the transfer function of the black hole case, see Fig. \ref{fig:trans}, due to the presence of an event horizon there. Here, the inner part of these curves is associated with the replacement of an event horizon by a naked singularity whose effective potential repels every light ray at a finite distance from it.}
	\label{fig:trans_naked}
\end{figure}

In Fig. \ref{fig:rt_naked} we depict the result of the ray-tracing procedure for these horizonless configurations, both above (left figure) and below (right figure) the critical impact parameter $b_{c}$. Given the absence of an event horizon, every single light ray (for any value of the impact parameter $b$) gets to a closest distance from the central naked singularity before being deflected by it. As can be seen from  Fig. \ref{fig:trans_naked}, the transfer function for the $n=1$ and $n=2$ photon ring emissions is much larger and its slope sharper than on its black hole counterpart. Furthermore, there are two branches for each type of emission (compare to Fig. \ref{fig:trans}). Additionally, trajectories with higher $n$ are no longer negligible, suggesting the presence of multiple additional rings in the optical appearance of these objects not present in the black hole case discussed in Sec. \ref{C:V}. 

\begin{figure*}[t!]	\includegraphics[width=4cm,height=3.2cm]{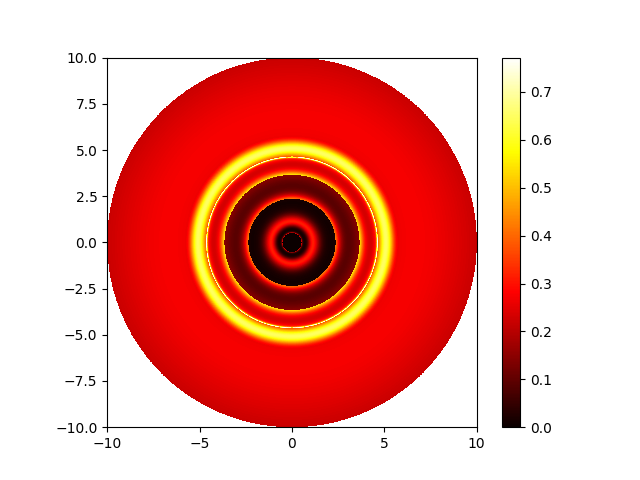}	\includegraphics[width=4cm,height=3.2cm]{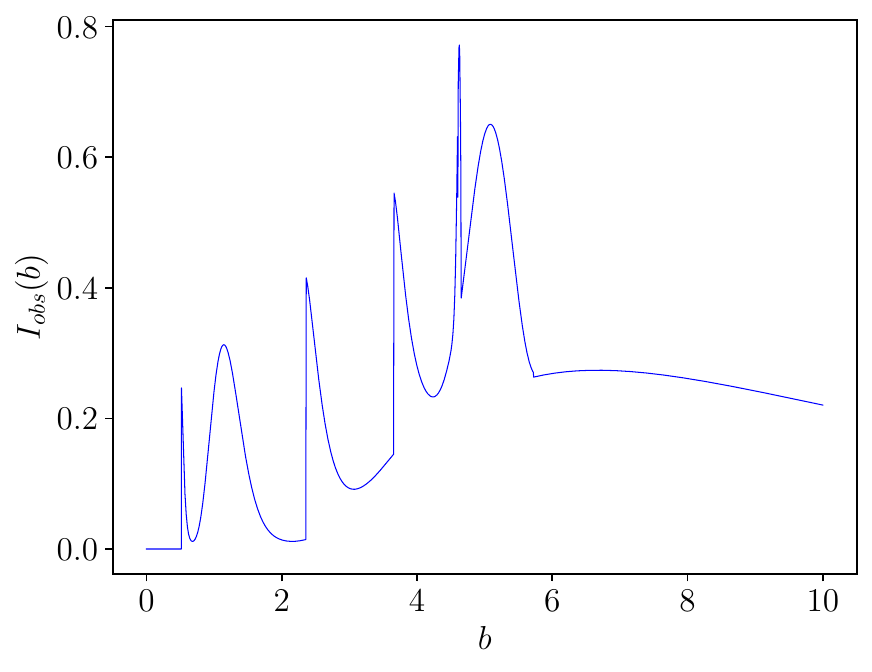} 	\includegraphics[width=4cm,height=3.2cm]{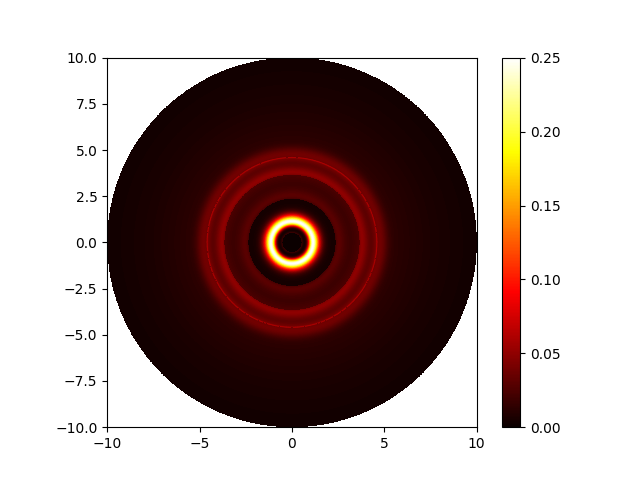}	
\includegraphics[width=4cm,height=3.2cm]{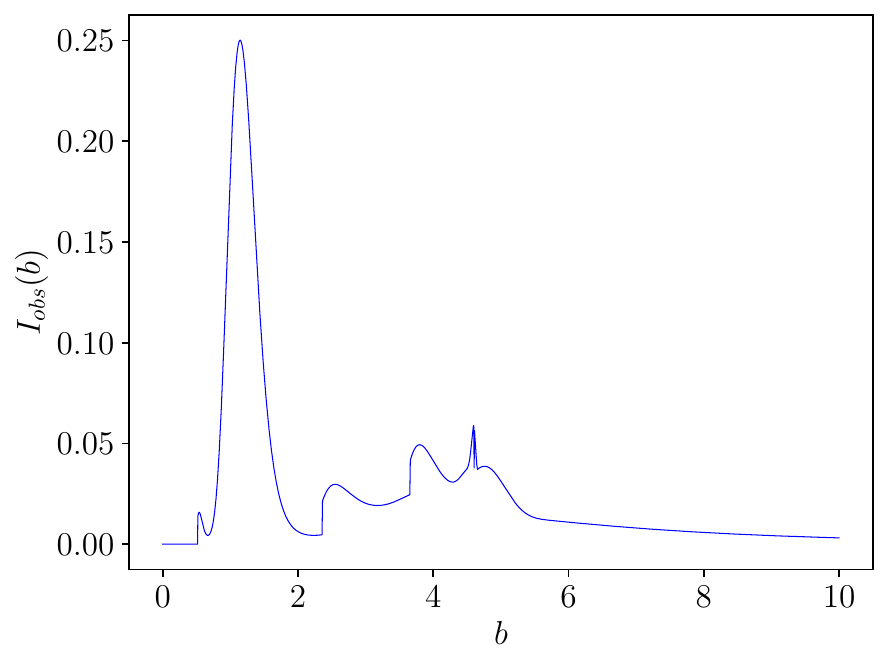}	\includegraphics[width=4cm,height=3.2cm]{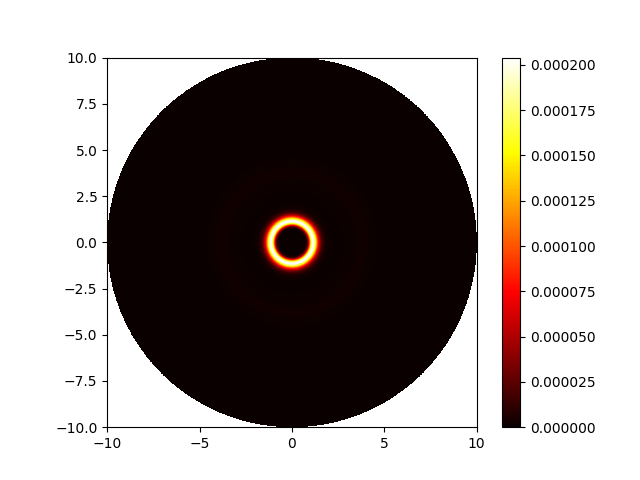}	\includegraphics[width=4cm,height=3.2cm]{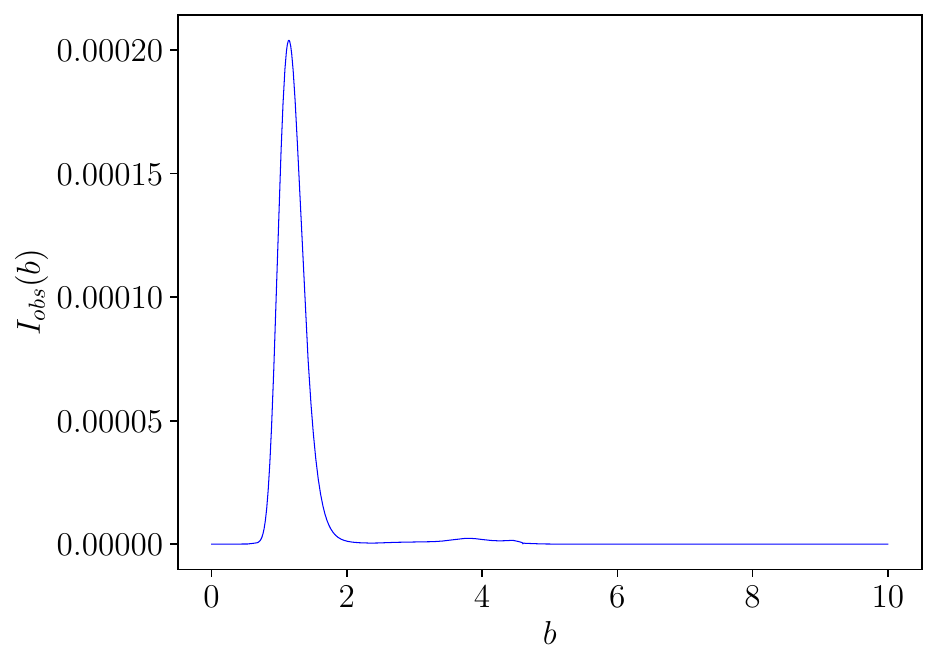} 	
\caption{The optical images and the observed intensity $I_{ob}(b)$ for SU1 (top left figures), SU2 (top right figures) and SU3 (bottom figures), 
for the KRZ solution with $\eta=-1.5$ representing a naked singularity.}
\label{fig:im_naked}
\end{figure*}

To confirm this expectation, in Fig. \ref{fig:im_naked} we display the optical appearance for these horizonless compact objects and a few of the SU models. As we can see, for every emission model there is at least one separate photon ring that may break into many others depending on the pick for the intensity profile. In this sense, for the SU1 model there are six clearly defined peaks of intensity, corresponding to a similar number of photon rings in the optical images. On the other hand, for the SU2 model there is a large peak of intensity for small impact parameter and five smaller and more spread peaks. And for the SU3 model there is only one peak of intensity, located at quite small values of the impact parameter. It is worth mentioning that, despite the absence of a horizon, this naked singularity object still features a central brightness depression in all SU cases corresponding to the inner region to the innermost photon ring. The size of such a shadow is, however, strongly reduced from the black hole case.

The bottom line of the analysis of these configurations is that, similarly to what happens in other horizonless objects studied in the literature, the multi-ring structure and strongly-reduced shadow size of KRZ naked singularities are hardly consistent with current observed images. This problem could be circumvented, though, by the recently noted fact that the extreme strong redshift present in the innermost regions of horizonless compact objects may act to significantly darken the additional photon rings so as to effectively create a shadow \cite{Chen:2024ibc}.

\section{Inclined black hole images} \label{S:VI}

\begin{figure*}
    \centering
    \includegraphics[width=0.23\linewidth]{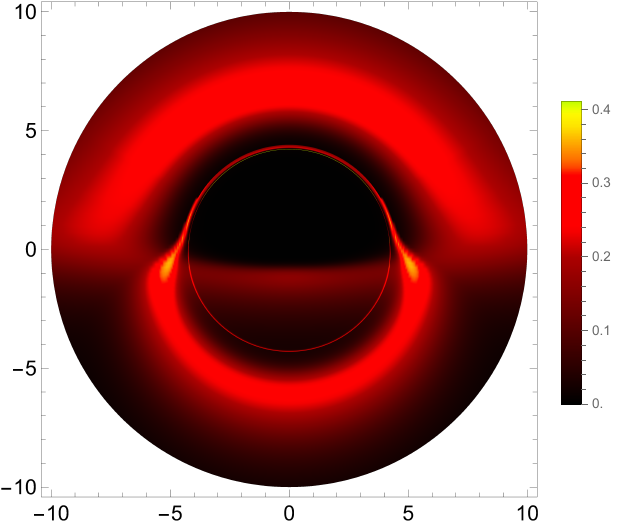}\quad
    \includegraphics[width=0.23\linewidth]{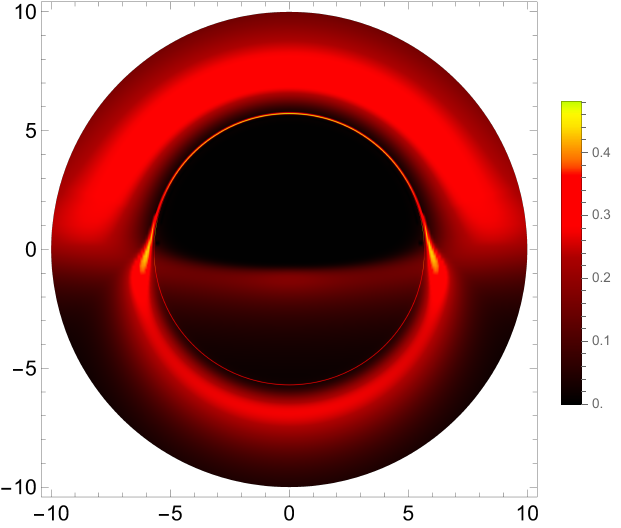}\quad
    \includegraphics[width=0.23\linewidth]{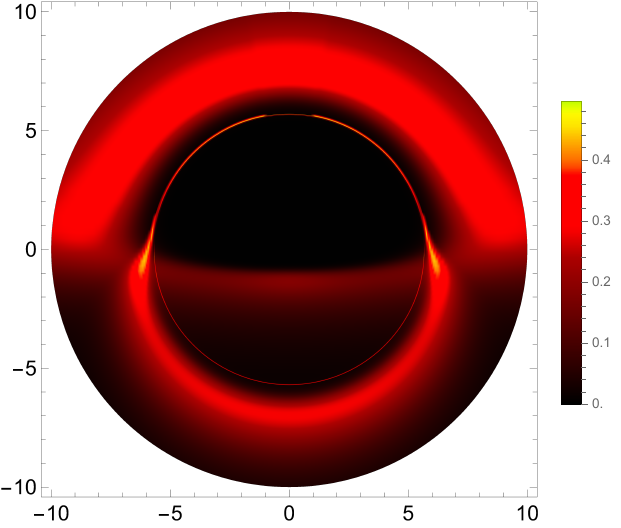}\quad
    \includegraphics[width=0.23\linewidth]{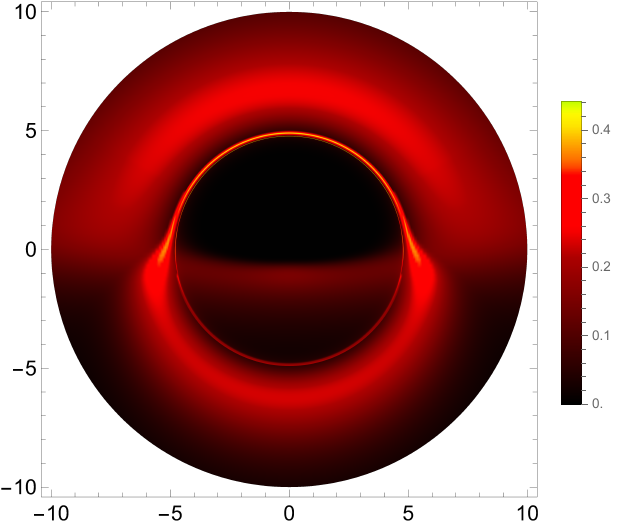}\\
    \includegraphics[width=0.23\linewidth]{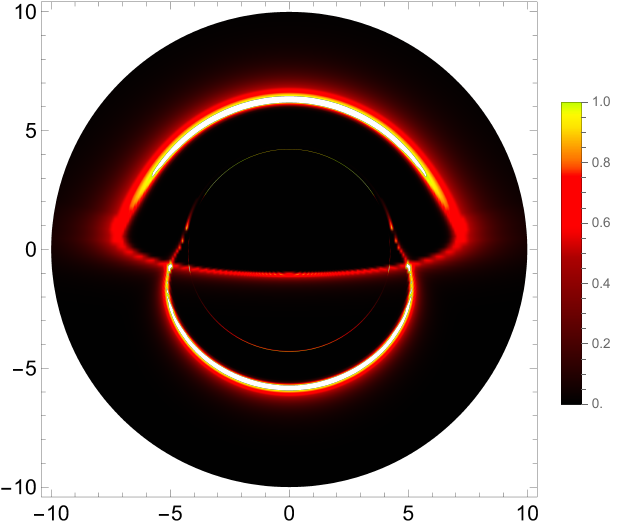}\quad
    \includegraphics[width=0.23\linewidth]{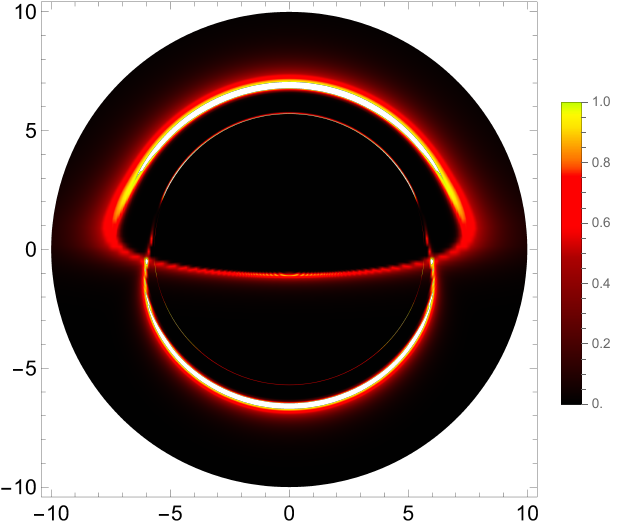}\quad
    \includegraphics[width=0.23\linewidth]{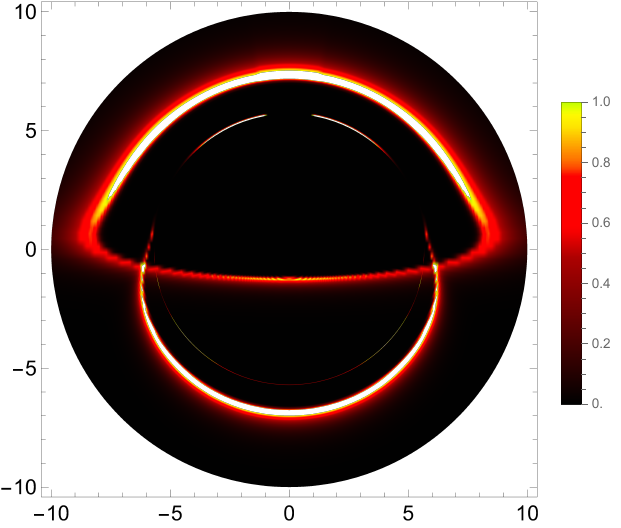}\quad
    \includegraphics[width=0.23\linewidth]{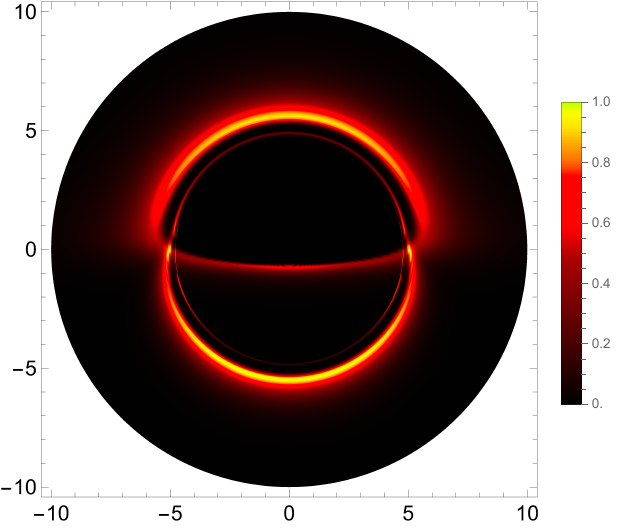}\\
    \includegraphics[width=0.23\linewidth]{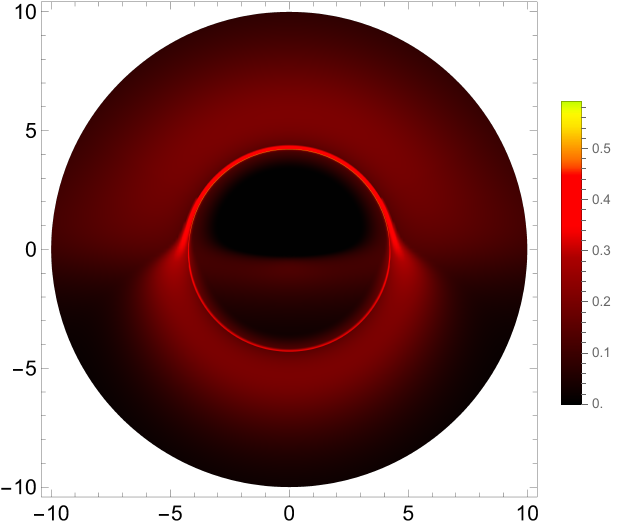}\quad
    \includegraphics[width=0.23\linewidth]{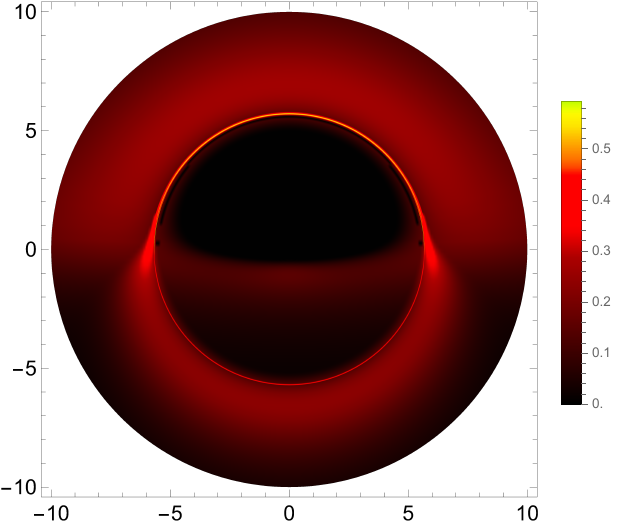}\quad
    \includegraphics[width=0.23\linewidth]{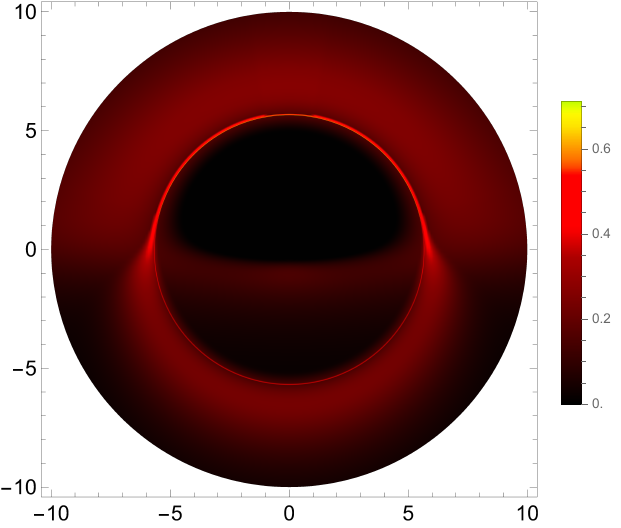}\quad
    \includegraphics[width=0.23\linewidth]{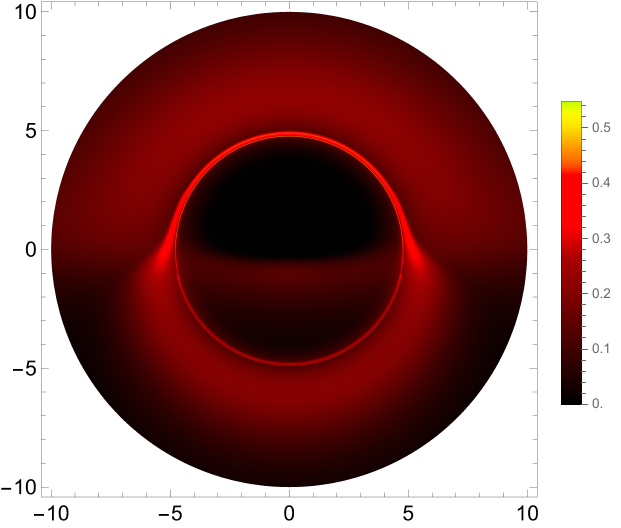}\\
    \includegraphics[width=0.23\linewidth]{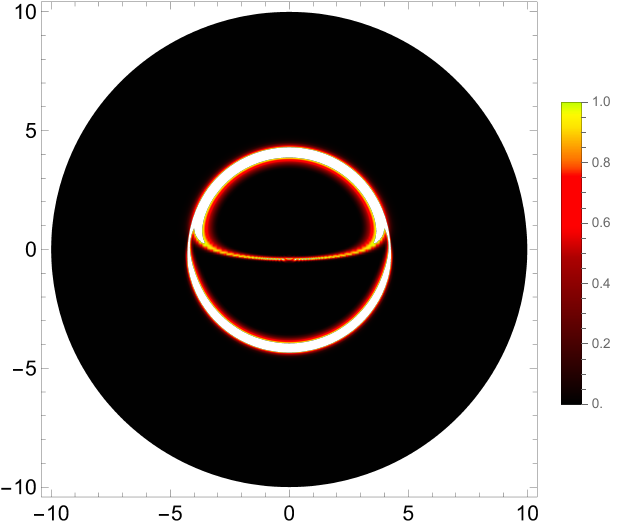}\quad
    \includegraphics[width=0.23\linewidth]{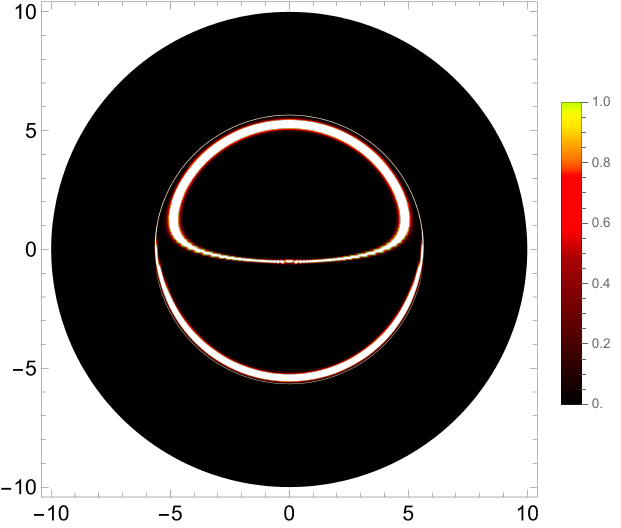}\quad
    \includegraphics[width=0.23\linewidth]{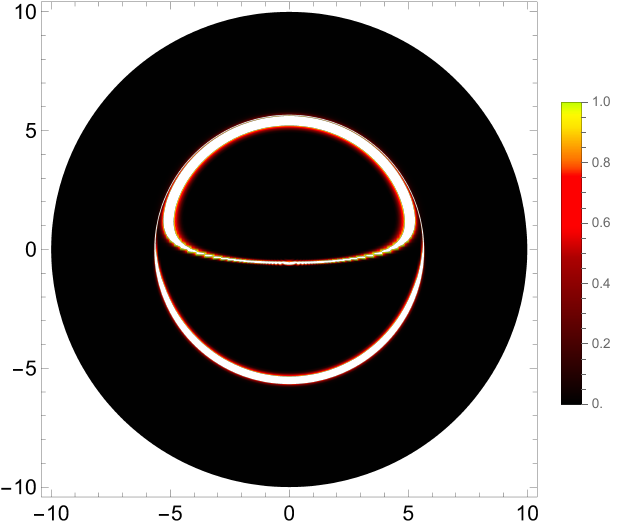}\quad
    \includegraphics[width=0.23\linewidth]{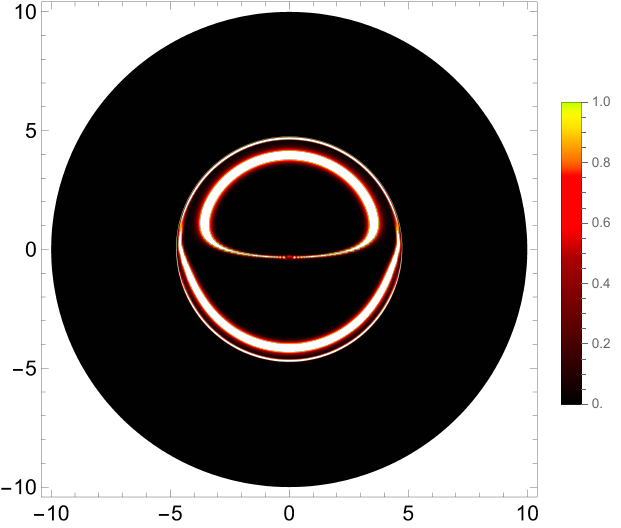}\\
    \includegraphics[width=0.23\linewidth]{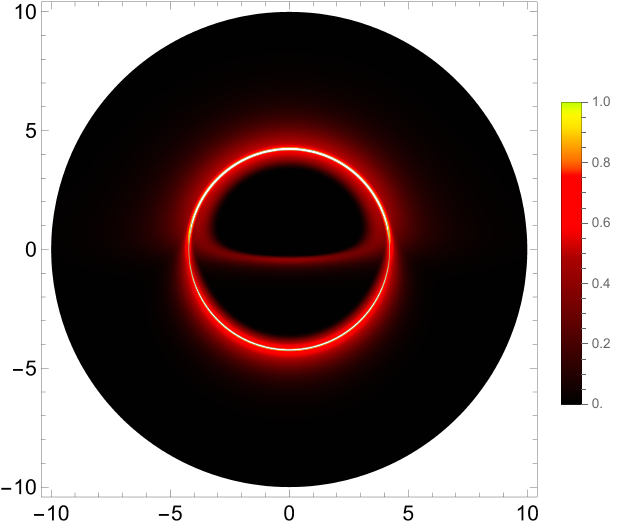}\quad
    \includegraphics[width=0.23\linewidth]{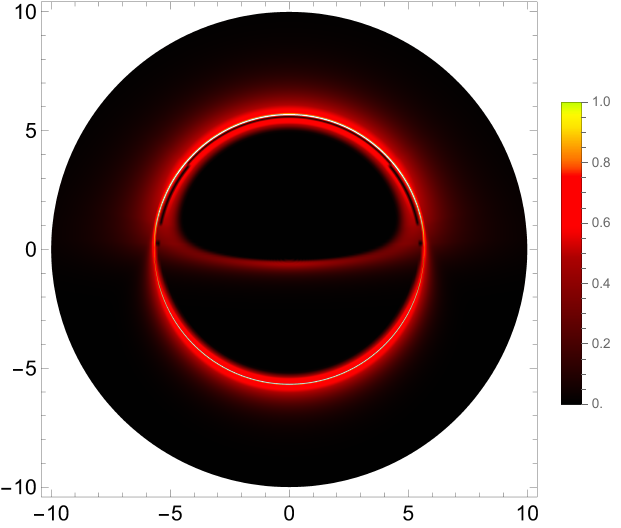}\quad
    \includegraphics[width=0.23\linewidth]{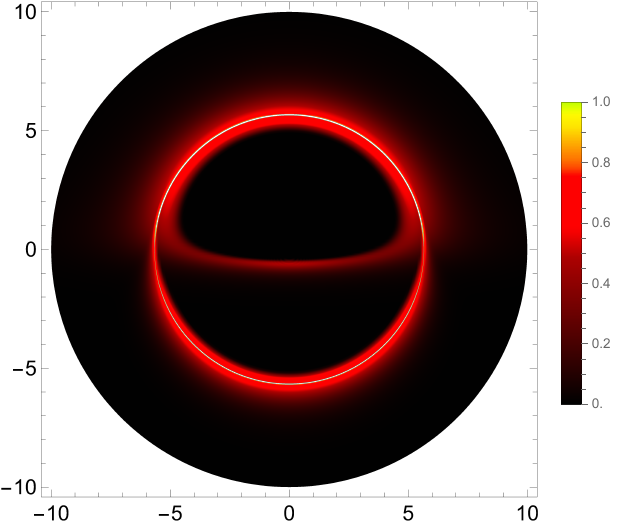}\quad
    \includegraphics[width=0.23\linewidth]{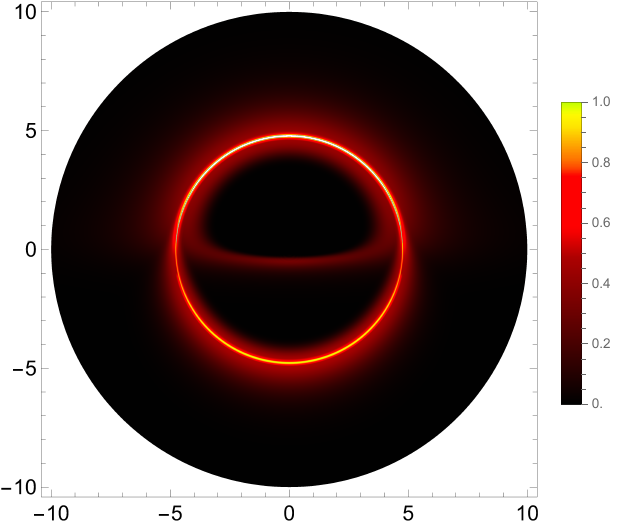}
    \caption{Optical images for the JPp (left column) JPn (middle left column), KRZp (middle right column), KRZn (right column), for the SU1 (top row) to the SU5 (bottom row) emission models, with an observation inclination of $80^\circ$ degrees.}
    \label{fig:inclinedone}
\end{figure*}

\begin{figure*}
    \centering
    \includegraphics[width=0.23\linewidth]{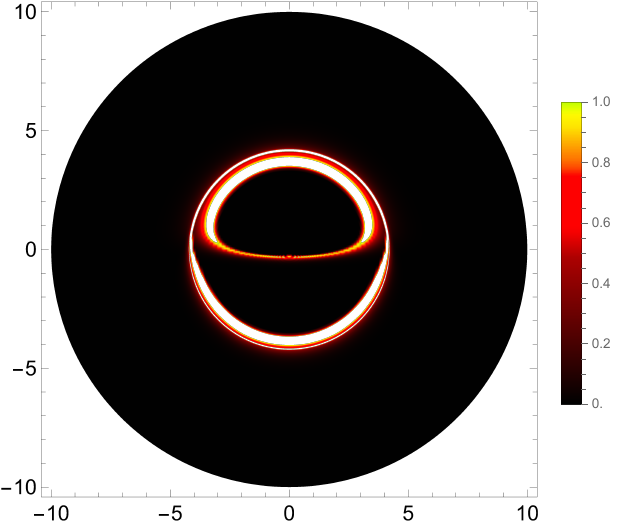}\quad
    \includegraphics[width=0.23\linewidth]{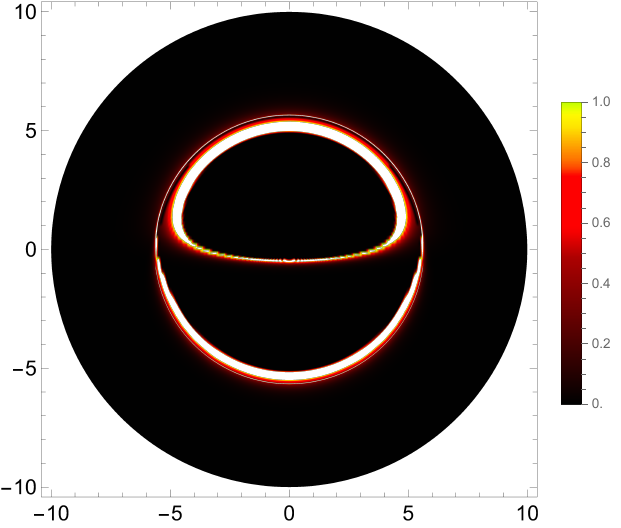}\quad
    \includegraphics[width=0.23\linewidth]{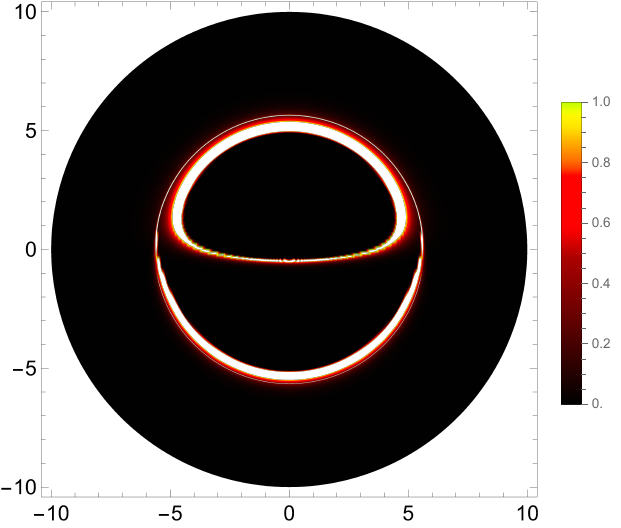}\quad
    \includegraphics[width=0.23\linewidth]{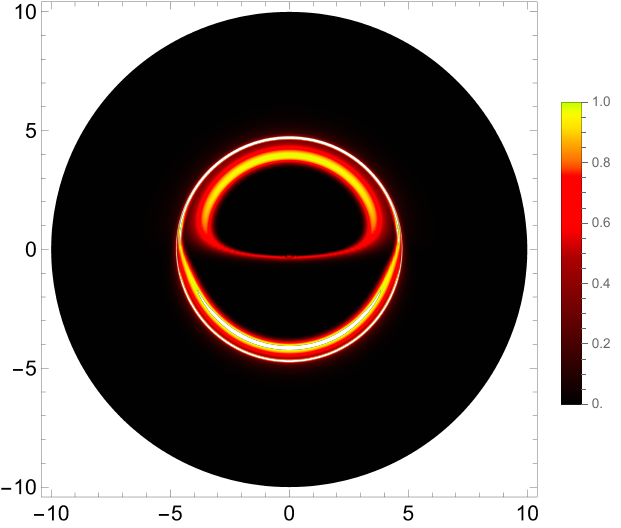}\\
    \includegraphics[width=0.23\linewidth]{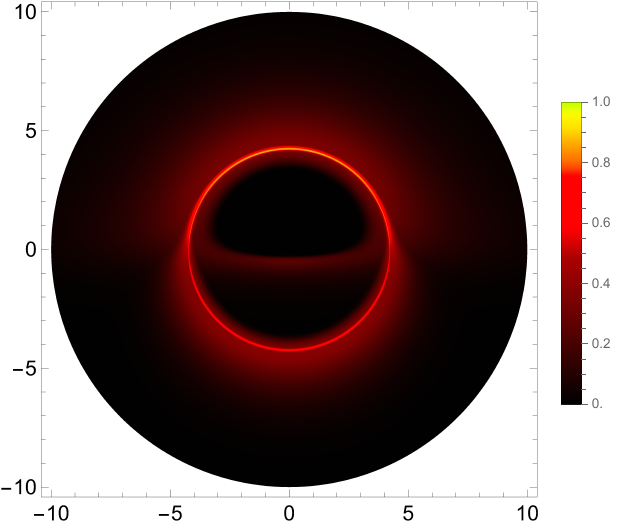}\quad
    \includegraphics[width=0.23\linewidth]{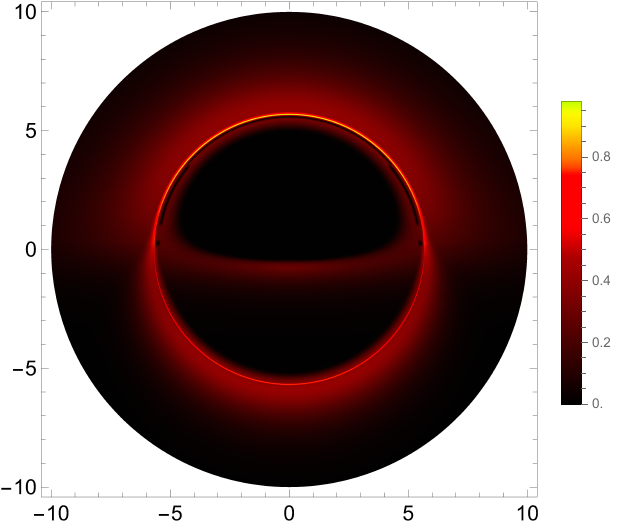}\quad
    \includegraphics[width=0.23\linewidth]{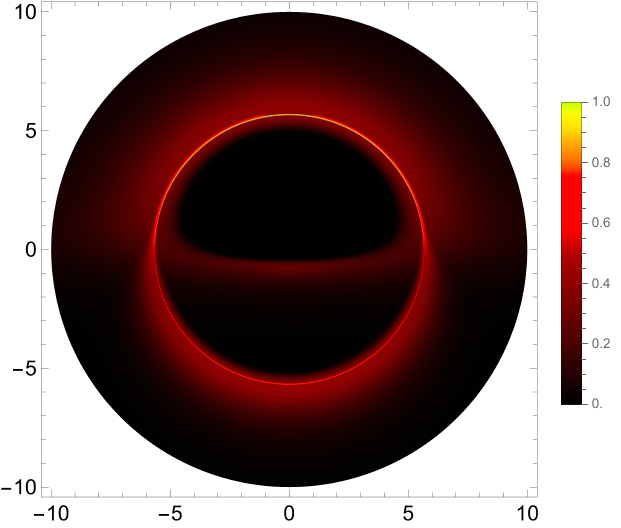}\quad
    \includegraphics[width=0.23\linewidth]{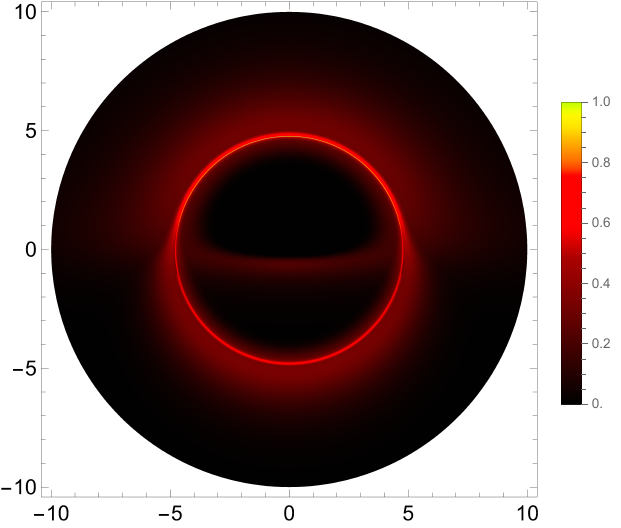}\\
    \includegraphics[width=0.23\linewidth]{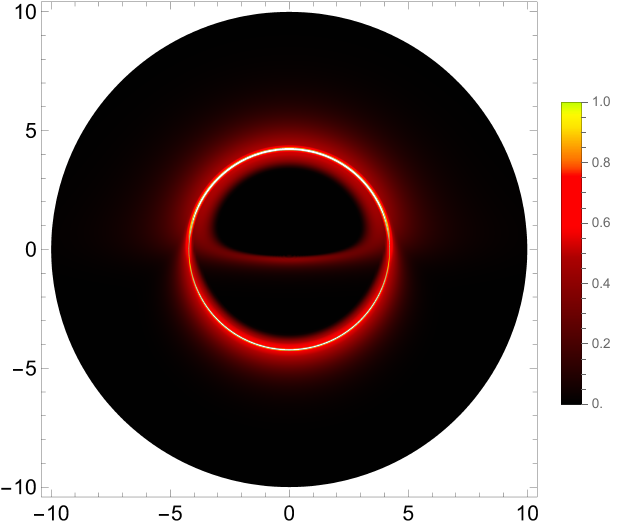}\quad
    \includegraphics[width=0.23\linewidth]{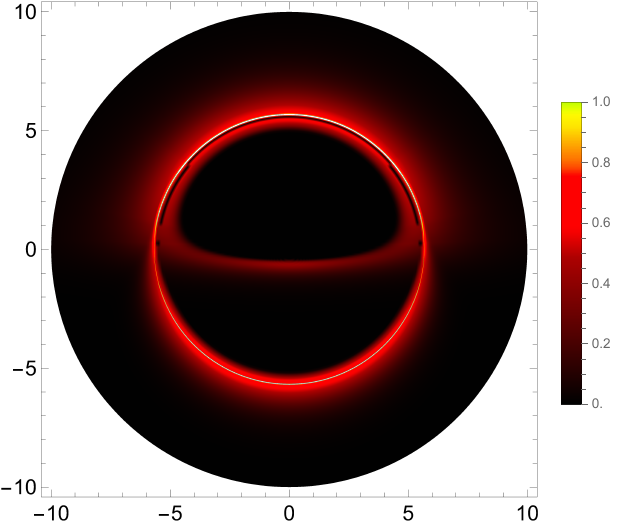}\quad
    \includegraphics[width=0.23\linewidth]{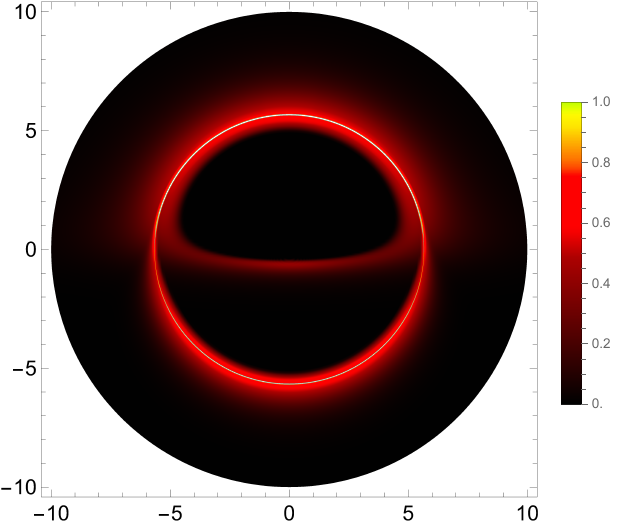}\quad
    \includegraphics[width=0.23\linewidth]{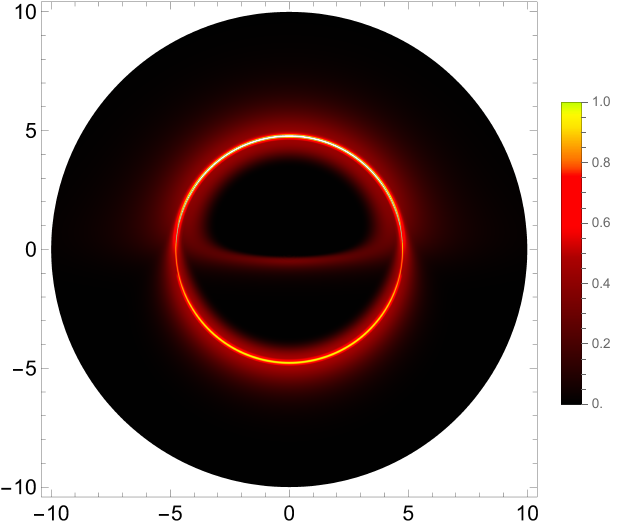}\\
    \includegraphics[width=0.23\linewidth]{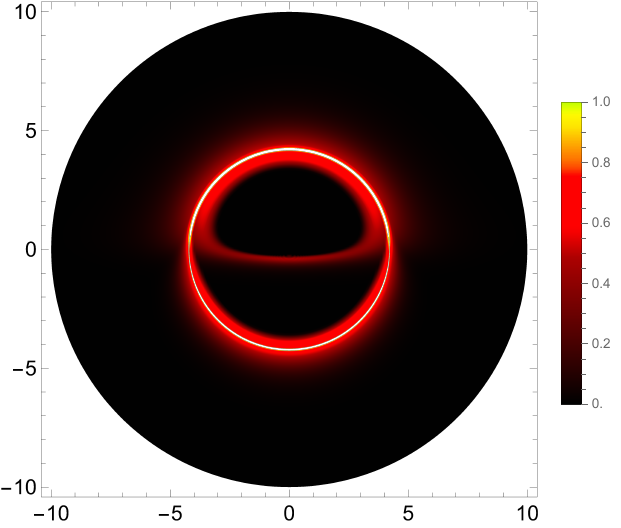}\quad
    \includegraphics[width=0.23\linewidth]{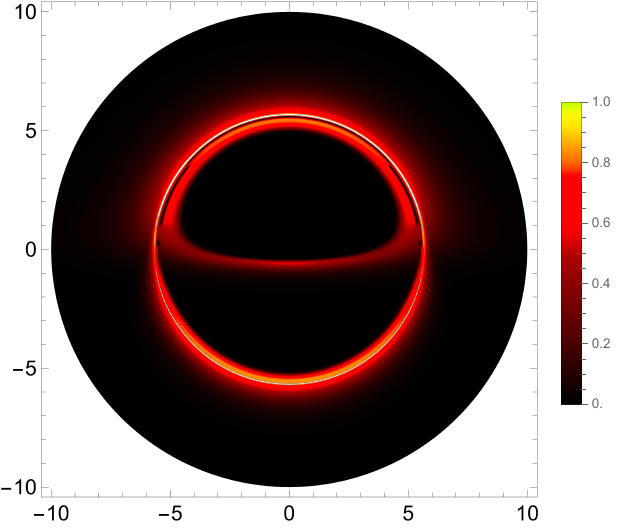}\quad
    \includegraphics[width=0.23\linewidth]{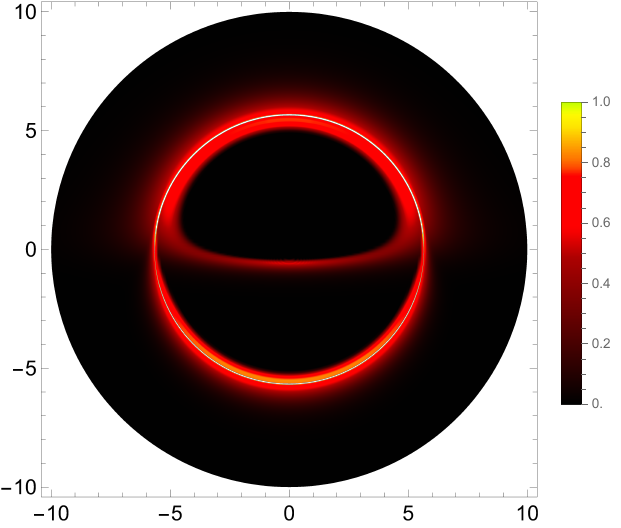}\quad
    \includegraphics[width=0.23\linewidth]{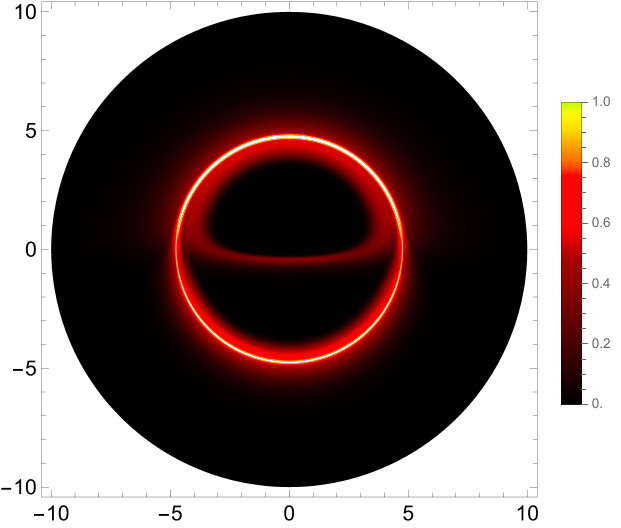}\\
    \includegraphics[width=0.23\linewidth]{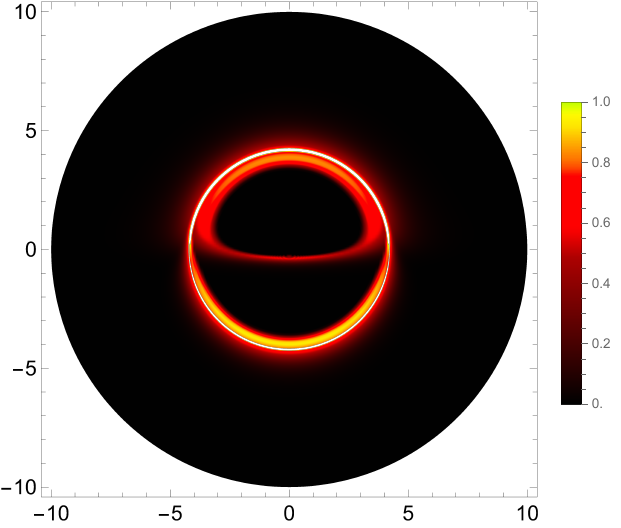}\quad
    \includegraphics[width=0.23\linewidth]{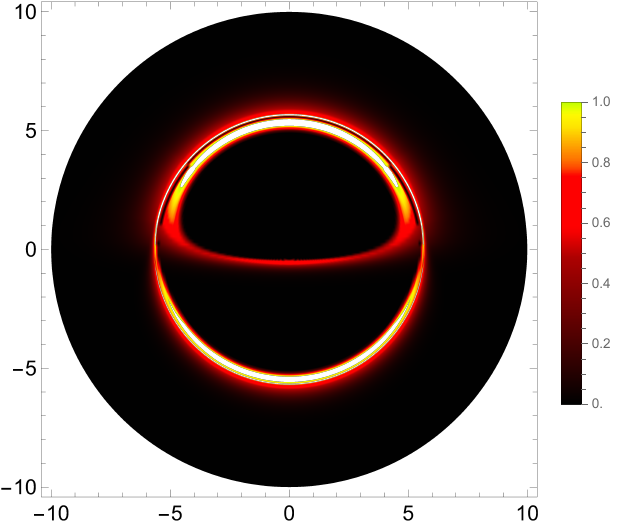}\quad
    \includegraphics[width=0.23\linewidth]{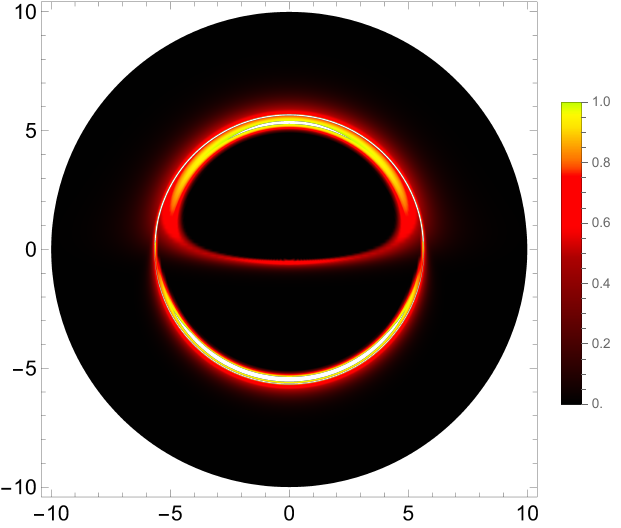}\quad
    \includegraphics[width=0.23\linewidth]{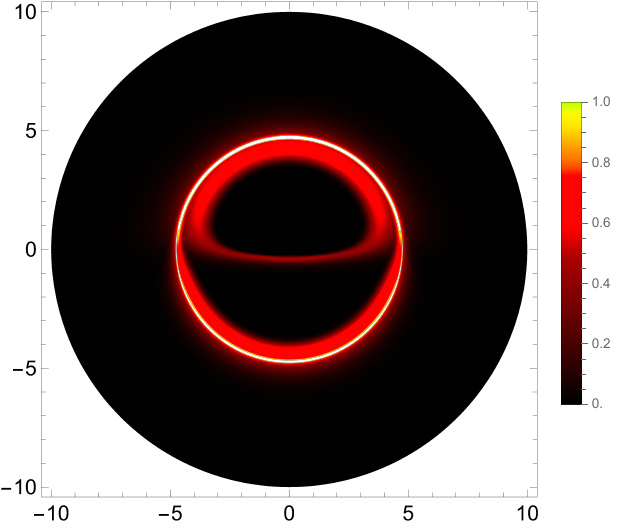}
    \caption{Optical images for the JPp (left column) JPn (middle left column), KRZp (middle right column), KRZn (right column), for the SU6 (top row) to the SU10 (bottom row) emission models, with an observation inclination of $80^\circ$ degrees.}
    \label{fig:inclinedtwo}
\end{figure*}

Our analysis of the images in the previous sections made use of face-on inclination between the axis of the disk and the observer's line of sight. However, current observed images of M87 and Sgr A$^*$ display some nonzero inclination (around $17$ degrees for M87) so it is worth considering the full picture of how parametrized black holes look like under different inclinations. In Figs. \ref{fig:inclinedone} and \ref{fig:inclinedtwo} we display inclined images at quite an extreme inclination of $80^\circ$ for JPn, JPp, KRZp, and KRZn solutions (left to right) for the models SU1 to SU5 (top to bottom) and for models SU6-SU10, respectively. These images serve to see how these four types of configuration can be distinguished in observations at very large inclinations, in agreement with our qualitative and quantitative analysis above for the face-on orientation. 

The same effects observed in face-on observations are still present in the inclined images. One can still label the JPn and KRZp solutions as ``enhancing" and the JPp and KRZn solutions as ``diminishing" the shadow, with the diminishing solutions presenting a wider and more intense $n=2$ light ring contribution. This result is in agreement with those found in other works \cite{Rosa:2023qcv,Rosa:2023hfm} where it was shown that, once an object is compact enough to develop unstable bound photon orbits, the observational properties of such object cease to be qualitatively influenced by inclination, e.g. the size of the shadow remains approximately constant, and the number of image contributions in the observer's screen remains unchanged, independently of the observation inclination angle. The same is true in the absence of an event horizon. It is worth mentionining that differences between the images for each SU models are remarkably more marked than in the face-on case, particularly when we move to the sequence of SU6-SU10 models belonging to the category of peaking intensity near the horizon, suggesting that inclined observations can reveal more details about the disk's intensity profile than their face-on counterparts (for a complementary analysis of this issue see \cite{Salehi:2024cim}).

\section{Conclusion} \label{C:VIII}

The current imaging of supermassive black holes by the EHT Collaboration, and the promises for the development of new technologies and strategies within very long baseline interferometry, has bolstered our confidence in the potential of black hole imaging to explore new regimes of the gravitational interaction beyond those tested so far. This new opportunity can be employed either to check the consistency of the Kerr solution of GR to accurately describe present and future observations of this kind, or to explore the compatibility of the theoretical images generated by any alternative black hole or horizonless compact objects with them, for instance, via the addition of matter fields or via those framed within the context of gravitational extensions of GR. 

To tackle this challenge, the best tool at our disposal is the output of both GRMHD simulations and semi-analytical generation of images which consistently and systematically report the presence of a bright ring of radiation (the photon ring) surrounding a central brightness depression (the shadow). Both features can be accounted for using our current theoretical framework: the photon ring is associated to nearly-bound unstable photon orbits, while the shadow is related to the brightness deficit caused by the shorter paths of light trajectories that terminate on the black hole horizon. Nonetheless, there are several uncertainties in the optical, geometrical, and emission features characterizing the accretion disk surrounding the shadow caster which play a major role in the generation of the observed images and the features of their photon ring and shadow, posing a challenge to properly make use of this opportunity.

In this work, instead of considering specific, hard-won solutions of a given theory of gravity plus matter fields, we have followed another approach relying on solutions parametrized by an arbitrarily large number of coefficients, which may be engendered in different such theories but without specifying them, later to be re-constructed using a reverse-engineering procedure. This way, these parametrized black holes can be used as a proxy for the features of wide families of solutions when compared with shadow observations and the constraints derived upon their parameters. Following this trail of thought, we have considered two such parametrized black holes, corresponding to the Johanssen-Psaltis and  Konoplya-Zhidenko solutions. After imposing asymptotic flatness and compatibility with solar system observations, we have used the recent correlation raised by the EHT between the size of the bright ring of radiation and the central brightness depression in black hole images to constrain the next non-trivial coefficient in these two parametrizations. Subsequently, we have taken two solutions on each parametrized black hole, which enhance and diminish such a size to its maximum and minimum bounds, respectively. This way, we have generated four solutions that have the greatest chance of producing large enough differences with respect to the Schwarzschild solution. 

For our simulations to generate images, we ran a ray-tracing procedure and employed an optically-thin (up to the $n=2$ photon ring) and infinitesimally-thin accretion disk with a monochromatic emission in the reference frame of the disk, and with an intensity profile given by ten picks for the coefficients of Johnson's Standard Unbound profile, the latter previously employed in the literature to reproduce the results of certain scenarios of GRMHD simulations. We successfully reproduced the bright ring of radiation and central brightness depression features for all images. Interpreting the resulting images, we conclude that the way the photon rings are weaved together with the direct emission of the disk is mostly driven by the accretion disk, something already spotted in many works in the field. To clarify this point further, we split the emission models into three categories according to whether the peak of emission is located outside the photon sphere, in the intermediate region between event horizon and photon sphere, or close to the event horizon. This aspect seems to be the main ingredient in generating the optical appearance of each black hole, with the spread of the emission profile outwards playing a minor (but non-negligible) role. 

We subsequently discussed the structure of the $n=1$ and $n=2$ photon rings for all SU models and the four background geometries. Based both in terms of their qualitative structure, location, as well as their respective luminosity ratio, the latter using an approximation of the Lyapunov exponent of nearly bound geodesics, we concluded there are moderate differences (up to roughly a factor two in the relative luminosity of photon rings) between enhancing and diminishing solutions, and also between each of them and the Schwarzschild solution. This way, and considering this method alone (though alternative ones do exist in the literature) these geometries could be potentially distinguishable via VLBI imaging of the features of their photon rings if the hopes placed in future technological developments are met\footnote{An overall discussion on the chances to using photon rings in order to test beyond-GR black hole physics can be found at \cite{Staelens:2023jgr}.}. 

Opportunities within this field are currently present on many fronts. Among them we underline the development of a shadowgraphy of every possible shape of the effective potential for a compact object, a better understanding of the accretion disk physics in observationally accessible scenarios, further methods to analyze and discuss the features of photon rings and central brightness depression, and comparison of images of modified black holes and horizonless compact objects. And last but not least, we underline the development of strong-field multimessenger tests of compact objects using both gravitational waves and VLBI imaging, using the correspondence between these two phenomena uncovered in the last few years \cite{Pedrotti:2024znu,Pedrotti:2025upg}. We hope to further report on some of these topics in more detail soon. \\

\section*{Acknowledgements}

This work is supported by the Spanish National
Grants PID2020-116567GB-C21, PID2022-138607NBI00, PID2023-149560NB-C21, CNS2024-154444, and CEX2023-001292-S, funded by MICIU/AEI/10.13039/501100011033 (``ERDF A way of making Europe" and ``PGC
Generaci\'on de Conocimiento") and FEDER, UE.  The authors would like to acknowledge the contribution of the COST Action CA23130 (``Bridging high and low energies in search of quantum gravity (BridgeQG)").

\end{document}